\def\papermode{preprint}
\providecommand{\papermode}{submit}
\newif\iftwocol
\def\pmpreprint{preprint}\edef\pmgiven{\papermode}
\ifx\pmpreprint\pmgiven \twocoltrue \fi
\iftwocol
  \documentclass[twocolumn]{aastex701}
\else
  \documentclass{aastex701}
\fi

\usepackage{amsmath}
\iftwocol \newcommand{\colfigheight}{\textheight} \else \newcommand{\colfigheight}{0.42\textheight} \fi
\usepackage{xcolor}
\usepackage{booktabs}

\ifdefined\acknowledgments\else
  \newenvironment{acknowledgments}{\section*{Acknowledgments}}{}
\fi
\providecommand{\software}[1]{\par\medskip\noindent\textit{Software:} #1\par}

\ifdefined\theoremstyle\else
  \ifdefined\openbox\let\openbox\relax\fi
  \ifdefined\proof\let\proof\relax\fi
  \usepackage{amsthm}
\fi
\ifdefined\theorem\else
  \theoremstyle{plain}
  \newtheorem{theorem}{Theorem}[section]

  \theoremstyle{definition}

  \theoremstyle{remark}
  
\fi

\providecommand{\R}{\mathbb{R}}
\providecommand{\C}{\mathbb{C}}
\providecommand{\Z}{\mathbb{Z}}
\providecommand{\N}{\mathbb{N}}
\providecommand{\Q}{\mathbb{Q}}
\providecommand{\E}{\mathbb{E}}

\providecommand{\Msun}{\ensuremath{M_\odot}}
\providecommand{\Rsun}{\ensuremath{R_\odot}}

\providecommand{\Mearth}{\ensuremath{M_\oplus}}
\providecommand{\Rearth}{\ensuremath{R_\oplus}}

\providecommand{\kms}{\ensuremath{\mathrm{km\,s^{-1}}}}

\providecommand{\apjl}{ApJL}
\providecommand{\apjs}{ApJS}
\providecommand{\ao}{Appl.\ Opt.}

\providecommand{\aap}{A\&A}

\providecommand{\tabpar}[2]{\parbox[t]{#1}{\raggedright #2\strut}}

\iftwocol
\else
\fi

\shorttitle{Shaderous}
\shortauthors{Guillochon}

\begin{document}

\title{Shaderous: A Browser-based Hydrodynamics Code for Physics Research and Teaching}

\author[0000-0002-9809-8215]{James Guillochon}
\affiliation{Independent Researcher}
\email{guillochon@gmail.com}

\begin{abstract}
Shaderous is a three-dimensional code for compressible gas dynamics and magnetohydrodynamics that runs in WebGPU compute shaders in a web browser on any vendor's GPU. It combines an unsplit MUSCL-Hancock scheme and HLLC fluxes with a multigrid Poisson solver, block-structured adaptive mesh refinement with per-block timesteps, constrained transport with HLLD fluxes, point masses with sinks, gray radiation transport with the M1 closure, implicit pressure for low Mach numbers, tabulated equations of state from degenerate electrons to rock and iron, cooling, viscosity, conduction, resistivity, mixtures of gases, and nuclear burning by a neural network trained on a 13-isotope reaction network. A setup is shared as a link and runs on a student's own laptop or phone, so the code suits teaching at every level. A run is limited to one GPU: the largest adaptive mesh so far held about 175 million cells on an NVIDIA H200. It therefore suits parameter studies, run as sweeps from the command line, rather than the largest single simulations. Three aspects of the design are, as far as we can find, new. First, the hydrodynamic solver stores no face flux. The two cells that share a face compute bit-identical fluxes, so cells at boundaries between levels and between timesteps recompute their neighbors' fluxes instead of keeping a flux register. Mass, momentum and energy are conserved to round-off on the adaptive mesh, and compensated sums keep the divergence of the magnetic field at round-off. Second, the whole step runs on the GPU: the timestep and the assignment of blocks to time bins are reduced there, and ghost cells are computed as they are loaded rather than exchanged. Third, the code computes in single precision yet reproduces a range of astrophysical systems accurately, from stars and atmospheres in hydrostatic balance to shocks, gravitational collapse and magnetized flows, in 291 validation tests.
\end{abstract}


\section{Introduction}
\label{sec:intro}

Grid codes for astrophysical gas dynamics, such as FLASH \citep{Fryxell2000}, RAMSES \citep{Teyssier2002} and Athena++ \citep{Stone2020}, are large native programs. A user builds them with a compiler, an MPI library and, on graphics processors, a vendor toolchain, then runs them on a workstation or a cluster and inspects the output afterward. That workflow suits production science. It is a poor fit for teaching, for exploring a problem before committing to a production run, and for handing a colleague a result that they can reproduce and modify in a minute.

Graphics processors entered numerical gas dynamics through their graphics interfaces. Before CUDA, solvers were written as fragment programs that read and wrote textures: linear algebra and multigrid \citep{Bolz2003, Goodnight2003, Kruger2003}, incompressible cloud dynamics \citep{Harris2003}, and the compressible Euler equations with high-resolution finite-volume schemes in two and three dimensions \citep{Hagen2006}. \citet{Owens2007} surveyed that period. General-purpose toolchains then replaced the graphics route, and astrophysics gained a generation of GPU codes, among them GAMER \citep{Schive2010, Schive2018}, the adaptive solver of \citet{Wang2010}, the Euler solver of \citet{Kestener2010}, Cholla \citep{Schneider2015}, FARGO3D \citep{BenitezLlambay2016}, Astaroth \citep{Pekkila2017}, K-Athena \citep{Grete2021}, Quokka \citep{Wibking2022}, Idefix \citep{Lesur2023}, Kratos \citep{Wang2025}, AthenaK \citep{Stone2026}, AGILE \citep{Porth2026} and RAMSES-GPU \citep{Teyssier2026}. These codes are fast and scale to many devices, but each is a native application built with CUDA, HIP or a portability layer.

WebGPU, a graphics and compute interface that browsers now ship, brings compute shaders to every desktop and many phones without an installation step. Its shading language, WGSL, compiles to whatever the local driver accepts. Simulators that use the GPU from a browser exist for reaction-diffusion and cardiac models \citep{Kaboudian2019}, for the Gross-Pitaevskii equation \citep{Baggaley2026}, and for plasma turbulence: \citet{Mallet2026} wrote a WebGPU solver for reduced MHD that runs in the browser, a port of their pseudospectral code taranis.\footnote{\url{https://alfredmallet.github.io/taranis/webgpu/rmhd2d.html}} We found none that solves the compressible Euler or MHD equations with shocks, self-gravity or adaptive refinement (Section~\ref{sec:prior}).

This paper describes Shaderous, a code that does. All fluid and gravity computation runs in WGSL compute shaders, and the code runs at a web address. Nothing in the scheme is specific to astrophysics. It solves the compressible Euler equations, or the equations of magnetohydrodynamics (MHD), ideal or resistive, with optional gravity, diffusion, turbulence driving, heating, cooling, gray radiation transport and nuclear burning. An implicit treatment of the pressure lets it follow nearly incompressible flow, such as a liquid, without a timestep set by the speed of sound. The code serves the same purpose in a course on fluid dynamics or for a problem in another field. We demonstrate it on astrophysical problems because they exercise every part of the code at once: strong shocks, self-gravity, stars that must stay in hydrostatic balance, point masses that accrete, magnetized disks, detonations, and equations of state as far from an ideal gas as degenerate electrons, dissociating molecular hydrogen and molten rock.

The physics is conventional and we credit its sources throughout. What is new is how the scheme is mapped onto an interface designed for graphics. WebGPU forbids a pass from reading and writing the same four-channel floating-point texture, has no double precision, limits the number of storage bindings, and returns data to the CPU only asynchronously. We turned each restriction into a design choice. The result is a code that conserves to round-off on an adaptive mesh without storing a single hydrodynamic face flux, that never exchanges ghost cells, that keeps every step on the GPU, and that keeps hydrostatic objects in balance and magnetic fields free of divergence in single precision.

The code is also built to be driven by AI agents. A programming interface and a server for the Model Context Protocol expose the operations of the page, and every run is defined by a setup document that holds only numbers and names. A user can therefore describe a simulation in words to an agent, which writes the setup, checks it against the code's own rules, runs it and reads back the results (Section~\ref{sec:agents}).

Section~\ref{sec:prior} compares Shaderous with earlier work. Section~\ref{sec:webgpu} describes the constraints of WebGPU and the data layout that follows from them. Section~\ref{sec:hydro} gives the hydrodynamic scheme, including the equations of state, mixtures, nuclear burning and the implicit pressure for low Mach numbers (Section~\ref{sec:implicit}). Section~\ref{sec:conservation} gives the flux evaluation that conserves without stored fluxes, Section~\ref{sec:amr} the adaptive mesh and its local timesteps, Section~\ref{sec:gravity} gravity, point masses and sinks, Section~\ref{sec:mhd} magnetic fields, and Section~\ref{sec:radiation} radiation transport. Section~\ref{sec:interface} describes the interface, the problems, the setup files that define them, the file format, the interfaces for scripts, and how an agent builds a setup from a description. Section~\ref{sec:education} covers the uses of the code in teaching, Section~\ref{sec:validation} the validation suite, and Section~\ref{sec:performance} performance. Section~\ref{sec:discussion} discusses limitations.

\section{Relation to earlier work}
\label{sec:prior}

We searched for earlier codes in three literatures: general-purpose computation in graphics shaders, GPU codes for astrophysical fluids, and scientific simulators that run in a browser.

The shader-era solvers are the closest in spirit. \citet{Hagen2006} solved the compressible Euler equations on uniform grids in fragment programs, and \citet{Goodnight2003} and \citet{Bolz2003} ran multigrid in the same way. Adaptive data structures in textures also came from computer graphics: \citet{Lefohn2006} built page tables and octrees in textures for random access on the GPU. Shaderous uses compute shaders rather than fragment programs, but it inherits the same model of state held in textures and updated by one invocation per cell. It adds what those solvers did not have: self-gravity, adaptive refinement with conservative coupling between levels, local timesteps, and the physics of astrophysical problems.

Among GPU codes for astrophysics, GAMER \citep{Schive2010, Schive2018} is the closest in structure: it refines in patches of a fixed size, as Shaderous does. The adaptive solver of \citet{Wang2010} instead runs on the grids of arbitrary size of the Enzo framework. GAMER follows \citet{Berger1989} in correcting the coarse side of a refinement boundary with fluxes stored from the fine side. The hydrodynamic solver of Shaderous stores no fluxes at all (Section~\ref{sec:conservation}).

Several of these codes also solve ideal MHD with the constrained transport of \citet{Evans1988}, among them K-Athena, Idefix and AthenaK, which follow the unsplit integrators of Athena \citep{GardinerStone2005, Stone2008, StoneGardiner2009}. The MHD solver of Shaderous uses the same integrator (Section~\ref{sec:mhd}), and like those codes it stores its face fluxes and keeps a flux register at refinement boundaries.

The browser simulators of \citet{Kaboudian2019} and \citet{Baggaley2026} show that interactive computing on the GPU through WebGL and WebGPU is practical. They solve reaction-diffusion and Schr\"odinger-type equations on uniform grids, where no conservation law has to hold across a shock or a refinement boundary.

The closest earlier code is the WebGPU solver for reduced MHD of \citet{Mallet2026}, which ports the pseudospectral algorithms of their JAX code taranis to the browser. Like Shaderous, it runs in WebGPU compute shaders in single precision, and it solves a magnetized fluid in two and three dimensions. Reduced MHD assumes a strong guide field and a nearly incompressible plasma, so shocks cannot form, and the pseudospectral method needs a uniform periodic grid. Within those limits, it follows Alfv\'enic turbulence with little numerical dissipation and offers tools that Shaderous lacks: energy spectra and Els\"asser variables. Three parts of Shaderous follow that page: its random forcing in a band of wavenumbers, its driven-turbulence problems and several of its magnetized test problems (colliding Alfv\'en waves, Els\"asser packets, tearing and island coalescence). Shaderous solves the full compressible equations with shocks, gravity and adaptive refinement, but its Godunov scheme dissipates more at a given resolution.

\section{Constraints of WebGPU and the data layout}
\label{sec:webgpu}

Four properties of WebGPU shaped the code. Each is a property of the interface, not of the hardware.

\emph{No read-write access to four-channel floating-point textures.} A compute pass can write an \texttt{rgba32float} texture only as a write-only storage texture. Every update therefore reads one copy of the state and writes another. Single-channel \texttt{r32float} textures can be read and written in one pass, which the red-black smoother of the Poisson solver uses.

\emph{Limited storage buffers.} The default limits allow few storage buffers per shader stage and cap a single binding at 128~MB. We hold the state in 3D textures instead: density and momentum in one \texttt{rgba32float} texture, and total energy, internal energy and a passive scalar in a second. We measured the same flux stencil on storage buffers and found it about 1.4 times slower than on textures.

\emph{No double precision.} WGSL has no 64-bit floating-point type. Every quantity is a 32-bit float, and the scheme has to hold hydrostatic balance, track internal energy in fast flows, and keep the magnetic field free of divergence at that precision (Sections~\ref{sec:robustness}, \ref{sec:wellbalanced} and~\ref{sec:ct}).

\emph{Asynchronous readback.} Data return to the CPU only through a mapped buffer, some frames later. A code that reads the timestep back before every step would stall. Shaderous computes the timestep on the GPU, writes it to a small buffer, and lets the following passes read it there. The CPU reads diagnostics and refinement indicators without waiting for them.

WGSL also has no dynamic dispatch. A small preprocessor assembles each shader from modules: the equation of state, the Riemann solver, the slope limiter, the cooling function and the diffusive fluxes in use are chosen when the shaders are built, and changing one rebuilds the affected pipelines. A module that is switched off maps to a stub, so species, cooling and diffusion cost nothing in a run that does not use them.

Reductions run in stages. The timestep, the diagnostics and the refinement indicators all use them. Each stage folds blocks of $4^3$ records from the stage before, so sums are formed hierarchically, which limits the round-off of single precision. Maxima of non-negative values are accumulated with an atomic maximum on their bit patterns, which order the same way as the floating-point values.

No browser reports how much memory its GPU has, and a phone that runs out kills the page before WebGPU reports an error. Shaderous therefore estimates a memory budget from what the browser does report: 2.5~GB for a discrete desktop GPU, 2~GB where the GPU shares the system's memory, and 1~GB on a phone. The user can change it. Before each large allocation the page writes a record to the browser's local storage. A record that is still there when the page next loads means that the page crashed holding that much memory, and the budget is then lowered below it for good. A run without a person at the page, on a server, takes its budget from the command line (Section~\ref{sec:interface}).

\section{Hydrodynamics}
\label{sec:hydro}

Shaderous solves the Euler equations with gravity,
\begin{equation}
\begin{aligned}
\partial_t \rho + \nabla\cdot(\rho\mathbf{v}) &= 0,\\
\partial_t (\rho\mathbf{v}) + \nabla\cdot(\rho\mathbf{v}\mathbf{v} + p\,\mathsf{I}) &= -\rho\nabla\Phi,\\
\partial_t E + \nabla\cdot\bigl[(E + p)\mathbf{v}\bigr] &= -\rho\mathbf{v}\cdot\nabla\Phi + \mathcal{H} - \mathcal{C},
\end{aligned}
\label{eq:euler}
\end{equation}
where $\rho$ is the density, $\mathbf{v}$ the velocity, $p$ the pressure, $E$ the total energy density, $\Phi$ the gravitational potential, $\mathsf{I}$ the identity, and $\mathcal{H}$ and $\mathcal{C}$ the heating and cooling rates per unit volume. A passive scalar is advected with the gas, and the internal energy density $e$ is evolved alongside $E$ (Section~\ref{sec:robustness}).

\subsection{Scheme}
\label{sec:scheme}

\begin{figure*}[t]
\centering
\includegraphics[width=\textwidth]{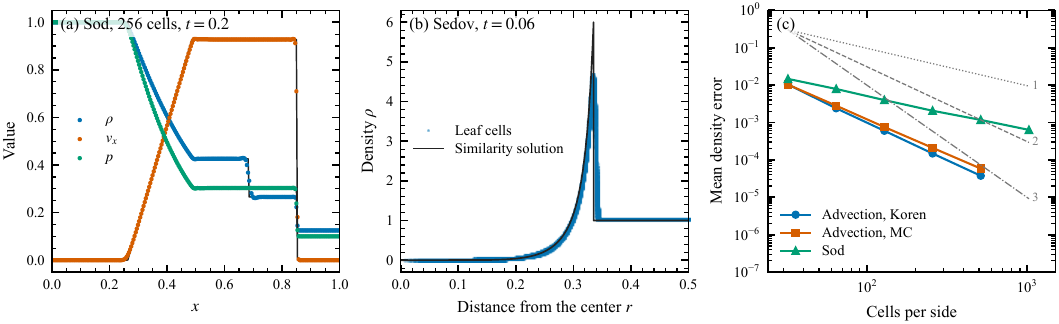}
\caption{Accuracy of the hydrodynamics. (a)~Density, velocity and pressure of the Sod shock tube at $t = 0.2$ along the middle row of a $256^2$ grid (points), and the exact solution (lines). (b)~Density against distance from the center for the leaf cells of a three-dimensional Sedov blast on the adaptive mesh (finest resolution 256 cells, three levels) at $t = 0.06$, every 33rd of the 6.5 million cells (points), and the similarity solution (line). (c)~Mean density error against the number of cells per side for a Gaussian advected once around the periodic box along a diagonal, with the default limiter of \citet{Koren1993} and the monotonized central limiter, and for the Sod tube; gray lines show first-, second- and third-order convergence.}
\label{fig:hydro}
\end{figure*}

The scheme is an unsplit MUSCL-Hancock method \citep{vanLeer1979, vanLeer1984}, in the form given by \citet{Toro2009}. A predictor pass advances each cell half a step from its own reconstructed slopes. An update pass then reconstructs face states from the half-step values, solves a Riemann problem at each of the cell's six faces, and applies the flux differences and the gravitational source. With the passes that compute the signal speed and reduce the timestep, a step is four passes. On the uniform grid the predictor and the update convert the conserved state to primitive variables as they load it, unless the equation of state is costly to invert (Section~\ref{sec:performance}). This is cheaper because the flux update is limited by texture fetches (with all of its arithmetic removed, it still takes 85~percent as long) and a pass that only rewrites the state costs more than repeating the conversion. Figure~\ref{fig:hydro} shows the accuracy of the scheme on a shock tube, a blast and a smooth flow.

The timestep is the smallest of three limits: $C\,\Delta x/\max\sum_i(|v_i| + c_{\rm s})$, $C\sqrt{\Delta x/\max|\mathbf{g}|}$, and the limit of each pair of point masses (Section~\ref{sec:sinks}). Here $C = 0.8$ is the CFL number, $\Delta x$ the cell size, $v_i$ the component of the velocity along axis $i$, $c_{\rm s}$ the sound speed, and $\mathbf{g}$ the gravitational acceleration. The sum runs over the axes, because an unsplit update adds the fluxes of every axis in one step, and the maxima run over all cells. The step is also shortened so that it ends exactly on the next output time. Viscosity and conduction add a signal speed (Section~\ref{sec:diffusion}). Cooling does not limit the step (Section~\ref{sec:cooling}).

The fluxes come from the HLLC solver of \citet{Toro1994}, the HLL solver of \citet{Harten1983}, or a low-Mach variant of HLLC. Wave speeds are the estimates of \citet{Davis1988}, {\relpenalty=200 $S_{\rm L} = \min(u_{\rm L} - c_{\rm L},\, u_{\rm R} - c_{\rm R})$} and {\relpenalty=200 $S_{\rm R} = \max(u_{\rm L} + c_{\rm L},\, u_{\rm R} + c_{\rm R})$}. Here $u$ is the velocity normal to the face, $c$ the sound speed, and L and R denote the two sides of the face. These estimates need only the pressure and the sound speed on each side, so the solvers do not depend on the equation of state. Barotropic gas uses HLL, since it has no contact wave for HLLC to resolve.

The low-Mach variant follows \citet{Thornber2008}: the velocity jump handed to HLLC is scaled by the local Mach number, so that slow flow is not damped at a rate set by the sound speed:
\begin{equation*}
\begin{aligned}
\mathbf{v}_{\rm L,R} &\rightarrow \tfrac{1}{2}(\mathbf{v}_{\rm L} + \mathbf{v}_{\rm R}) \pm \tfrac{z}{2}(\mathbf{v}_{\rm L} - \mathbf{v}_{\rm R}),\\
z &= \min\bigl[1,\ \max\bigl(z_{\rm min},\ |\mathbf{v}_{\rm L}|/c_{\rm L},\ |\mathbf{v}_{\rm R}|/c_{\rm R}\bigr)\bigr].
\end{aligned}
\end{equation*}
We set the floor $z_{\rm min}$ to 0.3. With one predictor and one update and the unlimited third-order reconstruction below, the scheme relies on some of the solver's damping, and a Gresho vortex \citep{Gresho1990} at Mach 0.01 gains kinetic energy without bound for a floor of 0.1 or less. With the floor at 0.3, that vortex on $64 \times 64$ cells loses 6~percent of its kinetic energy in two time units, against 15~percent with plain HLLC. Scaling the acoustic wave speeds instead, as \citet{Fleischmann2020} do, also removes the damping of pressure jumps and was unstable in this scheme at Mach 0.03.

\subsection{Reconstruction}
\label{sec:recon}

The default reconstruction gives each face the value of the parabola through the cell and its two neighbors, which is third-order accurate on the three-cell stencil that a slope limiter already uses. The face value is $q_{\rm c} + (\delta_{\rm far} + 2\delta_{\rm near})/6$, where $q_{\rm c}$ is the value in the cell, $\delta_{\rm near}$ the difference to the neighbor on the side of the face, and $\delta_{\rm far}$ the difference to the neighbor on the other side. A three-cell limiter cannot tell a smooth extremum from a jump and flattens both. Following the idea of \citet{Cada2009}, we apply a limiter only where a quantity changes by more than 1~percent of its natural scale across a cell: the density, the pressure, or the sound speed for the velocities. If we call that scale $Q$, the face value blends from the unlimited value to the limited one as $(\delta_{\rm near}^2 + \delta_{\rm far}^2)/(0.01\,Q)^2$ rises from 0.5 to 1.5. The limiter is that of \citet{Koren1993} with the bounds of the monotonized central limiter \citep{vanLeer1977}. A smooth peak then keeps its height: the largest error in the advection test is 2.5 to 5 times smaller than with the monotonized central limiter (Figure~\ref{fig:hydro}c). Kinks and jumps smaller than 1~percent go unlimited, and shock-tube errors are a few percent higher on axis-aligned tubes. The monotonized central, \citet{vanLeer1974} and minmod \citep[e.g.,][]{Roe1986} limiters remain available.

\subsection{Robustness in single precision}
\label{sec:robustness}

Single precision loses the pressure of gas whose kinetic energy dominates its total energy. We evolve the internal energy alongside the total energy and use it where the kinetic energy dominates, after the dual-energy formalism of \citet{Bryan1995}. The internal energy that we evolve replaces the difference of the total and kinetic energies wherever that difference is below $10^{-3}$ of the total energy, or below the least internal energy that the equation of state allows. Where we use it, we rebuild the total energy from it, except in degenerate gas. There we keep the total energy as the conservative update left it and take only the pressure from the internal energy, because a degenerate gas has a pressure, at least that of its cold electrons, even where the total energy minus the kinetic energy is negative. In a collision of two white dwarfs that difference turns negative in the cell ahead of each strong shock. Rebuilding the total energy there added 2.3~percent to it at the moment of greatest compression. Keeping it leaves a change of 0.6~percent through the collision.

A face state with a nonpositive density or pressure drops its cell to first order for that step. Next to a cell that a shock detector flags, the flux blends from HLLC toward HLL, which suppresses the carbuncle instability \citep{Quirk1994}. The detector flags a cell where the flow converges and the pressures of its two neighbors along some axis differ by more than half of the smaller one. It measures that difference against hydrostatic balance (Section~\ref{sec:wellbalanced}). Density and pressure floors, by default $10^{-10}$ and $10^{-14}$ in code units, catch what remains, and the number of floored cells is reported as a diagnostic.

\subsection{Well-balanced reconstruction}
\label{sec:wellbalanced}

In hydrostatic balance the pressure is not smooth across a stratified cell, but its balance with gravity is. Following \citet{Kappeli2014, Kappeli2016}, we limit and reconstruct each neighbor's deviation from that balance,
\begin{equation}
\delta p = (p_{\rm n} - p_{\rm c}) + \tfrac{1}{2}(\rho_{\rm c} + \rho_{\rm n})(\Phi_{\rm n} - \Phi_{\rm c}),
\label{eq:wb}
\end{equation}
where the subscripts c and n denote the cell and its neighbor. The balanced part of the step to the face, $-\rho_{\rm c}\,\Delta\Phi$, is added back unlimited, where $\Delta\Phi$ is half of the potential difference to the neighbor across that face. Where the cells obey the trapezoid rule of Equation~(\ref{eq:wb}) with $\delta p = 0$, the two cells that share a face reconstruct the same pressure at it. A cell's two face pressures then differ by exactly the momentum that the gravity kick adds, because the kick takes its acceleration from the centered difference of the same potential. The predictor uses the same slopes, so it leaves the gas at rest. With HLLC, whose mass flux vanishes between two states at rest with equal pressures, the state then holds to round-off however steep it is. HLL keeps a mass flux wherever the two densities at a face differ, so it does not hold such a state exactly. The shock detector measures pressure jumps against the same balance, so that a steep atmosphere at rest is not flagged as a shock.

\begin{figure}[t]
\centering
\includegraphics[width=\columnwidth,height=\colfigheight,keepaspectratio]{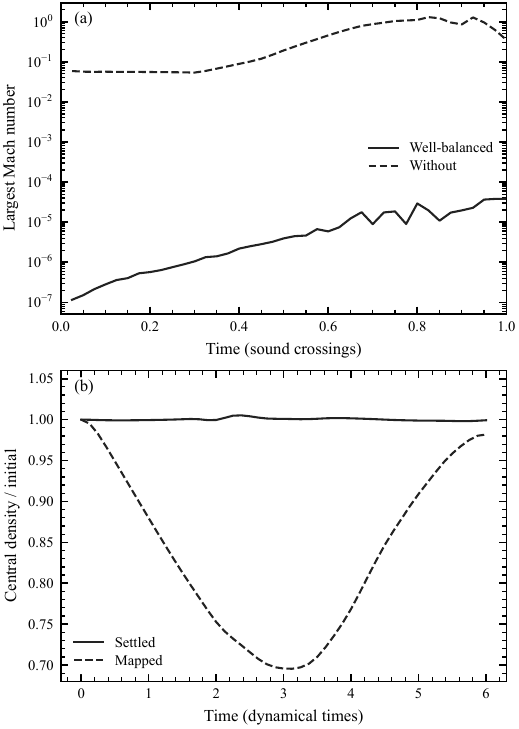}
\caption{Hydrostatic balance in single precision. (a)~Largest Mach number in an isothermal column whose density falls by a factor of $e$ every four cells, over one sound crossing, with the well-balanced reconstruction (solid) and without it (dashed). (b)~Central density of a white dwarf of $1\,\Msun$ on the adaptive mesh (finest resolution 256 cells, two levels), settled against the grid's own potential (solid) and mapped from its analytic profile (dashed), relative to its starting value.}
\label{fig:balance}
\end{figure}

An isothermal column whose density falls by a factor of $e$ every four cells keeps its largest Mach number at $4 \times 10^{-5}$ after a sound crossing, against 0.35 without the balanced reconstruction (Figure~\ref{fig:balance}a). The background of the Rayleigh-Taylor problem is 240 times quieter, and the error in the growth rate of the Jeans instability is 12 times smaller.

\subsection{Equations of state and cooling}
\label{sec:eos}

Every equation of state implements the same four functions (pressure, sound speed, internal energy from pressure, temperature), so the hydrodynamic passes are independent of it. Shaderous provides an ideal gas, a barotropic gas (isothermal or polytropic), and an ideal gas plus radiation pressure in local thermodynamic equilibrium, whose temperature comes from a Newton iteration with a fixed number of steps. Each WGSL module has a double-precision mirror in TypeScript for testing. Four further equations of state use the composition of the gas, which a problem with physical units gives as a list of species and their mass fractions.

\subsubsection{Degenerate electrons}
\label{sec:eos-degenerate}

This equation of state adds the pressure of fully degenerate electrons at any degree of relativity \citep{Chandrasekhar1935} to an ideal gas and radiation, for white dwarfs. Its density scale follows from the number of electrons per nucleon. Each part is thermodynamically consistent on its own, so the square of the sound speed is the derivative of the degenerate pressure plus $\Gamma_1 p_{\rm th}/\rho$ of the thermal part, where $\Gamma_1$ is the adiabatic index and $p_{\rm th}$ the pressure of that part. Where the Fermi momentum is below half of $m_{\rm e} c$, the pressure and energy of the degenerate gas are summed as series, because their closed forms cancel to nothing in single precision.

\subsubsection{Ionization of hydrogen}
\label{sec:eos-saha}

This equation of state lets hydrogen ionize in Saha equilibrium \citep{Saha1920}, so that the mean molecular weight and the internal energy, which includes 13.6~eV per ionization, follow the temperature. Its adiabatic index is computed exactly and dips below 1.3 in the ionization zone. Its temperature comes from Newton's method kept inside a bracket.

\subsubsection{Electrons and positrons}
\label{sec:eos-pairs}

This equation of state is for white dwarf matter that burns. It treats the electrons and positrons at any degeneracy and temperature, in the manner of \citet{TimmesSwesty2000}. The equation of state of fully degenerate electrons gives every electron a thermal energy of $\frac{3}{2}kT$ on top of the degenerate gas, which is far more heat than degenerate electrons take up. It therefore gives shocked degenerate fuel a temperature two to five times too low: at $10^7$~g~cm$^{-3}$, gas at $10^9$~K comes out at $4.5\times10^8$~K. Burning rates rise as $T^{20}$ or faster, so the error determines whether a detonation ignites.

We evaluated the Fermi-Dirac integrals of the electron-positron gas in double precision and tabulated its thermal free energy, less that of the pairs at zero net charge, with its first derivatives. The table runs in $\log_{10}(\rho Y_{\rm e})$, where $Y_{\rm e}$ is the number of electrons per nucleon, and in $\log_{10} T$ at intervals of 0.02~dex. Pressure and energy are first derivatives of the free energy, so a bicubic Hermite interpolation keeps them thermodynamically consistent. The pairs at zero charge depend on the temperature alone and are a Chebyshev series in it. As in the equation of state of \citet{TimmesSwesty2000}, the ions are an ideal gas and radiation is added. The ions also include the Coulomb energy of a one-component plasma, from the fits of \citet{YakovlevShalybkov1989}. The Coulomb term lowers the cold pressure, so a carbon-oxygen white dwarf of $1\,\Msun$ is 6~percent smaller than without it, and the white dwarfs of the problems are built on the same cold curve. Against the exact gas, the double-precision mirror gives the energy and the pressure to $7\times10^{-6}$ and the square of the sound speed to $2\times10^{-4}$ (99th percentile). The GPU matches the mirror to $2\times10^{-5}$ in energy and pressure, which is what single precision allows for a temperature found by iteration.

The hydrodynamic state gives the energy or the pressure, so the temperature comes from Newton's method in $\ln T$. Each step of the iteration reads four neighboring texels. It starts from a coarse table of the inverse at three mean nucleon numbers, and it converges in 2.7 iterations on average, against 9.3 from a guess that treats the gas as classical. The table, its inverse and the pair series fill one texture of 4.3~MB.

Every tabulated equation of state shares one module, which loads a table of a free energy, binds it to the GPU, interpolates it, and finds the temperature from it by Newton's method. The module has a mirror in TypeScript, and its interpolation has a copy in Python that gives the reference burns of Section~\ref{sec:burning} their temperatures. Each equation of state adds its own terms outside the table. With the burning network running, a carbon-oxygen detonation took 15~percent longer per step than with the equation of state of fully degenerate electrons. We also fitted small neural networks to the free energy and to its inverse. On the GPU, the table with Newton's method was three times faster than the best inverse network and about a thousand times more accurate.

\subsubsection{Hydrogen and helium}
\label{sec:eos-hhe}

This equation of state is for hydrogen and helium, from molecular clouds to the envelopes of stars. There most of the heat capacity of the gas goes into turning one species into another: 4.48~eV dissociates a molecule of H$_2$, 13.6~eV ionizes a hydrogen atom, and 24.6 and 54.4~eV ionize helium once and twice. We follow the model of \citet{Tomida2013}, an ideal mixture of H$_2$, H, H$^+$, He, He$^+$, He$^{2+}$ and electrons in chemical equilibrium, which we describe by one free energy $F$ per unit mass. Each species has its translational states and its internal ones. For H$_2$ these are the rotation, with ortho and para hydrogen held at 3:1, and the vibration. We count energies from the ground states of H$_2$ and He, so that the energy vanishes at zero temperature, and we use the textbook statistical weights. Those of \citet{Tomida2013} weight H$_2$ 4.6 times more, which moves its dissociation about 6~percent hotter at a given density. The abundances that minimize $F$ at fixed numbers of nuclei satisfy the dissociation and Saha equations, which reduce to one equation for the electron density. Because $F$ is at a minimum in the abundances, its first derivatives do not involve the derivatives of the abundances. With $h = F m_{\rm u}/kT$,
\begin{equation*}
\frac{p}{\rho} = \frac{kT}{m_{\rm u}}\frac{\partial h}{\partial\ln\rho},\qquad \varepsilon = -\frac{kT}{m_{\rm u}}\frac{\partial h}{\partial\ln T},
\end{equation*}
where $\varepsilon$ is the internal energy per unit mass and $m_{\rm u}$ the atomic mass unit.

We tabulated $h$ for two compositions, solar gas (hydrogen mass fraction 0.7381, with its metals counted as helium) and pure hydrogen. Each table runs in $\log_{10}\rho$ from $10^{-20}$ to $10^3$~g~cm$^{-3}$ at intervals of 0.1~dex and in $\log_{10}T$ from 10 to $10^8$~K at intervals of 0.01~dex, and it holds the exact first derivatives of $h$ and its cross derivative. The dissociation and ionization fronts are sharp in temperature and broad in density, which is why the intervals differ. The interpolation and the solve for the temperature are those of the white dwarf table, with two changes. First, $h$ runs from $-100$ to 100 across the table, and single precision would round it to $10^{-4}$ of the pressure and $10^{-2}$ of the sound speed. Only the derivatives of $h$ are ever needed, so the table holds the differences of $h$ between neighboring nodes in place of $h$. Second, Newton's method starts from a table of the inverse at the full resolution of the table. In dissociation and ionization the energy rises steeply at nearly constant temperature, and a starting guess that treats the gas as classical would stall there. In the unit tests the solve takes 1.1 evaluations on average and 3 at most. Each table is one texture of 5.2~MB. Between the nodes, the table gives the pressure of the model to $6\times10^{-5}$, its energy to $4\times10^{-4}$ and the square of its sound speed to 0.4~percent at worst, on the fronts at the lowest densities. On an Apple GPU, the round trip from pressure to energy and back agrees to $7\times10^{-6}$, and the energy and the sound speed match the double-precision mirror to $2.5\times10^{-5}$.

The adiabatic index $\Gamma_1$ is 5/3 below 50~K, where H$_2$ does not rotate, and 7/5 at a few hundred kelvin. It falls to 1.1 where H$_2$ dissociates (2000~K at $10^{-8}$~g~cm$^{-3}$) and again where hydrogen ionizes. Both dips are below 4/3, the least index with which a self-gravitating ball can support itself, and the first is what starts the second collapse of a protostar \citep{Larson1969}. The adiabat of gas that became opaque at $10^{-13}$~g~cm$^{-3}$ and 10~K reaches 2060~K at $10^{-8}$~g~cm$^{-3}$. In the calculation of \citet{Masunaga2000}, where the gas became adiabatic near $10^{-12}$~g~cm$^{-3}$, the center reaches 2000~K near $10^{-7}$~g~cm$^{-3}$.

The model has no degenerate electrons and no Coulomb terms. Pressure ionization enters through a stand-in term, which adds $kT\rho/\rho_{\rm pi}$ to the free energy of each bound electron, with $\rho_{\rm pi} = 0.33$~g~cm$^{-3}$, the density of hard spheres of a Bohr radius. We did not calibrate it against anything. Because it is part of the free energy, the pressure and the energy stay consistent. It leaves gas below 0.01~g~cm$^{-3}$ unchanged and ionizes hot gas above about 1~g~cm$^{-3}$, as the interior of a star is ionized. Without it, the Saha equations leave a third of the hydrogen at the center of the Sun neutral. The table is therefore valid below about 0.1~g~cm$^{-3}$ at any temperature, and at higher densities where the gas is hot enough to be ionized. In cool gas denser than about 1~g~cm$^{-3}$ the pressure need not rise with temperature or density, and the square of the sound speed is kept above $0.01\,p/\rho$. The radiation term $aT^4$ is analytic and optional. The interior of a star needs it. A cloud does not, because its radiation escapes: in a cloud at $10^{-18}$~g~cm$^{-3}$, radiation in equilibrium with gas at 100~K would hold a hundred times the energy of the gas.

\subsubsection{Rock, iron and water}
\label{sec:aneos}

For rock, iron and water, Shaderous reads tables of ANEOS, the analytic equation of state of \citet{ThompsonLauson1972} with the later changes of \citet{Melosh2007} and \citet{Stewart2020}. We ran four published input decks through pyaneos \citep{StewartPyaneos2026}, the Python port of M-ANEOS \citep{ThompsonMANEOS2019}: pyrolite for a mantle \citep{StewartPyrolite2022}, the iron alloy Fe85Si15 for a core \citep{StewartFe85Si15_2020}, water \citep{StewartWater2024}, and forsterite \citep{StewartForsterite2019}, the rock of older calculations. At the nodes of the published tables, pyaneos reproduces them to $5\times10^{-9}$, the precision to which they are printed. All four materials share one grid of 401 densities from $10^{-8}$ to 100~g~cm$^{-3}$ and 173 temperatures from 50~K to $10^6$~K, uniform in the logarithms at intervals of 0.025~dex. ANEOS is built as a Helmholtz free energy, but in the ionized fluid the pressure and entropy that it reports are not the derivatives of its free energy. At $2\times10^4$~K in the iron alloy they differ from those derivatives by 1 to 13~percent. A table of the free energy that takes the pressure and entropy as its slopes is then inconsistent with itself, and a first version that we built that way was wrong by 3~percent at the median in the fluid.

We therefore tabulated the pressure and the energy that ANEOS reports, each as a bicubic Hermite interpolant with its first and cross derivatives at every node. The pressure is tabulated as $\operatorname{asinh}(p/P_{\rm s})$ with $P_{\rm s} = 10^9$~dyn~cm$^{-2}$, which is linear in the pressure of a vapor and logarithmic in condensed matter. The energy is tabulated as the logarithm of its excess over a cold curve, the free energy along the lowest isotherm plus a constant. The slopes at the nodes are limited as those of a monotone cubic are \citep{FritschCarlson1980}, so that the interpolated energy rises with temperature wherever the nodes do. Against ANEOS at 6,000 random states per material between the nodes, the median error in the liquid and the gas is $10^{-6}$ in pressure and $10^{-5}$ in energy, and it is smaller in the solid. Inside the vapor dome the saturation pressure rises by up to a factor of 20 from one temperature node to the next, and there the table places a state to about half a node, 3~percent in temperature. The table of one material fills 2.2~MB and is fetched when a run uses that material.

The hydrodynamic state gives the energy or the pressure, so the temperature comes first. A bisection over the 173 rows of the table at the cell's density finds the two rows that bracket it, in eight steps, and Newton's method on the cubic between them finds the temperature, with a step of false position wherever a limited slope is zero. The square of the sound speed is $(\partial p/\partial\rho)_T + \Gamma_{\rm G}\,(\partial p/\partial\ln T)_\rho/\rho$, where $\Gamma_{\rm G} = (\partial p/\partial T)_\rho/(\rho\,\partial\varepsilon/\partial T)$ is held between 0 and 5, so that every term is positive. The exact sound speed of a gas whose pressure is a function of density and energy multiplies a latent heat by the ratio of two slopes that both vanish in cold gas. With that form, cold dust had a sound speed of 26~km~s$^{-1}$ and set the timestep of a whole run. For solid and liquid rock and iron at or above their reference densities, our form matches the sound speed of the published tables \citep{StewartFe85Si15_2020, StewartPyrolite2022} at their own nodes to $3\times10^{-6}$ at the median and $2\times10^{-4}$ at the 90th percentile. It differs by more than 1~percent at 2 to 3~percent of the nodes. On the GPU, the pressure recovered from its own energy agrees with the float64 mirror to $10^{-4}$ at worst, and the sound speed to $3\times10^{-4}$ (99th percentile). A run holds one material or two, which are its species (Section~\ref{sec:species}). In a cell that holds both, each material takes the volume that brings the two to one pressure at the cell's temperature. Shaderous solves this closure in the page when a run first uses the pair, in about a second, and stores the result as three more tables at mass fractions of 1/4, 1/2 and 3/4. A cell blends, texel by texel, the two tables on either side of its own mass fraction, so the shader reads a mixture exactly as it reads one material. At the tables' own mass fractions, the pressure of the mixture matches each material's pressure at its own density to 0.5~percent (90th percentile, both materials condensed).

\subsubsection{Cooling}
\label{sec:cooling}

Optically thin cooling is either a power law in temperature with uniform heating, or the interstellar cooling function of \citet{Koyama2002} with the correction of \citet{VazquezSemadeni2007}. In the second, the internal energy per unit volume changes at the rate $n\Gamma\,[1 - n\Lambda(T)/\Gamma]$, where $n$ is the number density of hydrogen atoms in cm$^{-3}$, $\Gamma$ the heating rate per atom, $T$ the temperature in kelvin, and
\begin{equation*}
\begin{aligned}
\frac{\Lambda(T)}{\Gamma} ={}& 10^{7}\exp\left(\frac{-1.184\times10^{5}}{T + 1000}\right)\\
&+ 1.4\times10^{-2}\sqrt{T}\,\exp\left(\frac{-92}{T}\right).
\end{aligned}
\end{equation*}
Cooling is applied to the internal energy at fixed density at the end of the update, in up to 12 second-order substeps that each change the energy by at most a tenth. A substep that would cross thermal balance stops at the balance point, so cooling times far shorter than the timestep are stable and do not overshoot. The timestep is not shortened for cooling. A cell can therefore cool faster than the 12 substeps can follow. Such a cell stops short of the end of the step, having lost at most about 70~percent of its energy, and continues in the next step.

\subsection{Viscosity and conduction}
\label{sec:diffusion}

Shaderous can add a viscous stress and a conductive heat flux to Equation~(\ref{eq:euler}). The momentum flux gains $-\boldsymbol{\tau}$ and the energy flux gains $-\boldsymbol{\tau}\cdot\mathbf{v} + \mathbf{q}$, with
\begin{equation}
\begin{aligned}
\boldsymbol{\tau} &= \rho\nu\left[\nabla\mathbf{v} + (\nabla\mathbf{v})^{\rm T} - \tfrac{2}{3}(\nabla\cdot\mathbf{v})\,\mathsf{I}\right],\\
\mathbf{q} &= -\kappa\rho\nabla T,
\end{aligned}
\label{eq:diffusion}
\end{equation}
where $T$ is the temperature, and the kinematic viscosity $\nu$ and the coefficient $\kappa$ are constants. The temperature of an ideal gas with adiabatic index $\gamma$ then diffuses at $(\gamma - 1)\kappa$.

Both fluxes are evaluated explicitly from the state at the start of the step and added to the Riemann flux of each face. Derivatives along the normal of a face are differences of its two cells. Derivatives across it are the mean of the centered differences in the two cells. The density in $\rho\nu$ and $\kappa\rho$ on a face is the smaller of the densities of its two cells. With their mean, a cell of thin gas beside dense gas would diffuse faster than any timestep allows, by the ratio of the two densities. Because the diffusive terms are part of the face flux, the properties of that flux carry over to them. The two cells that share a face compute the same bits, and the matching of Sections~\ref{sec:reflux} and~\ref{sec:timebins} conserves momentum and energy across refinement boundaries and between blocks on different timesteps. On the adaptive mesh the transverse differences on the rim of a block's face would need ghost cells along the edges of the block, which Shaderous never computes. Both blocks that share the face use one-sided differences along the rim instead. A face of an obstacle is a wall at rest to which the gas sticks and through which no heat passes. The heat that the stress dissipates is added to the separately evolved internal energy (Section~\ref{sec:robustness}).

The timestep includes diffusion as an additional signal speed of $2.5\,d\,D/\Delta x$, where $d$ is the number of dimensions, $\Delta x$ the cell size, and $D$ the larger of $4\nu/3$ and the thermal diffusivity. The step therefore falls as $\Delta x^2$ where diffusion dominates. There is no implicit solve, so strong diffusion on a fine mesh is slow. Viscosity lengthens the flux update by a factor of 1.6 in two dimensions and 1.8 in three, because the transverse derivatives need the cells diagonally across from the one being updated. Conduction alone adds no measurable time.

\subsection{Nearly incompressible flow}
\label{sec:implicit}

In slow flow the sound speed, not the flow, sets the step of an explicit scheme. At a Mach number $\mathcal{M}$ the scheme takes $1 + 1/\mathcal{M}$ times the steps that the flow alone would need, and most of that work goes into sound waves that carry little of the physics. Shaderous offers two remedies, which can be combined: a liquid whose sound speed is chosen, and a step that treats the pressure implicitly.

The liquid has the weakly compressible equation of state $p = c_0^2(\rho - \rho_0)$, where $\rho_0$ is a reference density and the sound speed $c_0$ is a constant chosen to be about ten times the fastest flow. This follows the weakly compressible smoothed particle hydrodynamics of \citet{Monaghan1994}, whose equation of state is a stiffer one of the same kind. The density then departs from $\rho_0$ by about $\mathcal{M}^2$, and this departure is the error of treating the flow as incompressible. An isothermal gas with the same sound speed would serve as well in exact arithmetic. Its pressure $c_0^2\rho$, however, is a hundred times the differences of order $\rho v^2$ that drive a flow at Mach 0.1, and the linear form stores only the departure from $\rho_0$, which single precision resolves. The shared passes give a barotropic gas the HLL solver (Section~\ref{sec:scheme}), which has no shear wave. A Gresho vortex at Mach 0.1 on $64^2$ cells lost 8.4~percent of its kinetic energy in four turns with HLL. The liquid therefore uses the low-Mach HLLC solver, with which it lost 2.2~percent, as much as an ideal gas at the same Mach number. The slot of the internal energy holds the elastic energy of compression, so that the total energy is conserved as far as the scheme conserves it.

The implicit step follows \citet{Kwatra2009}. Advection, viscosity and turbulence driving stay explicit, at the step that the flow allows, and the acoustic terms are treated implicitly. In backward Euler form, for the liquid,
\begin{equation}
\begin{aligned}
\rho^{n+1} &= \rho^n - \Delta t\,\nabla\cdot\mathbf{m}^{n+1},\\
\mathbf{m}^{n+1} &= \mathbf{m}^* - \Delta t\,\nabla p^{n+1},\\
p^{n+1} &= p(\rho^n) + c_0^2\,(\rho^{n+1} - \rho^n),
\end{aligned}
\label{eq:implicit}
\end{equation}
where the superscripts $n$ and $n+1$ mark the start and the end of the step and $\mathbf{m}^*$ is the momentum density that the explicit update left. Eliminating $\rho^{n+1}$ and $\mathbf{m}^{n+1}$ gives one equation for the new pressure,
\begin{equation}
\frac{p^{n+1}}{c_0^2\,\Delta t^2} - \nabla^2 p^{n+1} = \frac{p(\rho^n)}{c_0^2\,\Delta t^2} - \frac{\nabla\cdot\mathbf{m}^*}{\Delta t},
\label{eq:helmholtz}
\end{equation}
which becomes the Poisson equation for the pressure in a projection method \citep{Bell1989} as $c_0 \to \infty$. For an ideal gas the first term on each side is divided by $\rho c_{\rm s}^2$ rather than $c_0^2$, the Laplacian becomes $\nabla\cdot(\rho^{-1}\nabla p^{n+1})$, $p(\rho^n)$ becomes the pressure advected over the step, and $\mathbf{m}^*$ becomes the velocity. We solve for the change from the pressure of the previous step. At Mach 0.001 the round-off of the pressure itself in single precision is as large as the pressure differences of the flow.

The predictor of Section~\ref{sec:scheme} advances only the advection over the half step, with the gradient of the previous step's pressure in place of the acoustic terms. Equation~(\ref{eq:helmholtz}) is solved by the geometric multigrid of Section~\ref{sec:poisson}, warm-started, with three V-cycles a step. A face that nothing crosses, such as a wall of the box or the surface of an obstacle, has a coefficient of zero, and a coarse face takes the mean of the fine faces that tile it, so obstacles are represented on every level without a mask. The correction changes the momentum by differences of face pressures and the density by differences of face mass fluxes, so mass and momentum are conserved to round-off. For an ideal gas the face pressure is $(\rho_{\rm R}\,p_{\rm L} + \rho_{\rm L}\,p_{\rm R})/(\rho_{\rm L} + \rho_{\rm R})$, as in \citet{Kwatra2009}. The work of the pressure on the total energy is also a difference of face values, so energy is conserved too. We add one thing to the published scheme. Advection carries the gas on the predictor's face velocities, which are free of divergence only to first order in the step. In a liquid the pressure corrects the resulting density error, but in a gas the error is an error in entropy and remains. Each face therefore also moves the upwind state times the difference between its corrected velocity and the predictor's. In a float64 copy of the scheme, this reduced the density error of a Gresho vortex from 7~percent to 0.07~percent.

Backward Euler damps sound that the step resolves, and it makes the whole step first order in time. The step is therefore time-centered, at no extra cost. Its pressure force is that of $p_\theta = \theta p^{n+1} + (1-\theta)p^n$, and the divergence in the solve is that of $\theta\mathbf{m}^{n+1} + (1-\theta)\mathbf{m}^n$. Eliminating as before gives the operator of Equation~(\ref{eq:helmholtz}) with its first coefficient divided by $\theta^2$, so the step still takes one solve of three V-cycles. Each cell takes $\theta = \max(\chi, \min(1, \tfrac12 + 0.1\,C))$, where $C = c_{\rm s}\Delta t/\Delta x$ is its acoustic Courant number and $\chi$ is a shock switch, defined below. Sound that the step resolves is then carried nearly undamped, as by the Crank-Nicolson scheme, and sound that it does not resolve is damped as before, which at a low Mach number is all of the sound. The switch $\chi = \min(1, \tfrac12 + 4\,\Delta t\max(0, -\nabla\cdot\mathbf{v}))$ gives backward Euler to the cells that a shock compresses, behind which Crank-Nicolson rings.

A gas is carried at the mean of the face velocities at the start and the end of the step, and the pressure and the fields do their work at that mean as well. With the corrected velocity alone, the step was first order even in nearly incompressible flow. Where the step resolves sound, the predictor's pressure also takes the compression over the half step. In the float64 copy, a smooth wave in moving gas then converges at second order in the step, with a sixth of the error of backward Euler at the default Courant number and a twelfth at half of it. A sound wave at an acoustic Courant number of 0.5 keeps 98~percent of its amplitude over a period, against 86~percent.

On the adaptive mesh the liquid takes one $\theta$ for the whole mesh, that of its finest cell, so that the coarse levels of the multigrid still solve the leaves' equation. A face between two levels carries the gas with the larger of its two cells' $\theta$. There the velocity at the start of the step is formed from the two cells' centers rather than as the face's own velocity is, and the difference is sound that the step must damp. The Sod tube on two levels then has an L1 error in the density of 0.0061, against 0.0093.

Gravity acts on the faces. A uniform field kicks each face velocity by $\Delta t\,g$, so that a face is at rest where the trapezoid rule of Equation~(\ref{eq:wb}) holds. Self-gravity kicks each face by the difference of the potential across it. The point masses of Section~\ref{sec:sinks} kick the faces by the difference of their softened potential, and they receive the sum of the kicks that the gas was given, so the momentum of gas and masses together is conserved. A settled polytrope on $64^3$ cells kept its central density to 0.5~percent over two dynamical times in 27 steps, where the explicit scheme took 436.

On the adaptive mesh the equation is solved over the levels of the tree. Where a fine cell borders a coarser one, the fine cell's ghost pressure lies on the line between the two centers. Through that face, the coarse cell takes the mean of the fluxes of the fine faces, so mass and momentum are conserved across levels to round-off. Each level relaxes the plain operator, with the difference from the conservative operator added to its right-hand side, in the full approximation scheme of \citet{Brandt1977}. Density and pressure are nearly uniform in a liquid, so the refinement indicator of Section~\ref{sec:amr} also takes the second derivative of the velocity. One solve couples every cell at one time, so all blocks take the same timestep. A Gresho vortex on three levels, with the refinement boundaries running through it, took 273 steps to $t = 1$ at Mach 0.1, 0.01 and 0.001 alike, against 12,900 for the explicit scheme at Mach 0.1.

The step of a gas is also limited by the acceleration of the solved pressure. It is at most the solution of $(v + a\,\Delta t)\,\Delta t = C\,\Delta x$, where $v$ and $a$ are the largest speed and acceleration on the grid. A liquid Gresho vortex on $64^2$ cells took 212 steps to $t = 2$ at Mach 0.1, 0.01 and 0.001 alike, where the explicit scheme took 3,411 and 32,198 at the first two. The split is not meant for shocks, where the switch returns it to backward Euler. The Sod tube on 128 cells took 38 steps against 111, with an L1 error in the density of 0.0097 against the explicit scheme's 0.004. On an Apple M4 Pro, a step of the Gresho vortex of an ideal gas cost as much as 6.6 explicit steps at $512^2$ cells (2.9 against 0.44~ms) and 4.2 at $128^3$ cells (18 against 4.4~ms). A step replaces about $1 + 1/\mathcal{M}$ explicit ones at a Mach number $\mathcal{M}$, so the implicit step is cheaper than the explicit scheme below a Mach number of about 0.2 in two dimensions and 0.3 in three. It works for the liquid and the ideal gas in Cartesian coordinates. It runs without sinks, cooling, accelerating frames, a shearing box, mirror symmetry or burning. Rotating frames and damping zones are not available with a gas in this mode, and self-gravity and point masses are not available with the liquid.

\subsection{Bodies in a stream}
\label{sec:bodies}

An obstacle is a set of solid cells. Each face between a gas cell and a solid one is a reflecting wall, and a no-slip wall with viscosity. The flux through a wall face is the momentum that the gas gives the wall there, so the force on a body is the sum of the fluxes through its wall faces, which the update forms anyway. It is exactly the momentum that the gas lost. In a periodic box, the change of the gas's momentum balances the summed force to $7\times10^{-6}$ or better, on both grids and with explicit and implicit pressure. The diagnostics give the force and torque on each body, and its drag and lift coefficients. A body is a shape of the setup or a triangle mesh read from an STL, OBJ, PLY or glTF file, which becomes a signed distance field on the grid. One part of a setup makes a wind tunnel: an inlet, an outlet with a damping zone, and matching faces of the box, with the liquid and implicit pressure below Mach 0.3. A rotor can turn in its own frame (Section~\ref{sec:rotating}).

On the uniform grid a body can also move: spin, slide, swing or follow a path. Each step, the cells that it covers pass their gas to their neighbors, and the cells that it uncovers take gas from theirs. Both are gathers, so mass is conserved to the last bit. The wall moves with the body. A slab that slides through a closed box at Mach 0.085 does work on the gas that equals the gain in its energy to 0.03~percent. The force on a moving body, however, is dominated by the jumps of its staircase as it crosses cells. A cylinder that moves through still liquid has 87~percent more drag than the same cylinder at rest in a stream (Table~\ref{tab:validation-bodies}).

A staircase of whole cells is a poor surface for a body only a few cells thick. For bodies at rest in the liquid with implicit pressure, cut cells can replace it on the uniform grid. Each cell holds gas in a share of its volume, each face is open in a share of its area, and the state redistribution of \citet{BergerGiuliani2021} keeps the small cells stable. With 16 cells along its chord, a NACA 0012 airfoil at $5^\circ$ and a Reynolds number of 1,000 has a lift coefficient of 0.27 with cut cells and 0.16 with the staircase, against 0.25 and 0.23 with 64 cells.

\subsection{Turbulence driving}
\label{sec:driving}

A random force can stir the gas at the largest scales of the box. We follow the forcing of \citet{EswaranPope1988}, in which the Fourier modes of the force in a band of wavenumbers have amplitudes that follow Ornstein-Uhlenbeck processes, with the projection of \citet{Federrath2010b} that sets the share of the force that is free of divergence. The feature and the problems that use it follow the WebGPU solver of \citet{Mallet2026}. The modes are the wavevectors $2\pi\mathbf{n}/L$, where $L$ is the side of the box and $\mathbf{n}$ an integer vector whose length lies in the band, at most 64 of them. Each cell sums the modes at its own position, so the force needs no Fourier transform and works on every level of the adaptive mesh. Each amplitude decays over a correlation time and receives random kicks, of the size that keeps the mean square acceleration over the box at the value that the problem sets. The pass that sets the timestep on the GPU (Section~\ref{sec:webgpu}) also advances the amplitudes, so that every grid and every way of stepping uses the same force. The random numbers come from a hash of the seed, the step count and the mode, so a run is the same from the same seed. The update adds the force to the momentum as a kick and adds the kinetic energy of the kick to the total energy. The acceleration has zero mean over the box, but the force $\rho\mathbf{a}$ does not where the density varies, and the code does not remove it. After four crossing times at Mach~1, the box moves at about a tenth of the root-mean-square speed.

\subsection{Species}
\label{sec:species}

A problem can carry any number of species, as mass fractions $X_k$ that sum to one. They are stored apart from the hydrodynamic state, as $\rho X_k$ in four-channel textures, and the hydrodynamic passes never read them. The update records the mass flux through every face, and a separate pass then moves every species with it. The flux of a species through a face is the mass flux times the fraction of that species in the upwind cell at the half step, extrapolated to the face with its limited slope, as for the passive scalar.

Each species has its own monotonized central slope, so a sharp front in one species does not flatten the others. The slopes then need not sum to zero, while the fractions on a face must sum to one. We therefore take the excess back from the slopes of the sign that has too much, each in proportion to its size. This only shrinks slopes, so no face value leaves the range of the neighboring cells, and a species that is uniform stays uniform. Scaling the face fractions so that they sum to one, as in the consistent multi-fluid advection of \citet{Plewa1999}, would change such a species. Because the face fractions sum to one, $\sum_k \rho X_k$ moves exactly as $\rho$ does, and every fraction stays between zero and one. What rounding leaves is removed by scaling the new values to the new density. On the adaptive mesh a coarse cell takes the average of the fine cells' species fluxes, as it does for the mass (Section~\ref{sec:reflux}). Local timesteps are not yet available with species.

When every species is a nuclide, or dry air taken as one species for laboratory flows, the equation of state follows the composition of each cell, and each species may have its own adiabatic index. Each equation of state needs at most four numbers from the composition, and each is a sum over the species weighted by mass fraction, so they mix linearly. A texture beside the state holds them. For an ideal gas they are the inverse of the mean molecular weight and the heat capacity, so that the adiabatic index of a mixture follows from the heat capacities of its gases summed by mass, the mixing rule of ideal gases at one temperature.

A conservative scheme that carries mass fractions makes the pressure oscillate where two gases with different adiabatic indices meet \citep{Abgrall1996, Karni1996}. In Shaderous the cause is that the density and the mass fractions are reconstructed in different passes, so that at a face they disagree on how much of each gas crossed it. A contact between helium and air at one temperature, carried across the box, erred by 3.7~percent in pressure. Where $\gamma - 1$ differs between neighbors by more than 1~percent, the predictor and the species passes therefore blend their face values toward first order, fully so at a difference of 5~percent. The same contact then errs by 0.2~percent. The blend is continuous because an on-off switch turned round-off into differences of 5~percent between mirror-image cells of the shock-bubble problem (Section~\ref{sec:problems}). On the adaptive mesh a jump of 0.1 in any mass fraction between neighbors also refines a block, because an interface smeared over a few coarse cells is too smooth to raise the indicator of Section~\ref{sec:amr}. Without this criterion, the helium bubble of \citet{QuirkKarni1996} drifted onto coarse blocks, and 24~percent of its helium ended at the finest level, against 99~percent with it.

\subsection{Nuclear burning}
\label{sec:burning}

A problem can burn its species in one of two ways, in a pass of its own after the species pass, on the uniform grid and on the adaptive mesh alike. The first is the single-step Arrhenius reaction that is the standard model in the theory of detonations \citep{Fickett1979}. The fraction $f$ of fuel falls as $\mathrm{d}f/\mathrm{d}t = -K f\exp(-E_{\rm a}/T)$, where $K$ is a rate constant, $E_{\rm a}$ the activation temperature and $T$ the temperature, and each unit of mass that burns releases an energy $q$ into the gas at fixed density. The pass splits the step into up to eight substeps and solves each exactly at its starting temperature, so the burn never overshoots however stiff it is. Each of the first seven releases at most a fifth of the cell's internal energy, and the last takes the rest of the step, so the burn always covers the whole step. The timestep is not shortened for burning. On the adaptive mesh, blocks are refined where the burning length is shorter than 80 cells. The burning length is the distance that sound travels while the burn releases the cell's internal energy.

A problem also sets whether cells that the shock detector flags may burn. Where the front is far thinner than a cell they should not, because a captured shock is a few cells of numerical smearing through states that no real gas passes through, and burning there lets the scheme rather than the reaction set the speed of the front \citep{Colella1986}. Where the reaction zone is resolved they should. A stable detonation started from its exact steady solution, with 12.8 cells per half-reaction length, then runs within 1~percent of the Chapman-Jouguet speed, on the uniform grid and on the adaptive mesh. With burning suppressed in shocks it falls 5~percent short, an error that halves with each doubling of the resolution.

The second is the 13-isotope $\alpha$-chain of \citet{Timmes2000}, for white dwarf matter. Its reference is the aprox13 network as pynucastro builds it \citep{SmithClark2023}, from the rates of REACLIB \citep{Cyburt2010} with the screening that pynucastro applies. The reference folds the captures of $\alpha$-particles that pass through a proton, ($\alpha$,p)(p,$\gamma$), into the direct captures. Our first reference left these channels out, and it burned carbon 1.8 times and oxygen 3 to 4 times more slowly. We integrated the reference at constant density with the heat that the burning releases, which the tabulated equation of state of electrons and positrons (Section~\ref{sec:eos-pairs}) turns into temperature as the simulator does, with the Coulomb energy of the ions.

An implicit step of the reference in every cell is costly on the GPU, so the step is taken from tables of the reference's own burns where they apply and from a small neural network elsewhere. The network takes the logarithms of the density, temperature and timestep, the mass fractions, and the logarithm of every screened rate term of the $\alpha$-chain. The rate terms are inputs because they rise with temperature far more steeply than a network of this size could resolve from the temperature alone. The output is built so that mass is conserved exactly and no fraction falls below zero, whatever the weights: each species loses a share of itself, and a softmax distributes what is lost among the species. The energy released is the change in nuclear binding energy and is not predicted, so energy and composition always agree.

The network has ten hidden layers of 128 units, and each layer after the first adds its result to its input. We trained it for 600{,}000 steps on 351{,}763 burns of the reference, each sampled at 74 times. The training set includes the states that earlier networks reached when they were applied step after step, labeled with the reference, as in the DAgger method of \citet{Ross2011}. We also ran further rounds of DAgger on the final network, and we distilled larger networks into networks of 128 units per layer. Neither gave a smaller error than the final network. On the GPU, the cells of a workgroup are evaluated together: their activations are held in shared memory, and each thread computes one unit. Cells whose whole step the tables below cover skip the network. On a detonation at $1024^2$, 600 steps took 6.2 to 6.5~s with the tables and 13.4~s with the network alone. On the GPU, the hybrid step agrees with our implementation of it in double precision to $2\times10^{-4}$ in the summed error defined below.

The tables follow the stages of burning that \citet{Townsley2016} described. Oxygen burns on paths that the reference followed at constant density with its own heating, with the remaining oxygen as the clock. Silicon then burns toward equilibrium along paths at constant temperature, with the fraction in the iron group as the clock, and the step is taken at the mean of the temperatures before and after it. A cell on a path takes the change in composition along the path over the step. On a path of silicon burning, its mixture of intermediate elements also relaxes toward the path's mixture: in each step it moves a fraction $1 - \exp(-\Delta t/\tau)$ of the way, with $\tau = 10^{-4}$~s. We first replaced the composition with the path's in every step, which is the limit $\Delta t \gg \tau$. In small steps, this stopped dense carbon and oxygen short of equilibrium: in steps of 3~$\mu$s, gas at $4.7\times10^7$~g~cm$^{-3}$ from $4.1\times10^9$~K stopped at a mass fraction of $^{56}$Ni of 0.41, where the reference reaches 0.55. Gas that would burn less than $10^{-6}$ of its mass in carbon over a step does not change. Carbon burning is left to the network, which matched the reference more closely than the paths of carbon burning did. Each table's step is blended with the rest of the step by a weight that falls smoothly to zero at the edges of the table, so the step changes continuously with the state.

The error of a burn at a given time is the sum over the isotopes of the absolute difference between its mass fractions and those of the reference. On 600 burns of fuel and mixtures in steps of 1~ms, the 99th percentile of this error at 3~ms, 30~ms, 0.1~s and 1~s was 0.21, 0.19, 0.19 and 0.21. In steps of 10~$\mu$s it was 0.20 at 30~ms. We also burned 100 mixtures of carbon and oxygen and 100 of the ash of silicon and sulfur, at $10^{6.5}$ to $10^8$~g~cm$^{-3}$ and 3 to $5\times10^9$~K, in steps of 3~$\mu$s to 1~ms. For each burn, we took the spread of the mass fraction of $^{56}$Ni at 10~ms across the step sizes. Its 90th percentile was 0.02 for carbon and oxygen and 0.03 for the ash, against 0.02 and 0.23 for the network without the tables. Carbon and oxygen at $10^7$~g~cm$^{-3}$ from $3.5\times10^9$~K made a mass fraction of $^{56}$Ni at 0.1~s of 0.65 in every step from 10~$\mu$s to 1~ms, and 0.69 in steps of 3~$\mu$s, where the reference makes 0.65. At $2.1\times10^7$~g~cm$^{-3}$ from $4.4\times10^9$~K, a single step of 0.12~ms made 0.15, and steps of 3 to 30~$\mu$s made 0.18 to 0.19, where the reference makes 0.19. Only gas hotter than $10^8$~K and denser than $10^4$~g~cm$^{-3}$ burns, the range of the training set, and a hydrodynamic step longer than 10~ms is split into network steps.

Following \citet{Kushnir2013}, a run can limit the rate at which a cell takes in energy from burning to $f_{\rm b}\,e/t_{\rm s}$, where $e$ is the internal energy density, $t_{\rm s} = \Delta x/c_{\rm s}$ the time that sound takes to cross the cell, and $f_{\rm b}$ a chosen factor. Each network step is then scaled down as a whole, which keeps mass conserved and every fraction positive. The limiter is off by default. It spreads the burning over a length of about $\Delta x/f_{\rm b}$, and it describes a detonation correctly where that length is much smaller than every other length of the problem \citep{KushnirKatz2020}. A detonation must then be lit over a region about $\Delta x/f_{\rm b}$ across. A small hot spot ignites without the limiter only through the numerical instability that the limiter removes \citep{Kushnir2020}. We tested this in the tube of carbon and oxygen at $10^7$~g~cm$^{-3}$ on cells of 2~km, which do not resolve the burning of carbon or oxygen. With $f_{\rm b} = 0.05$ and 0.1, the default igniter, a layer at $4\times10^9$~K, drove only a shock at 5000 to 6000~km~s$^{-1}$ for every thickness of the layer from 15 to 200~km. With $f_{\rm b} = 0.2$ it failed only for the layer of 15~km. A layer at $8\times10^9$~K lit a detonation at every $f_{\rm b}$ from 0.05 to 0.4 and every thickness, and the front moved within 1~percent of the speed of the unlimited front from the same layer.

The runs with burning use the tabulated equation of state of electrons and positrons (Section~\ref{sec:eos-pairs}). With it, the head-on collision of two white dwarfs of $0.64\,\Msun$ on cells of 940~km made $0.63\,\Msun$ of $^{56}$Ni. The yield depends on the timestep and on the resolution. At Courant numbers of 0.6 and 0.4 instead of the default 0.8, the collision made 0.68 and $0.73\,\Msun$. On cells of 1900~km it made only $0.04\,\Msun$, and on cells of 470~km it made $0.40\,\Msun$. The run therefore demonstrates that the network burns a collision, but its yield has not converged. \citet{Kushnir2013} found $0.41\,\Msun$ for this pair on cells of 4~km, where the ignition in the shocked gas is resolved, and \citet{Raskin2010} found $0.51\,\Msun$ with smoothed particle hydrodynamics. With the limiter at $f_{\rm b} = 0.1$ and 0.4, the collision made 0.06 and $0.19\,\Msun$. On these cells, $\Delta x/f_{\rm b}$ at $f_{\rm b} = 0.1$ is larger than the stars. The small yields therefore show that the run does not resolve the detonation, not that the limiter suppresses a real one.

\subsection{Special relativity}
\label{sec:relativity}

A run can treat the gas as special-relativistic, at any speed below light, in units in which the speed of light is one. It runs on the uniform grid and on blocks with local timesteps, in Cartesian, $(R, z)$ and $(r, \theta)$ coordinates, for the ideal gas with $\gamma \le 2$, with species and optically thin cooling. There is no gravity, because Newtonian gravity is not consistent with relativistic gas. The state holds $D = \rho W$, $\mathbf{S} = \rho h W^2 \mathbf{v}$ and $\tau = \rho h W^2 - p - D$, where $W$ is the Lorentz factor and $h$ the specific enthalpy. The energy $\tau$ leaves out the rest mass, so that slow gas keeps its pressure in float32. Beside them the state stores the proper internal energy density $e$ that the last recovery of the primitive variables found. With $e$ the primitive variables follow from the state in closed form, because $(\tau + D + p)^2 - S^2 = (\rho h W)^2$, and the passes that load a cell use this form. The iterative recovery runs only where a step makes a new state: in the predictor, in the update and in a re-mesh. It is the bracketed root finder of \citet{Galeazzi2013}, eight safeguarded Newton steps that start from the closed form. Slopes are limited in $\rho$, $p$ and $\mathbf{u} = W\mathbf{v}$, so that no face state moves faster than light, and the predictor takes its half step in conserved form. The flux is the relativistic HLLC solver of \citet{MignoneBodo2005}, blended toward HLLE \citep{Schneider1993} at the cells that the shock detector flags.

Single precision sets two limits. The recovered specific internal energy is a small difference of large terms, with an error close to $1.5\times10^{-7}u^2$. We therefore cap the Lorentz factor at 50: at 100, rounding the state alone moves the exact recovery of hot gas. A cell held at the cap keeps the lesser of the heat that its $\tau$ implies and the heat that it had, so that the cap only removes energy. Where the recovered energy is below $10^{-3}u^2$ and the cell is not at a shock, the pressure comes instead from the entropy $p/\rho^\gamma$, which the fluxes carry with $D$. A cold beam at $W = 10$ with a specific internal energy of $10^{-6}$ then keeps its pressure to $3\times10^{-3}$, where the energy alone leaves it 100 times off. On blocks, a prolongation of $D$, $\mathbf{S}$ and $\tau$ with separate slopes is ill-conditioned in fast hot gas, whose heat is a fraction of about $1/(2W^2)$ of them. A refined cell therefore keeps its interpolated state only if it recovers within the range of its parent and the parent's neighbors, and otherwise takes the state of its parent. Without this check, a blast on three levels made single cells 30 to 90 times hotter than their neighbors in its emptied center, and these expanded as small blasts.

The references are a float64 copy of the scheme and an exact Riemann solver with tangential velocities \citep{Pons2000}. On the blast waves of \citet{MartiMuller2003} at 512 cells, the density errors on the GPU are 0.022 and 0.076, those of the float64 copy, and the thin shell of the second peaks at 8.2 of its exact 10.42. With a tangential velocity of 0.9 ahead of it, the shock runs within 0.5~percent of its exact speed. After 20 steps of a two-dimensional blast the GPU matches the float64 copy to $1.4\times10^{-6}$ in density. On three levels of refinement the second blast wave reproduces the error, shell and shock position of the uniform grid. In $(R, z)$ and $(r, \theta)$ the geometric source of the momentum is built from $\mathbf{S}\mathbf{v}$ and the pressure of the cell. A spherical blast then places its shell at the same radius in one dimension, in $(R, z)$, in $(r, \theta)$ and in three-dimensional Cartesian coordinates, to 1.3 Cartesian cells. Tables~\ref{tab:validation-relativity} and~\ref{tab:validation-relativity-2} list the tests. The problems include these tests, a relativistic Kelvin-Helmholtz layer, light jets at $W = 7$ and 10, planar and round, and a jet that bores through a star and breaks out of its surface, as in the simulations of \citet{Zhang2003}, \citet{Mizuta2006} and \citet{Morsony2007}. Cooling acts in the frame of the gas, over its proper time, and the energy that it radiates carries momentum away with it. Gas moving at 0.9 then follows a float64 integration of its energy and momentum to $7\times10^{-4}$.

\section{Conservation without stored fluxes}
\label{sec:conservation}

\begin{figure*}[t]
\centering
\includegraphics[width=\textwidth]{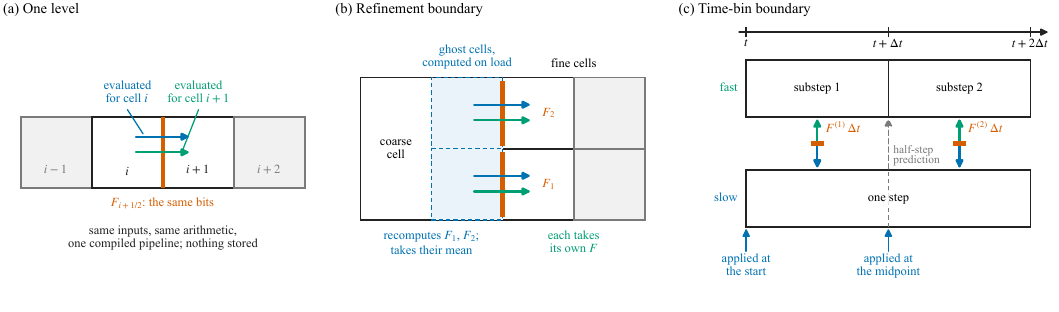}
\caption{How the hydrodynamic solver conserves without storing a face flux. The shared face is orange, and each arrow across it is one evaluation of its flux, colored by the side that makes it. (a)~Two cells of one level each evaluate the flux of the face between them, from the same inputs with the same arithmetic, and get the same bits. (b)~At a refinement boundary, shown in two dimensions, each fine cell evaluates the flux of its own face against a ghost cell that is computed as it is loaded. The coarse cell evaluates the same fine fluxes again and takes their mean. (c)~Between time bins, with time running to the right, the slower block takes the fluxes of the faster block's two substeps, the first when its own step starts and the second at its midpoint. The faster block reads the slower one from its half-step prediction (dashed).}
\label{fig:fluxes}
\end{figure*}

Most finite-volume codes on GPUs evaluate each face flux once, store it, and difference the stored fluxes. The hydrodynamic solver of Shaderous gathers instead: each cell needs the fluxes on its own six faces, and a face that two cells share is evaluated wherever either of them needs it. Conservation then requires that the evaluations agree bit for bit. They do if they read identical inputs and perform identical arithmetic in the same order. Every quantity that enters a face flux, including the potential differences of the well-balanced reconstruction, is formed from the same values whichever cell evaluates the flux. Figure~\ref{fig:fluxes} shows the three places where a face is shared: within a level, across a refinement boundary (Section~\ref{sec:reflux}), and between blocks on different timesteps (Section~\ref{sec:timebins}).

Identical source code is not enough. WGSL lets a compiler fuse and reassociate floating-point operations, and the Metal backend of Chrome compiles shaders in a relaxed mode that does so. On an Apple M4 Pro, a cell's lower face and its neighbor's upper face, computed by two calls of the same function, differed in the last bit on thousands of faces. Every face that two cells share is therefore computed at one place in the code and in one compiled pipeline. For the same reason, compensated sums pass their intermediate values through an exclusive-or with a zero that is unknown at compile time, which keeps the compiler from simplifying the compensation away.

Within a workgroup, the update computes each face of the workgroup's box of cells once, into workgroup memory, and each cell reads its six faces from there. The neighboring workgroup computes the faces on the edge of the box again, so no flux leaves the pass. This made the largest uniform grids 1.3 to 1.5 times faster on an Apple M4 Pro ($256^3$ and $2048^2$ cells). Gathering still evaluates the faces on the edges of the boxes twice. In exchange there is no flux storage, no second pass to difference fluxes, and no race between invocations that write the same face. The validation suite checks the property directly: for three flows, the two cells that share each face compute bit-identical mass fluxes on every face. A uniform state advected ten times across the box conserves mass and every momentum component to the round-off of single precision.

The other stencil passes are compiled twice from one source. Cells at least two cells from every boundary use a version with the boundary mapping compiled out. The shell uses the general version, dispatched in slabs two cells thick so that it launches no more invocations than it has cells. The update is not split this way, so that every face comes from one pipeline, at a cost of about 10~percent. Boundaries are chosen per face (periodic, outflow, outflow without inflow, or reflecting) and are applied in the function that maps a cell index to a texel, so the uniform grid has no ghost cells.

\section{Adaptive mesh refinement}
\label{sec:amr}

The adaptive mesh is a tree of blocks of $16^3$ cells ($16^2$ in two dimensions). A block splits into eight children of half the cell size, and face neighbors differ by at most one level. Each field lives in one 3D texture atlas that holds every block, with two ghost layers around each. The tree itself is small and lives on the CPU, which uploads it to the GPU as flat tables whenever the mesh changes. Every few steps a pass reduces a refinement indicator per block, the CPU reads it back without blocking, adapts the tree, and one pass fills the new leaves by copying, prolonging or restricting from the old state. Restriction averages the conserved variables of the children. Prolongation adds limited linear slopes of the conserved variables to the value of the parent, and the slopes cancel over its children, so both conserve mass, momentum and energy. The total energy is not given a slope of its own. A slope in it that is independent of the slopes in density and momentum leaves each child with a difference between its total and kinetic energy that has nothing to do with its heat: in cold, fast gas that difference was eight orders of magnitude above the internal energy, and the dual-energy scheme treated it as heat. Each child instead takes its own kinetic energy plus its prolonged internal energy, and one offset shared by the children keeps their mean total energy that of the parent.

Blocks are refined where the normalized second derivative of the density or the pressure exceeds 0.8. Sibling blocks are merged when it falls below 0.2 in all of them. This is the error indicator of \citet{Lohner1987} in the form that FLASH uses \citep{Fryxell2000}, including a filter that ignores ripples of about 1~percent of the local value. For a quantity $q$ in cell $i$, it is
\begin{equation*}
\chi^2 = \frac{\sum (q_{i+1} - 2q_i + q_{i-1})^2}{\sum (D_i + \epsilon A_i)^2},,
\end{equation*}
where $D_i = |q_{i+1} - q_i| + |q_i - q_{i-1}|$, $A_i = |q_{i+1}| + 2|q_i| + |q_{i-1}|$, the neighbors are taken along each axis in turn, the sums run over the axes, and $\epsilon = 0.01$ is the filter. The indicator is close to 1 at a discontinuity and falls in proportion to the cell size across a smooth profile. An indicator that is based on first differences refines smooth gradients as readily as shocks, and in our tests it kept most of the disk of the binary problem at the finest level. With self-gravity, blocks are also refined where the Jeans length falls below 16 cells. A problem can also require that every block near its point masses be kept at the finest level (Section~\ref{sec:sinks}), or every block with gas above a given density. The white dwarf problems use the second, because the smooth interior of a star does not raise the indicator.

\begin{figure*}[t]
\centering
\includegraphics[width=\textwidth]{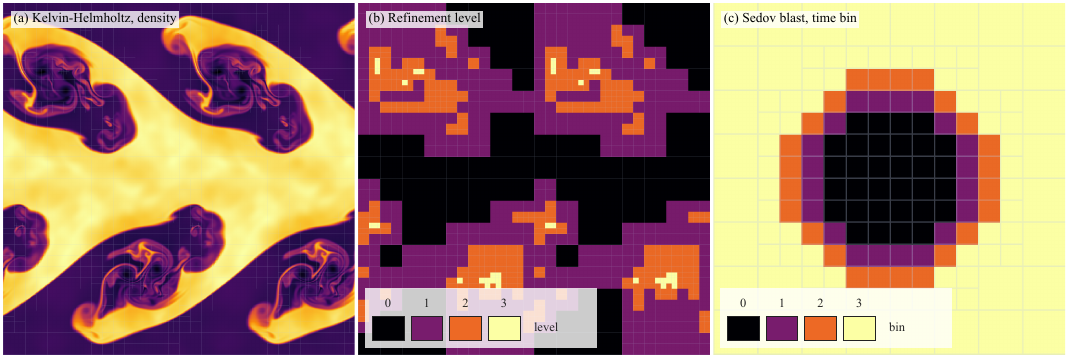}
\caption{The adaptive mesh, drawn by the code's own renderer. (a)~Density of a two-dimensional Kelvin-Helmholtz instability at $t = 1.5$, at an effective resolution of $2048^2$ with three levels of refinement. The faint lines outline the blocks of $16^2$ cells. (b)~Refinement level of the same state. (c)~Slice through a three-dimensional Sedov blast (finest resolution 256 cells, three levels) with local timesteps, each block colored by its time bin; a block in bin $k$ steps $2^k$ times the smallest step. The hot gas behind the shock takes the smallest steps, and the cold gas outside it the largest.}
\label{fig:amr}
\end{figure*}

The atlas holds a fixed number of blocks, which the memory budget of Section~\ref{sec:webgpu} sets. When a re-mesh needs more blocks than the atlas holds, the most strongly flagged blocks are refined first. Room for them is made by merging the sibling blocks with the lowest indicator, which are the closest to being merged anyway, as long as their indicator is below 0.8 of that of the block that needs the room. Without this margin, blocks that are flagged about equally would trade their slots back and forth. A block that is merged this way is not split again in the same re-mesh, because its new children would be filled from the slot of a parent, which holds no data. In a three-dimensional Sedov blast whose shock needed more blocks than the budget allowed, the total energy changed by $2\times10^{-7}$ of itself. Figure~\ref{fig:amr} shows meshes that the indicator builds in two and three dimensions.

\subsection{Ghost cells computed on load}
\label{sec:ghost}

Because a pass cannot read and write one \texttt{rgba32float} texture, ghost cells cannot be filled in place by copying from neighbors. Shaderous does not fill them. The pass that converts the conserved state to primitive variables computes every texel of a block, ghost or interior, directly from wherever its value lives: the same block, a neighboring block, the image of a cell across a physical boundary, the average of the finer cells that cover it, or an interpolation from a coarser block. There is no separate exchange step, and no ghost cell is ever stale.

\subsection{Matching fluxes across refinement boundaries}
\label{sec:reflux}

The adaptive solver of \citet{Berger1989} stores the fluxes that the fine side computes on a coarse face and corrects the coarse cell afterward, as does the GPU code GAMER that follows it \citep{Schive2010}. Shaderous extends the gathering of Section~\ref{sec:conservation} across levels instead. At a refinement boundary the coarse cell recomputes the fluxes of the fine faces that cover its face, from the same fine states, with the same arithmetic as the fine cells (Figure~\ref{fig:fluxes}b). Mass, momentum and energy are therefore conserved to round-off without storing any flux. The validation suite checks that the fluxes agree bit for bit between blocks on the same level. Across a refinement boundary it checks that the flux of the coarse cell agrees with the mean of the fluxes of the fine cells to $10^{-6}$, which allows for the rounding of the mean.

\subsection{Local timesteps}
\label{sec:timebins}

The subcycling of \citet{Berger1984} lets fine levels take more steps than coarse ones. Shaderous assigns timesteps per block rather than per level, using power-of-two time bins in the manner of particle codes such as GADGET-2 \citep{Springel2005}. Time advances in cycles of up to eight substeps of the smallest step. At the start of a cycle each block is put in a bin from its own signal speed, on the GPU: a block in bin $k$ advances $2^k$ substeps at a time. A first sweep then lowers every block to the bin of its fastest face neighbor, because fast gas next to a block may reach it before the cycle ends. This sweep runs once, so it widens each fast region by one block. Further sweeps enforce two rules: face neighbors are at most one bin apart, and a block is never slower than a coarser neighbor.

Across a face whose other side steps twice as often, a block takes the other side's fluxes, recomputed from stored states: half when its own step starts and half at its midpoint (Figure~\ref{fig:fluxes}c). Consider a slow block that takes one step of $2\Delta t$ beside a fast block that takes two steps of $\Delta t$. Every substep runs the same three passes: one assembles the primitive variables and the ghost cells of each block, one predicts the half-step state of each block that starts a step, and one updates those blocks.

In the first substep both blocks start. Each predicts from the state at the start of the step, the fast block to $\Delta t/2$ and the slow block to $\Delta t$. The ghost cells of the fast block that lie in the slow block hold the slow block's starting state, and the fast block predicts them with its own step. The fast block then updates to $\Delta t$, and the slow block to $2\Delta t$, so the stored state of the slow block is from then on that of the end of its step. On the shared face the slow block does not use its own flux. It evaluates the flux of the fast block instead, from the fast block's cells, ghost cells and half-step states, with the function that the fast block's own cells use, and applies it for half of its step.

In the second substep only the fast block starts a step. Its ghost cells in the slow block now come from the slow block's half-step prediction, which is the slow block's estimate of its state at $\Delta t$ and has been kept since its step began. The fast block predicts and updates to $2\Delta t$. In the same update pass, the cells of the slow block on the shared face evaluate the new flux of the fast block in the same way and add it for the other half of their step. Nothing else in the slow block changes, because its own fluxes, kicks and sources were applied when its step began. Both blocks have then counted the same two fluxes across the face. As at refinement boundaries, nothing is stored and conservation holds to round-off. With self-gravity the potential is solved on every block at the end of a cycle. In between, it is solved only on the blocks that finish a step and carried forward elsewhere at the rate at which it last changed. The potential that a block's gravitational work uses at its half step is extrapolated with the potential's curvature in time, taken from the rates over the block's last two steps, because in a collapse the mean of the potentials at the two ends of a step runs ahead of a linear extrapolation. The collapse of \citet{Evrard1988} then loses the same energy to $t = 0.8$ with local timesteps as with one, $5.7\times10^{-4}$ against $5.8\times10^{-4}$ of the total. With the linear extrapolation it lost twice as much.

The extra passes are not free. In three dimensions each block update costs about 10~percent more without gravity and 50 to 70~percent more with it, where every substep also relaxes the potential. Every few cycles, Shaderous therefore reads back the share of block updates that a cycle actually makes, and falls back to a single timestep while that share is above 80~percent (65~percent with gravity). A three-dimensional Sedov blast at a finest resolution of 256 cells with three levels makes 2.2 times fewer block updates with local timesteps and runs 1.9 times faster. An isothermal collapse at the same resolution makes 2.3 times fewer and runs 1.6 times faster. Figure~\ref{fig:amr}c shows how the bins follow a blast.

\section{Gravity, point masses and sinks}
\label{sec:gravity}

\subsection{Poisson solver}
\label{sec:poisson}

The potential is solved every step by geometric multigrid \citep{Brandt1977}, with red-black Gauss-Seidel smoothing, restriction by the average of eight cells, and trilinear prolongation, warm-started from the previous potential extrapolated linearly in time. A V-cycle makes two smoothing sweeps before each restriction, two after each prolongation, and eight on the coarsest level of $2^3$ cells. A step runs one V-cycle on the uniform grid: from the extrapolated guess one cycle leaves the solve closer to converged than two from the previous potential, which in an isothermal collapse let total momentum drift ten times as far. On the adaptive mesh it runs up to three, and the later ones only while they still reduce the residual. Periodic boundaries subtract the mean density from the source. For isolated boundaries a reduction computes the moments of the mass about the box center up to the fourth order, a pass moves them to the center of mass, and the boundary values of the potential come from the multipole expansion to $l = 4$,
\begin{equation*}
\Phi_{\rm b} = -G\sum_{l=0}^{4}\frac{1}{r^{l+1}}\sum m\,|\mathbf{y}|^l\,P_l(\hat{\mathbf{n}}\cdot\hat{\mathbf{y}}),
\end{equation*}
where the inner sum runs over the cells, $m$ is the mass of a cell and $\mathbf{y}$ its position relative to the center of mass, $r\hat{\mathbf{n}}$ is the position of the boundary point relative to the center of mass, and $P_l$ is the Legendre polynomial. Each term is a polynomial in the components of $\mathbf{y}$, so the sum follows from the moments. The dipole term vanishes about the center of mass. The terms of the expansion fall as $(|\mathbf{y}|/r)^l$, so the truncated sum is accurate only where the mass lies well inside the boundary, and a problem with isolated boundaries should keep its mass away from the faces of the box.

On the adaptive mesh the potential lives on every block, parents included. The density is averaged into the parents, a level's ghost cells come from its neighbors or are interpolated from the level above, and a multigrid cycle runs inside the blocks, at 16, 8, 4, 2 and 1 cells across. The coupling between levels is one-way: a fine level never corrects the coarse potential beneath it. A level's boundary values are trilinear in the coarser potential, extended linearly through the nearest coarse cell where a coarse cell that they need lies beyond the box or where the coarser level ends. On a uniform sphere the error is 0.03~percent with two levels at a finest resolution of 128 cells and 0.05~percent with three at 256, against 0.15~percent on $64^3$ blocks without refinement. A new mesh is solved level by level from the coarsest down. During a run all levels relax in the same dispatches, each against the boundary values that its coarser level had at the start of the cycle, so the cost does not grow with the depth of the tree.

Neither this lag nor a converged potential keeps the net force of the gas's gravity on itself at zero, as it is for an isolated body. Because of the lag, a body that moves across the mesh is pulled back toward where it was, so its momentum decays. A converged potential, differenced across refinement boundaries that lie unevenly around a moving body, pushes it forward by about a third as much. A white dwarf of 0.6~\Msun\ that moved at 182~\kms\ across a mesh of three levels, with cells of 156~km at the finest, lost 1.8~percent of its momentum in 27~s. On a uniform grid of $256^3$ cells the change was 0.02~percent. In a frame that turns with a binary, the drag toward rest in that frame moved the center of mass off the axis of rotation along an outward spiral, which e-folded every 1.8 orbits. With isolated boundaries, a reduction over the leaves therefore sums $\rho\,\mathbf{g}$ over the field of each step, and the next field subtracts this net force per unit mass as a uniform acceleration $\mathbf{a}$. The potential gains $\mathbf{a}\cdot\mathbf{x}$, so that the energy and the balanced reconstruction difference the same field that the kicks use. The white dwarf then changed its momentum by 0.02~percent, and the step took no measurable extra time.

\subsection{Coupling to the gas}
\label{sec:coupling}

Gravity enters the update as kick-drift-kick: half a kick with the old potential inside the update pass, and the other half after the new potential is solved. Applying the whole kick after the flux update leaks internal energy from gas in hydrostatic balance and slowly cools a star. The gravitational energy source defaults to a flux form, in which the work on a cell is the mass flux through each face times the difference of the potential across it, as in \citet{Springel2010}. Along one axis, the energy of cell $i$ changes by
\begin{equation*}
\begin{aligned}
\Delta E_i = -\frac{\Delta t}{2\Delta x}\bigl[&F_{i+1/2}\,(\Phi_{i+1} - \Phi_i)\\
&+ F_{i-1/2}\,(\Phi_i - \Phi_{i-1})\bigr],
\end{aligned}
\end{equation*}
where $F$ is the mass flux through a face and $\Phi$ is a potential at the start of the step (below), and the axes add. Summed over the cells, this work is the negative of the change in $\sum\rho\Phi$ that the same mass fluxes cause. In a fixed potential the sum of the gas energy and the gravitational energy is therefore conserved to round-off. With self-gravity the potential changes during the step, and exact conservation needs the mean of the old and new potentials. On the uniform grid the update computes the work with the old potential. Once the new potential is known, the pass of the second kick subtracts from each cell half its change of density times the change of its potential, which moves the summed work to that mean. The total is then conserved to the accuracy of the Poisson solve. On the adaptive mesh the potential is extrapolated to the half step from the last two solves.

\begin{figure}[t]
\centering
\includegraphics[width=\columnwidth,height=\colfigheight,keepaspectratio]{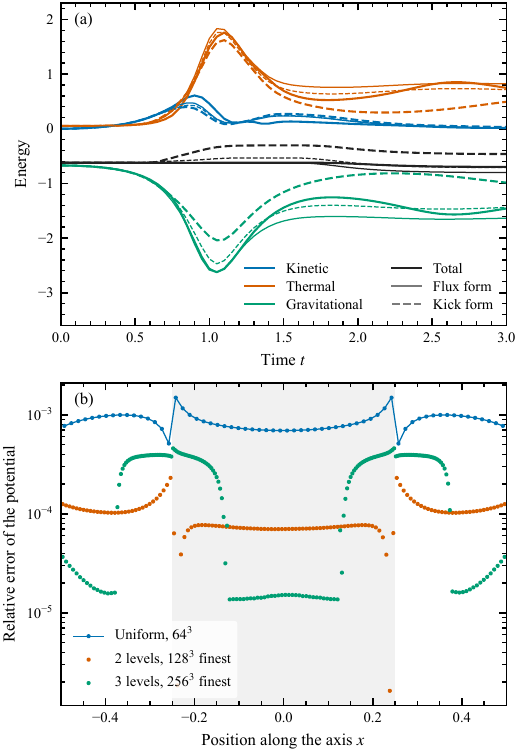}
\caption{Gravity. (a)~Kinetic, thermal, gravitational and total energy during the collapse of \citet{Evrard1988}, with the gravitational energy source in flux form (solid) and kick form (dashed), on a uniform grid of $64^3$ cells (thick lines) and on the adaptive mesh with a finest resolution of 128 cells and two levels (thin lines). After $t \approx 1.5$ the outgoing shock reaches the edge of the box and gas leaves it. (b)~Relative error of the potential of a uniform sphere with isolated boundaries, along the $x$ axis through its center, on a uniform grid and with two and three levels of refinement. The sphere fills the shaded interval.}
\label{fig:gravitytests}
\end{figure}

The alternative, the kick form, adds exactly the kinetic energy of each kick, so gravity never touches the internal energy, at the cost of total-energy conservation where the flow is under-resolved. Figure~\ref{fig:gravitytests}a compares the two forms in the collapse of \citet{Evrard1988}. The pull of the point masses (Section~\ref{sec:sinks}) and of an accelerating frame always takes the kick form. Debris torn from a star is cold and fast, and each step the tidal field changes its kinetic energy by hundreds of times its internal energy. In flux form, the mismatch between that work and the kick went into the internal energy. This mismatch is of the order of the truncation error. It cut the entropy of the leading half of a disrupted polytrope of index 3 a hundredfold. In kick form the entropy of the star holds to about 0.1~dex. Optionally, a pressure floor keeps the Jeans length resolved by a chosen number of cells, at least the four of \citet{Truelove1997}.

\subsection{Initial states in discrete balance}
\label{sec:settle}

A star set up from its analytic profile is in balance with the analytic potential, not with the one that the Poisson solver finds on the grid, and it rings in response to the mismatch. \citet{Zingale2002} met the same problem for stellar models mapped onto a grid, and put the mapped model into discrete hydrostatic equilibrium. Shaderous instead settles every star of a problem before the first step. The density is set to satisfy $h(\rho) + \Phi = C$ with the grid's potential and the star's mass, where $h$ is the specific enthalpy of the star's cold gas and $C$ a constant. The potential is then solved again, and the two steps are repeated until the density changes by less than $10^{-5}$, which takes six to eight rounds for a polytrope. Several stars settle together, in the same rounds. Each is balanced in its own potential: the grid's potential less the pull of the other stars, which outside a spherical star is that of a point of its mass. A star that moves keeps its velocity. A round changes only the density, and $C$ is found by regula falsi from its value in the previous round. For white dwarfs each round's change is over-relaxed by a factor of 1.6, and the rounds stop at a change of $10^{-3}$. They settle in 7 to 12 rounds, which take 1.6 to 5.5~s at finest resolutions of 256 and 512 cells.

Mapped without settling, a white dwarf of one solar mass rang in its fundamental mode, and its central density swung by 30~percent at a finest resolution of 256 cells (Figure~\ref{fig:balance}b). Its core is close to relativistic, with $\Gamma_1$ not far above 4/3, so the small mismatch meets little restoring force. Settled, the central density stays within 0.5~percent over six dynamical times. Settling leaves a star cold, so a problem that ignites a star adds the heat of its hot spot to the settled state. For a polytrope of index 1 the relation $h(\rho) + \Phi = C$ is exactly the trapezoid rule of Equation~(\ref{eq:wb}). On the adaptive mesh, at a finest resolution of 128 cells with two levels, the central density of the settled polytrope strays by 0.14~percent over three dynamical times, against 0.26~percent for the plain scheme, and its energy by 0.33~percent against 0.77~percent.

Planets of the ANEOS materials (Section~\ref{sec:aneos}) are not settled, because a core and a mantle of different materials do not follow one relation between pressure and density. A planet is given by its mass, the material of its mantle, optionally the material of a core and its share of the mass, and the temperature of its surface, by default 2000~K for a body still hot from its accretion. Each layer lies on its own adiabat, $\mathrm{d}e = p\,\mathrm{d}\rho/\rho^2$, so that a layer moved as a whole stays on it, and the pressure and temperature are continuous where the layers meet. Shaderous integrates the structure inward from the surface with fourth-order Runge-Kutta steps, and it adjusts the radius until the mass runs out at the center. A planet of $1\,\Mearth$ with 30~percent of its mass in a core of Fe85Si15 under pyrolite has a radius of $1.024\,\Rearth$, a central pressure of 351~GPa and a central density of 12.6~g~cm$^{-3}$, against 364~GPa and 13.1~g~cm$^{-3}$ at the center of the Earth \citep{Dziewonski1981}. Started on this structure at 16 cells to its radius, its central density oscillates by 3.6~percent.

A star can also have a core that is a point mass, as a giant has. Part of its mass is then a point mass $M_{\rm c}$ at its center, with the softened potential $-GM_{\rm c}/\sqrt{r^2 + \epsilon^2}$ of the point masses of Section~\ref{sec:sinks}, where $\epsilon$ is the softening length. The rest is a polytropic envelope of mass $M_{\rm env}$, index $n$ and radius $R$, in balance with its own gravity and with the pull of the core. We write the specific enthalpy of the envelope as $h = h_{\rm c}\vartheta$, with $h_{\rm c}$ its value at the center, and its mass inside the radius $r = xR$ as $M_{\rm env}\,\psi/A$. Then
\begin{equation*}
\frac{\mathrm{d}\vartheta}{\mathrm{d}x} = -\frac{1}{x^2}\left[\psi + \frac{A M_{\rm c}}{M_{\rm env}}\,\frac{x^3}{(x^2 + \epsilon^2/R^2)^{3/2}}\right],
\end{equation*}
\begin{equation*}
\frac{\mathrm{d}\psi}{\mathrm{d}x} = B x^2\vartheta^n,
\end{equation*}
with $\vartheta(0) = 1$, $A = GM_{\rm env}/(h_{\rm c}R)$ and $B = 4\pi G R^2\rho_{\rm c}/h_{\rm c}$, where $\rho_{\rm c}$ is the central density of the gas. The envelope ends at $R$ and holds $M_{\rm env}$, so $\vartheta(1) = 0$ and $\psi(1) = A$, and two nested bisections find $A$ and $B$. Without a core only their product matters, and the equations reduce to the Lane-Emden equation. The core joins the point masses of the run and moves with the star. The settling adds the potential of the core to that of the grid, and a star with a core settles on the CPU. We set up a giant of $1\,\Msun$ with a core of $0.4\,\Msun$, softened by a tenth of its radius, and resolved its radius by 32 cells of the finest level. It held its central density within 6~percent over six dynamical times, with a kinetic energy below $2\times10^{-3}$ of its internal energy.

A star on a circular orbit does not fall freely, so the stars of a circular binary are settled at rest in the frame that turns with the orbit at $\Omega$. Each is balanced in the Roche potential: that of the grid and the point masses, plus $-\tfrac12\Omega^2\varpi^2$, where $\varpi$ is the distance from the axis of rotation. A star thus starts tidally locked and stretched toward its companion, out to its inner Lagrange point at most. The tide pulls a star along the line of centers, so each round also adds a uniform pull to its potential that holds its center of mass in place. A setup can give the separation through the fraction of its Roche lobe that the first star fills, with the radius of the lobe from the approximation of \citet{Eggleton1983}. A single star in a rotating frame is settled spinning rigidly with the frame. A polytrope of index 1 at $\Omega = 0.4$ on $128^3$ cells then holds its central density to 0.2~percent and its axis ratio to $10^{-3}$ over four dynamical times.

\subsection{Point masses and sinks}
\label{sec:sinks}

Up to 32 point masses act on the gas through Plummer-softened gravity \citep{Plummer1911}. The texture of the gravity field holds self-gravity alone. Every pass that needs the pull of the masses (the kicks, the predictor, the timestep, the well-balanced reconstruction) adds it from their positions as the pass runs. The potential difference of a mass between neighboring cells is formed as $GM(q_1 - q_0)/[r_0 r_1 (r_0 + r_1)]$ rather than as the difference of two potentials thousands of times larger. Here $q = r^2 + \epsilon^2$, $r_0$ and $r_1$ are the softened distances of the two cells from the mass, and $\epsilon$ is its softening length.

Each step the masses drift. Then a reduction over the updated gas gives every mass two quantities: its pull on the gas, whose reaction moves the mass, and the mass that its sink takes. One reduction serves every four masses. The second half-kick pass removes exactly that amount from the cells. Inside a sink radius gas is removed at a rate that falls from $\sqrt{GM/r^3}$ at the center to zero at the radius, and the gas that is removed takes its momentum to the mass. Mass and momentum are conserved between gas and masses to $10^{-7}$ in the validation test. Sink particles were introduced in smoothed particle hydrodynamics by \citet{Bate1995} and brought to grid codes by \citet{Krumholz2004} and \citet{Federrath2010}. Our removal at a local orbital rate is closer to the sinks of circumbinary-disk studies, such as \citet{Farris2014}, who remove gas on a local viscous timescale. The pull between two masses is softened by the larger of their two lengths. By default the softening length and the sink radius of a mass are two cells. On the adaptive mesh these are cells of the finest level when the problem keeps the blocks near its masses at that level, as the binary problem does. Otherwise they are cells of the coarsest level, the only size that is certain to exist around the mass.

The timestep of the gas does not follow two masses that pass close to one another. Each pair of masses therefore limits the step to one sixteenth of the CFL number times the lesser of $r/v$ and $\sqrt{r^3/GM}$, where $r$ is the softened separation of the pair, $v$ its relative speed and $M$ its total mass. At the default CFL number of 0.8 a circular orbit takes about 125 steps. Two masses on an orbit of eccentricity 0.9 in near-empty gas keep the energy of the orbit to 0.7~percent over two periods with this limit. Without it they lose 94~percent.

The frame is inertial unless a problem specifies another. The tidal disruption problem uses the rest frame of the gas: as in the simulations of \citet{Guillochon2009} and \citet{Guillochon2013}, the frame accelerates with the center of mass of the gas, so the star stays at the box center. The orbit of the perturber is integrated alongside the gas. The binary problem uses the rest frame of the masses instead. That frame accelerates with the mean pull of the gas on the masses, and the gas feels the opposite acceleration. After each step we also set the velocity of the masses' center of mass back to zero, which removes the slow drift that rounding in their mutual pull and the momentum of accreted gas would cause. The box, its boundaries and the damping zone of the disk (Section~\ref{sec:interface}) then move with the binary.

\section{Magnetic fields}
\label{sec:mhd}

\begin{figure*}[t]
\centering
\includegraphics[width=\textwidth]{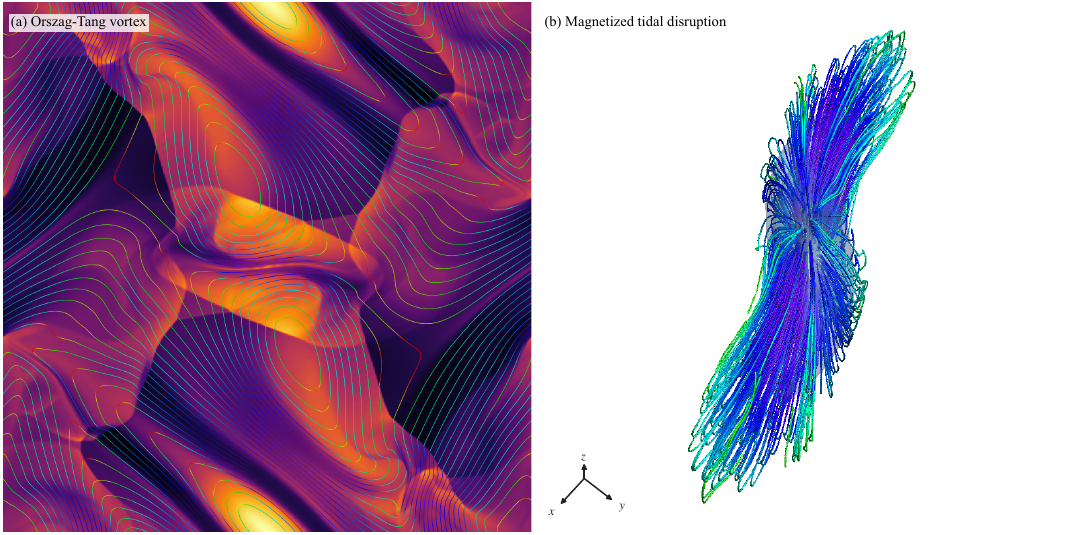}
\caption{Magnetic field lines, drawn by the code's own renderer and colored by the magnetic pressure on a logarithmic scale. (a)~The vortex of \citet{Orszag1979} on $512^2$ cells at $t = 0.5$. The background is the density, and the lines are 28 contours of the flux function, so the same magnetic flux passes between any two neighbors and the lines crowd where the field is strong. (b)~A star threaded by a dipole field at $t = 6$, about two dynamical times after it passed a black hole of $10^6\,\Msun$ at an impact parameter $\beta = 0.7$ ($256^3$ at the finest level, three levels of refinement). The lines are traced in three dimensions from 256 seeds and drawn as tubes. The stripped streams stretch their field straight, and the surviving core winds its field up. The arrows give the orientation of the axes.}
\label{fig:fields}
\end{figure*}

With a magnetic field, Shaderous solves the equations of MHD, ideal or with a constant resistivity (Section~\ref{sec:resistivity}), in units in which the magnetic pressure is $B^2/2$, so that the plasma beta, the ratio of the gas pressure to the magnetic pressure, is $\beta_{\rm M} = 2p/B^2$, and the total energy includes $B^2/2$. Gravity, point masses, sinks, cooling, viscosity, conduction, species and burning act on magnetized gas as they do without a field. A sink removes mass, momentum and the kinetic and internal energy of the gas that it takes, but leaves the field, which it could not remove without breaking constrained transport. Viscosity and conduction stay isotropic: the code has no transport that follows the direction of the field. Figure~\ref{fig:fields} shows the field lines of two of the magnetized problems of Section~\ref{sec:problems}, as the code draws them (Section~\ref{sec:interface}). Obstacles and inflows cannot be combined with a field, because the fluxes of the MHD solver have no treatment of a solid or held cell.

\subsection{Constrained transport}
\label{sec:ct}

The field lives on the faces of the cells, as in the constrained transport of \citet{Evans1988}. Each face is updated from the electric fields on its four edges, and every edge enters the faces around it with opposite signs, so the divergence of no cell changes. An initial field is the discrete curl of a vector potential sampled on the edges, so it starts free of divergence to rounding. The integrator is the van Leer predictor-corrector of \citet{StoneGardiner2009}, as in Athena++ \citep{Stone2020}. A first stage takes half a step with first-order fluxes. A second stage takes the whole step with fluxes from a limited linear reconstruction of the half-step state. Each stage computes every face flux once into a storage buffer, from which both cells read it, then computes the electric field on each edge, and then updates the cells and faces. These passes run in slabs along one axis, so the buffers hold a few layers of the grid at a time.

The electric fields on the edges follow the construction of \citet{GardinerStone2005}. It upwinds their gradients by the direction of the mass flux, which keeps intact a loop of field that a uniform flow carries. We also implemented the edge estimates of \citet{LeeDeane2009} and the upwind selection of \citet{Lee2013}. In the field-loop test these kept about 1~percent more of the magnetic energy, but they raised the overshoot of the peak field from 8~percent to 17 to 21~percent at $128^2$ cells and left more stray field outside the loop, so the default remains that of \citet{GardinerStone2005}. The fluxes come from the HLLD solver of \citet{Miyoshi2005}, which resolves contacts and Alfv\'en waves, or from HLL with the fast magnetosonic speeds in the estimates of \citet{Davis1988}. As HLLC does, HLLD falls back to HLL next to cells where the flow converges and the total pressure $p + B^2/2$ jumps by more than the threshold of the shock detector. The signal speed of the timestep is $|v_x| + |v_y| + |v_z| + d\,c_{\rm f}$, where $d$ is the number of dimensions and $c_{\rm f} = (c_{\rm s}^2 + B^2/\rho)^{1/2}$ the fastest magnetosonic speed.

Constrained transport keeps the divergence unchanged only in exact arithmetic. In single precision each face rounds its own sum of the impulses from its edges, and the rounding makes the divergence drift away from zero: in the vortex of \citet{Orszag1979} the largest $|\nabla\cdot\mathbf{B}|\,\Delta x/\max|B|$ reached $3.5\times10^{-6}$ after 750 steps. Each face therefore holds a pair of numbers, the field and a compensation in a second texture that keeps what the rounding of the field dropped, and the impulses are added with an error-free summation. The pair holds the exact sum of the face's impulses, and the divergence stays at a few roundings: $7\times10^{-8}$ after 751 steps and $9\times10^{-8}$ after 3483. The face field and its compensation, at both parities and at the half step, add 80 bytes per cell to the 112 that the uniform grid holds for the gas.

Next to the edge of a star in near-vacuum, the first stage can fill a nearly empty cell to hundreds of times its density. The second stage then pushes more gas out of the half-step cell through its other faces than the cell held at the start, and the update empties it. A cell whose density the first stage changed by more than a factor of four either way therefore takes the first-order HLL flux of the starting state on all of its faces, a step that keeps the density positive. A factor of four is more than a strong shock compresses gas. The face makes this decision from its own two cells, so the cells, slabs or blocks that share a face compute the same flux, and the edge fields made from these fluxes keep the divergence.

This test misses a cell that the second stage empties while the first stage barely changed its density, as in the thin gas above a magnetized disk. The second stage therefore also checks every cell after its update, as Athena++ does. A cell left with less than an eighth of its density, or with no internal energy, is listed. Its faces then take the first-order HLL fluxes of the starting state for mass, momentum and energy, and it and its neighbors are updated again. The fluxes of the field are left as they were, so the field and its divergence do not change. A cell updated again beside a corrected face takes its pressure from its separately evolved internal energy where the total energy would otherwise heat it to many times that energy. A neighbor that this leaves nearly empty is corrected in a second round.

On the uniform grid the list runs through all the slabs of a step, so slabs of any size give the same bits. On the adaptive mesh the update marks the cells in an unused channel of its output, a scan lists the marked blocks and their neighbors, and the flux and update passes of the second stage run again over that list. The corrected faces on a block's surface go into the flux register, which coarser and slower neighbors read. When nothing is marked, the extra passes end at once. On the RTX~3080 a three-dimensional MHD step on the uniform grid costs 3~percent more and a two-dimensional one 8~percent less, because the update's code compiles better in its new form. On the adaptive mesh the two rounds add seven small passes to a step, about 0.13~ms on the RTX~3080 however little they do. That is under 1~percent of a three-dimensional step, and 11.5~percent of a two-dimensional one of 1.1~ms. Where a magnetized problem enables it, a ceiling on the Alfv\'en speed adds mass at rest to cells whose $|B|/\rho^{1/2}$ would exceed it. This is not conservative, and only the magnetized disk uses it (Section~\ref{sec:problems}).

\subsection{Resistivity and walls}
\label{sec:resistivity}

With a constant magnetic diffusivity $\eta$, the electric field on each edge gains $\eta\mathbf{J}$, where $\mathbf{J} = \nabla\times\mathbf{B}$ is the current. The induction equation then gains $\eta\nabla^2\mathbf{B}$, and constrained transport keeps the divergence where it was. The current comes from the field at the cell centers, on an edge from the four cells around it. The energy flux through each face gains the matching Poynting flux $\eta(\mathbf{J}\times\mathbf{B})\cdot\hat{\mathbf{n}}$, so that the total energy is conserved to round-off, and the separately evolved internal energy gains the Ohmic heat $\eta J^2$. The diffusivity limits the timestep as viscosity does (Section~\ref{sec:diffusion}). A wave of the field at rest, held in balance by the gas pressure, loses its magnetic energy at the rate $2\eta k^2$ to within 0.6~percent, in two and three dimensions and on uniform grids and blocks.

The faces of the box can be periodic, open, open to outflow only, or reflecting walls. At a reflecting wall the field normal to the wall changes sign, so the field lies along the wall. A field whose lines cross the walls, such as that of a collapsing X-point, needs walls that hold the lines in place, like the closed boundaries of \citet{Thurgood2018} that conserve the flux through them. At such a line-tied wall the ghost cell reverses the whole velocity and keeps the field of the cell inside, so the gas does not slip. The electric field along the wall is zero, so the field through each face of the wall never changes. A moving wall slides along itself. The velocity of its ghost cell is $2\mathbf{v}_{\rm w} - \mathbf{v}$, where $\mathbf{v}_{\rm w}$ is the velocity of the wall, and the electric field along it is $-\mathbf{v}_{\rm w}\times\mathbf{B}_n$, so that the footpoints of the field lines move with it. Without a field the two kinds are a no-slip wall at rest and a moving lid. In a square cavity whose lid moves at unit speed at a Reynolds number of 100, the velocity along the vertical line through the center settles to within 0.0065 of the values of \citet{Ghia1982} at every height in their table. The rules for the ghost cells of each kind of face are written once and shared by every grid.

\subsection{Magnetic fields on the adaptive mesh}
\label{sec:mhdamr}

On the adaptive mesh every face of every leaf block is stored. With a field, the tree keeps neighbors across edges, and not only across faces, within one level of each other. At a refinement boundary a coarse cell takes the mean of the fine fluxes, which the fine blocks write to a flux register, and every edge takes the electric field of the finest blocks around it. Blocks of one level compute an edge that they share from identical inputs, and a coarser block takes the mean of the finer block's two values along the edge. Each coarse face then stays the mean of the fine faces on it, and the field stays free of divergence across levels. On a refined mesh, energy is conserved to $2\times10^{-7}$, and the magnetic energy of the rotor and of a blast is within 0.3 and 0.1~percent of a uniform grid at the finest resolution.

When the mesh changes, a copied block keeps its faces and their compensations, and a merged block's face is the mean of the old children's faces on it. A new child takes each fine face on the boundary of a coarse cell from an old leaf at its level where one exists, and otherwise from the coarse face, with limited slopes that keep its mean. The faces inside the coarse cell start from the mean of the outer faces in line with them. They are then corrected by the smallest change that makes every fine cell free of divergence, $-\mathbf{A}^{\rm T}\mathbf{L}^{+}\mathbf{r}$, where $\mathbf{r}$ is the divergence of the fine cells, $\mathbf{A}$ the incidence matrix of the interior faces on the cells, and $\mathbf{L} = \mathbf{A}\mathbf{A}^{\rm T}$ the graph Laplacian of the $2^d$ fine cells. The Walsh functions diagonalize $\mathbf{L}$, so applying its pseudo-inverse $\mathbf{L}^{+}$ costs a transform and a division. This prolongation meets the aims of \citet{TothRoe2002}: it is exactly free of divergence, it is conservative, and it keeps the fine faces that exist, but it does not use their formulas. Carried once around the box with 144 rebuilds of the mesh, a dense field loop keeps $|\nabla\cdot\mathbf{B}|\,\Delta x/\max|B|$ at $1.5\times10^{-7}$ and 96~percent of its magnetic energy, the same share as on a uniform grid at the finest resolution.

Per-block timesteps also work with a field. Neighbors across faces, edges and, in two dimensions, vertices are kept within one bin of each other. A slower block takes the fluxes and edge impulses of a faster neighbor's first substep during its own step, and those of the second substep at its midpoint, so both copies of a shared face receive the same impulses. The saving depends on the contrast in signal speed. A two-dimensional MHD blast at $t = 0.05$ makes 85~percent of the block updates of a single timestep, while problems whose signal speed varies by less than the factor of two between bins make 98 to 100~percent.

\section{Radiation transport}
\label{sec:radiation}

Shaderous can carry gray radiation as two fields of its own, the radiation energy density $E_{\rm r}$ and flux $\mathbf{F}_{\rm r}$, which it evolves with the M1 closure on the uniform grid and on the adaptive mesh, in Cartesian coordinates. The method follows the two-moment modules of Athena \citep{Skinner2013}, Fornax \citep{Skinner2019} and QUOKKA \citep{Wibking2022}, but it chooses the reduced speed of light cell by cell (Section~\ref{sec:rsla}). With radiation off, a run builds none of its passes and computes the same bits as before. Shaderous evolves the moments in the frame of the gas, to first order in $v/c$ and without the velocity terms of the flux equation, as in Fornax:
\begin{equation}
\begin{aligned}
\partial_t E_{\rm r} &+ \xi\bigl[\nabla\cdot(\mathbf{v}E_{\rm r}) + \nabla\cdot(\mathbf{v}\cdot\mathsf{P}_{\rm r}) + \nabla\cdot\mathbf{F}_{\rm r}\bigr]\\
&= \xi\bigl[c\kappa_{\rm P}\rho\,(a_{\rm r}T^4 - E_{\rm r}) - \kappa_{\rm R}\rho\,\mathbf{v}\cdot\mathbf{F}_{\rm r}/c\bigr],\\
\partial_t \mathbf{F}_{\rm r} &+ \xi\bigl[\nabla\cdot(\mathbf{v}\mathbf{F}_{\rm r}) + c^2\,\nabla\cdot\mathsf{P}_{\rm r}\bigr] = -\xi c\kappa_{\rm R}\rho\,\mathbf{F}_{\rm r},
\end{aligned}
\label{eq:m1}
\end{equation}
where $c$ is the speed of light, $a_{\rm r}$ the radiation constant, $T$ the temperature of the gas, $\kappa_{\rm P}$ the absorption opacity, $\kappa_{\rm R}$ the total opacity of absorption and scattering, and $\mathsf{P}_{\rm r}$ the radiation pressure tensor. The factor $\xi \le 1$ reduces the speed of light (Section~\ref{sec:rsla}). The gas gains momentum at the rate $\kappa_{\rm R}\rho\mathbf{F}_{\rm r}/c$ per unit volume and internal energy at the rate $c\kappa_{\rm P}\rho(E_{\rm r} - a_{\rm r}T^4)$. The two terms with $\mathsf{P}_{\rm r}$ and $\mathbf{v}\cdot\mathbf{F}_{\rm r}$ in the energy equation together make the compression work $\mathsf{P}_{\rm r}:\nabla\mathbf{v}$. They are split so that the energy that the radiation loses to the work of its pressure is the kinetic energy that the gas gains.

With radiation on, the equation of state of gas and radiation and that of degenerate electrons (Section~\ref{sec:eos}) leave their own radiation out, because $E_{\rm r}$ carries it. The transport runs with the ideal gas, gas and radiation, degenerate electrons and ionizing hydrogen, and not with the tabulated equations of state or the liquid. The closure of \citet{Levermore1984} gives the pressure tensor from the reduced flux $\mathbf{f} = \mathbf{F}_{\rm r}/(cE_{\rm r})$,
\begin{equation*}
\mathsf{P}_{\rm r} = \Bigl[\tfrac{1}{2}(1 - \chi_{\rm E})\,\mathsf{I} + \tfrac{1}{2}(3\chi_{\rm E} - 1)\,\hat{\mathbf{f}}\hat{\mathbf{f}}\Bigr]E_{\rm r},
\end{equation*}
\begin{equation*}
\chi_{\rm E} = \frac{3 + 4|\mathbf{f}|^2}{5 + 2\sqrt{4 - 3|\mathbf{f}|^2}},
\end{equation*}
where $\hat{\mathbf{f}}$ is the direction of the flux. The Eddington factor $\chi_{\rm E}$ runs from 1/3 for isotropic radiation to 1 for a beam. The same equations therefore describe diffusion through opaque gas and free streaming through transparent gas, and a beam casts a shadow behind an opaque obstacle. Six problems that ship with the code use the transport: a hot sphere that cools through its surface, a blast dominated by radiation, a beam with a shadow, a subcritical and a supercritical radiative shock, and a polytrope with a tenth of its pressure in radiation.

\subsection{The reduced speed of light}
\label{sec:rsla}

Explicit transport at the true speed of light would limit every step to $\Delta x/c$. The usual remedy replaces $c$ by a reduced speed $\hat{c}$ everywhere \citep{Skinner2013, Wibking2022}. This multiplies the weight of radiation in equilibrium with the gas by $c/\hat{c}$: in a closed box at $\hat{c} = c/10$, gas and radiation relaxed to a temperature 26~percent too low. Shaderous reduces the speed of light only in cells that are so transparent that the true speed would limit the step:
\begin{equation*}
\xi = \min\Bigl[1,\ \frac{\hat{c}}{c}\max\Bigl(1,\ \frac{\tau_{\rm c}}{\tau_0}\Bigr)\Bigr],
\end{equation*}
where $\tau_{\rm c} = \kappa_{\rm R}\rho\Delta x$ is the optical depth of a cell and $\tau_0 = 2.5/3$. In an opaque cell the step is limited by diffusion at $D = \xi c/(3\kappa_{\rm R}\rho)$, and at $\tau_0$ the signal speed of this diffusion equals the streaming speed $d\hat{c}$, where the signal speed is $2.5\,dD/\Delta x$ as for conduction (Section~\ref{sec:diffusion}). With $c/\hat{c} = 100$, every cell of optical depth above about 80 has the true speed of light, so the inside of a star has the right heat capacity. The one condition on $\hat{c}$ is $\hat{c} \gg |v| + c_{\rm s}$, and the diagnostics count the cells where it fails.

Where $\xi < 1$ the conserved total is $E + E_{\rm r}/\xi$. Every term of Equation~(\ref{eq:m1}) but the time derivative carries $\xi$, the advection with the gas included, so $\xi$ drops out of a steady state, in moving gas too. A reduced speed of light still cannot give a radiative structure that moves across the grid through cells with $\xi < 1$. The precursor of a shock that moves at speed $u_{\rm s}$ then has a head start of $\hat{c}/u_{\rm s}$ instead of $c/u_{\rm s}$. Behind the supercritical shock of Section~\ref{sec:radtests}, the temperature is 3.2~percent low at $\hat{c} = c/5$, 1.1~percent at $c/2.5$ and 0.1~percent at $c$. \citet{He2026} proposed a reduction factor that is also chosen cell by cell. Theirs follows from the stability of an implicit-explicit update and depends on the timestep: it returns to the true speed of light wherever the exchange with the gas damps the modes that a reduced speed would otherwise have to slow. Ours follows from the diffusion limit of the explicit sweeps and depends only on the optical depth of the cell. A setting restores the usual $\xi = \hat{c}/c$ everywhere.

\subsection{Transport and exchange}
\label{sec:radstep}

A step of the gas carries the radiation with it, and $N_{\rm r}$ sweeps (eight by default) then move the radiation between cells and exchange energy and momentum with the gas. The advection moves $E_{\rm r}/\rho$ and $\mathbf{F}_{\rm r}/\rho$ with the mass fluxes that the update recorded at every face, as for species (Section~\ref{sec:species}), so trapped radiation moves exactly as the gas does. In the sweeps, each cell computes the fluxes of all of its faces itself, so the two cells of a face compute the same bits (Section~\ref{sec:conservation}). $E_{\rm r}$ and $\mathbf{f}$ are reconstructed linearly, with the monotonized central limiter for $\mathbf{f}$ and a limit of 1.2 times either one-sided difference for $E_{\rm r}$. With the monotonized central limiter on $E_{\rm r}$ too, the surface of an opaque body sent out its light in pulses. The flux is the local Lax-Friedrichs flux with the largest wave speed of the M1 system normal to the face. Its dissipation is scaled by $\min(1, 1/\tau_{\rm f})$, where $\tau_{\rm f}$ is the optical depth of a cell at the face \citep{Skinner2019}, so that in opaque gas the scheme diffuses at $c/(3\kappa_{\rm R}\rho)$ whether or not the cells resolve the mean free path. Each cell multiplies the divergence of its fluxes by its own $\xi$. A sweep is a forward Euler stage, stable to a Courant number of 0.5, so its step is $\Delta t_{\rm s} = 0.5\,\Delta x/(d\hat{c})$.

After the transport, each cell exchanges energy and momentum with its gas implicitly and on its own. The flux first relaxes to $\mathbf{F}_{\rm r}' = \mathbf{F}_{\rm r}^*/(1 + \vartheta_{\rm R})$, with $\vartheta_{\rm R} = \xi c\kappa_{\rm R}\rho\Delta t_{\rm s}$, and the gas gains the momentum that the flux lost. Backward Euler for the internal energy density $e$ and for $E_{\rm r}$ then leaves one equation for the new temperature $T'$,
\begin{equation*}
e(T') + \zeta\,a_{\rm r}T'^4 = e + \zeta\,E_{\rm r}^*, \qquad \zeta = \frac{\vartheta_{\rm P}}{\xi(1 + \vartheta_{\rm P})},
\end{equation*}
with $\vartheta_{\rm P} = \xi c\kappa_{\rm P}\rho\Delta t_{\rm s}$, which Newton's method solves from a starting value that bounds the root from above. Both sides of the equation are sums of positive terms, and nothing subtracts $a_{\rm r}T^4$ from $E_{\rm r}$, so single precision is enough at any optical depth: at $\tau_{\rm c} = 10^{10}$, $E_{\rm r}$ stays within $8\times10^{-7}$ of $a_{\rm r}T^4$ over $10^4$ sweeps. The larger of $e'$ and $E_{\rm r}'/\xi$ takes up what the rounding of the root left, so the exchange conserves $e + E_{\rm r}/\xi$ exactly. The work that the radiation does on the gas is taken from $E_{\rm r}'$, and, summed over the sweeps, it is exactly the kinetic energy that the gas gains.

The opacity is constant (with absorption and scattering set apart), electron scattering with a small true absorption, or Kramers' free-free opacity plus electron scattering. With physical units, $c$, $a_{\rm r}$ and these opacities follow from the units and the composition. The luminosity is the energy that leaves through the open faces of the box, summed by a small pass after each step, so its history is the bolometric light curve. Snapshots carry $E_{\rm r}$ and $\mathbf{F}_{\rm r}$, and a run restarts from one with the same bits in its radiation.

\subsection{Radiation on the adaptive mesh}
\label{sec:radamr}

With radiation on, all blocks of the adaptive mesh take one timestep, and the local timesteps of Section~\ref{sec:timebins} are off. The advection and the sweeps advance every block together, as the passes of the species do. Before the advection and before every sweep, a pass assembles $E_{\rm r}$, $\mathbf{F}_{\rm r}$ and $\kappa_{\rm R}\rho$ for every leaf and two layers of ghost cells around it, which are filled as the gas's are (Section~\ref{sec:ghost}). Beside a coarser block a ghost cell takes the coarse cell beyond the face with minmod slopes, which keep $E_{\rm r}$ positive. Through a face that it shares with finer cells, a coarse cell takes the mean of their fluxes, so what leaves one level enters the other. Each cell keeps its own $\xi$, so a coarse cell may have the true speed of light where the fine cells beside it have a reduced one. When the mesh changes, a merged block takes the mean of its children and a new child takes its parent's cell with minmod slopes, so $E_{\rm r}$ and $\mathbf{F}_{\rm r}$ are conserved. Radiation adds 84 bytes per cell of the block atlas to the 152 that the gas uses.

\subsection{Validation and cost}
\label{sec:radtests}

A float64 version of the scheme in one dimension is the reference for the GPU, and Table~\ref{tab:validation-radiation} lists the 17 tests of the transport. They cover streaming and diffusion, a shadow, trapped radiation in moving gas, the non-equilibrium Marshak wave of \citet{SuOlson1996}, subcritical and supercritical radiative shocks, and a star held up partly by radiation, at rest and moving at ten times its sound speed, on the uniform grid and on three levels. On the RTX~3080 a step of $128^3$ cells takes 2.1~ms without radiation and 6.6~ms with eight sweeps, so a sweep costs about a quarter of a step of the gas. With $\hat{c}$ about 15 times the signal speed of the gas, a unit of simulated time costs 4.2 times as much as the gas alone.

\section{Frames and coordinate systems}
\label{sec:coordinates}

Besides a Cartesian box at rest, Shaderous runs in a rotating frame, with mirror symmetry, in comoving coordinates, in a shearing box, and in two dimensions as a body of revolution in cylindrical $(R, z)$ or spherical $(r, \theta)$ coordinates. Each one enters the passes through a small slot of WGSL that is compiled in only where it is used, so that every other run computes the same bits as before. Each has its own tests in the validation suite.

\subsection{Rotating frames}
\label{sec:rotating}

In a frame that turns at a constant $\boldsymbol\Omega$, the centrifugal force is added to the external potential of the point masses, and the well-balanced reconstruction balances it as it balances gravity. The Coriolis force turns the momentum in the plane normal to $\boldsymbol\Omega$ without changing its magnitude. We apply it as an exact rotation by half a step before and after the update, so that it does no work. In a periodic box with the Coriolis force alone, a uniform flow turns once in $\pi/\Omega$ with a phase error of $2.7\times10^{-7}$ and a change of kinetic energy of $5.6\times10^{-7}$. A binary that corotates with the frame keeps its separation to $7\times10^{-11}$.

A binary that transfers mass changes its orbit and drifts around a frame that turns at a fixed rate. The frame can instead follow the binary. After each step the GPU measures the axis of the quadrupole moment of the mass in the orbital plane and the rate at which that axis turns. It then changes $\Omega$ so that the axis returns to where it started, with critical damping, in about a third of an orbit. The gas also feels the Euler force $-\dot{\boldsymbol\Omega}\times\mathbf{x}$. In the problem of Roche lobe overflow (Section~\ref{sec:problems}), the orbit slows by 12~percent in two orbits as the donor loses gas. In the frame that follows the binary, the line of centers stayed within $4.5^\circ$ of where it started, while in a frame at the starting rate it turned by $32^\circ$. The liquid with implicit pressure also runs in a rotating frame (Section~\ref{sec:implicit}), so that it can model rotors.

\subsection{Mirror symmetry}
\label{sec:mirror}

A problem that is symmetric under reflection in one, two or three coordinate planes can be run on a half, a quadrant or an octant of the box. The lower face of a mirrored axis becomes a reflecting wall. The multigrid solver of Section~\ref{sec:poisson} treats it as a face of zero gradient, and the multipole boundary unfolds the moments of the computed region into those of the whole configuration. A magnetic field crosses a mirror as a pseudovector, keeping its normal component and reversing the others. When some axis of the box is not mirrored, a mirrored axis keeps the cell size of the full box and holds half as many cells, so the grid is rectangular. The diagnostics and the display show the whole configuration. A Sedov-Taylor blast on an octant matches the full box to $2.6\times10^{-6}$ in density. The Evrard collapse on an octant matches the full box in energy to $10^{-7}$, at an eighth of the cost.

\subsection{Comoving coordinates}
\label{sec:comoving}

For cosmological problems the state holds comoving densities and peculiar velocities in a box that expands with the scale factor $a$. The GPU advances $a$ with the clock, by one Runge-Kutta step of the Friedmann equation per step, so that the timestep still never passes through the CPU. Every gradient carries a factor $1/a$, and the passes work with the proper cell size $a\,\Delta x$. The drag of the expansion on the peculiar velocity and the internal energy is solved exactly over a step and applied in two halves around the update, in the slot of a non-inertial frame. The Poisson source is the comoving overdensity divided by $a$. In comoving coordinates the work of gravity on the gas is applied in kick form: with the flux form of Section~\ref{sec:gravity}, the first step of a cold Zel'dovich pancake heated the gas $10^5$-fold. The equations of state relate proper quantities, so every equation of state but the tabulated ones runs through wrappers that convert between comoving and proper values. A linear density mode in a matter-dominated universe grows to 0.992 of linear theory, and a sound wave in an expanding gas keeps its phase to $7\times10^{-4}$. On the uniform grid the pancake matches the analytic solution before caustic formation to 0.6~percent in density and 0.25~percent in velocity. On blocks the density error is 6.5~percent, which we have not explained.

\subsection{Shearing box}
\label{sec:shear}

The shearing box is the local model of a differentially rotating disk: a frame that turns at $\Omega$, with the tidal acceleration $2q\Omega^2 x$ along the radial axis $x$ and the background flow $v_y = -q\Omega x$, where $q = -\mathrm{d}\ln\Omega/\mathrm{d}\ln r$ is the shear parameter. Along $y$ the box is periodic. Along $x$ it is periodic with a shift that grows with the background flow, so a ghost cell is interpolated between the two cells that it overlaps. The two cells that share a shearing face compute its flux from different neighborhoods. We therefore average the face fluxes across the shearing boundary (and, with a magnetic field, the edge electric fields), so that mass, the momenta along $x$ and $z$, and the magnetic flux are conserved to round-off across it. A shearing wave in the gas follows the linear solution to 0.2~percent in amplitude as it swings from leading to trailing. A vertical field seeded with noise grows by the magnetorotational instability at 0.762 of the shear rate, against 0.75 from linear theory.

\subsection{Axisymmetric and spherical coordinates}
\label{sec:curvilinear}

In two dimensions a run can be a body of revolution, in cylindrical $(R, z)$ or spherical $(r, \theta)$ coordinates. The storage and the Riemann solvers are unchanged. The geometry enters as weights on the face fluxes, the ratios of each face's area to the cell's volume, and as one source term on the momentum. All of these are closed forms of the cell's indices, so the code never forms the areas and volumes as such, which in float32 would be small differences of large numbers. The cells next to the axis need two further measures. The total energy and the angular momentum cross the faces with different weights, so in fast-swirling cold gas the internal energy found by difference is mostly that mismatch. The dual-energy switch therefore takes the internal energy from its own equation where the swirl dominates. The timestep takes the larger of each cell's width and its volume over its outer face, which is half the width next to the axis or a pole. It is also held to the centrifugal acceleration, so that a swirl reaching the axis cannot empty the axis cells in one step.

In $(R, z)$ the radial reconstruction stays well balanced. The multigrid solver of Section~\ref{sec:poisson} solves self-gravity in two dimensions, on the uniform grid and on blocks, with the radial neighbors of a cell weighted by $R_\pm/R_{\rm c}$, the ratio of the radius of the shared face to that of the cell center. Magnetic fields run on the uniform grid, with constrained transport on areas and edge lengths in $(R, z)$. Stars sit on the axis and are settled as in Section~\ref{sec:settle}, with the volumes of rings. A Sedov-Taylor blast on the axis reaches the similarity radius to 1.5~percent along both axes, and the Noh problem reaches the post-shock density of 64 to 3~percent. The potential of a uniform sphere is within $7\times10^{-4}$ of the analytic one. The Evrard collapse keeps its energy to 0.11~percent through bounce, with the thermal energy peaking at $t = 1.125$ as in three dimensions. Bondi accretion onto a point mass matches the analytic rate to 3.5~percent, and a torus in equilibrium about a point mass keeps its peak density to $2.4\times10^{-4}$ over 62,700 steps.

In $(r, \theta)$ the grid runs from an inner sphere to an outer one, uniform or logarithmic in $r$, with both poles in the grid. Gravity is the monopole of the gas, with point masses at the origin. A run can also use the adaptive mesh of Section~\ref{sec:amr} with local timesteps. Each block takes its geometry from its global indices on its level, restriction and prolongation are weighted by volume, and a coarse face's flux is the area-weighted mean of the fine fluxes. The slice view maps the $(r, \theta)$ grid onto the meridional half-plane. A Sedov-Taylor blast reaches the similarity radius to 0.74~percent, with a spread of $1.3\times10^{-7}$ in its radius between the pole and the equator. A settled polytrope keeps its central density to $2\times10^{-4}$, and Bondi accretion matches the analytic rate to 1.9~percent. On blocks the blast reaches the similarity radius to 0.74~percent, and the polytrope keeps its central density to $5\times10^{-4}$ with one timestep and $5.6\times10^{-4}$ with local timesteps.

A run can also be one-dimensional: a single row of cells along $x$, or along $r$ or $R$. Across the axes that the run does not resolve, each cell is its own neighbor, so the solvers run unchanged. Self-gravity becomes a sum of the enclosed mass. On the uniform grid a row has at most 2048 cells, the largest side of a 3D texture. Longer rows and rows with refinement use the adaptive mesh, with blocks of 16 cells on a binary tree, and can have up to 262,144 cells. Magnetic fields and implicit pressure run on the uniform grid only. The view repeats the row across the unresolved axes, as a band in Cartesian runs and as rings in curved ones. A label marks the run as one-dimensional.

\section{Interface, input and output}
\label{sec:interface}

\begin{figure*}[t]
\centering
\includegraphics[width=\textwidth]{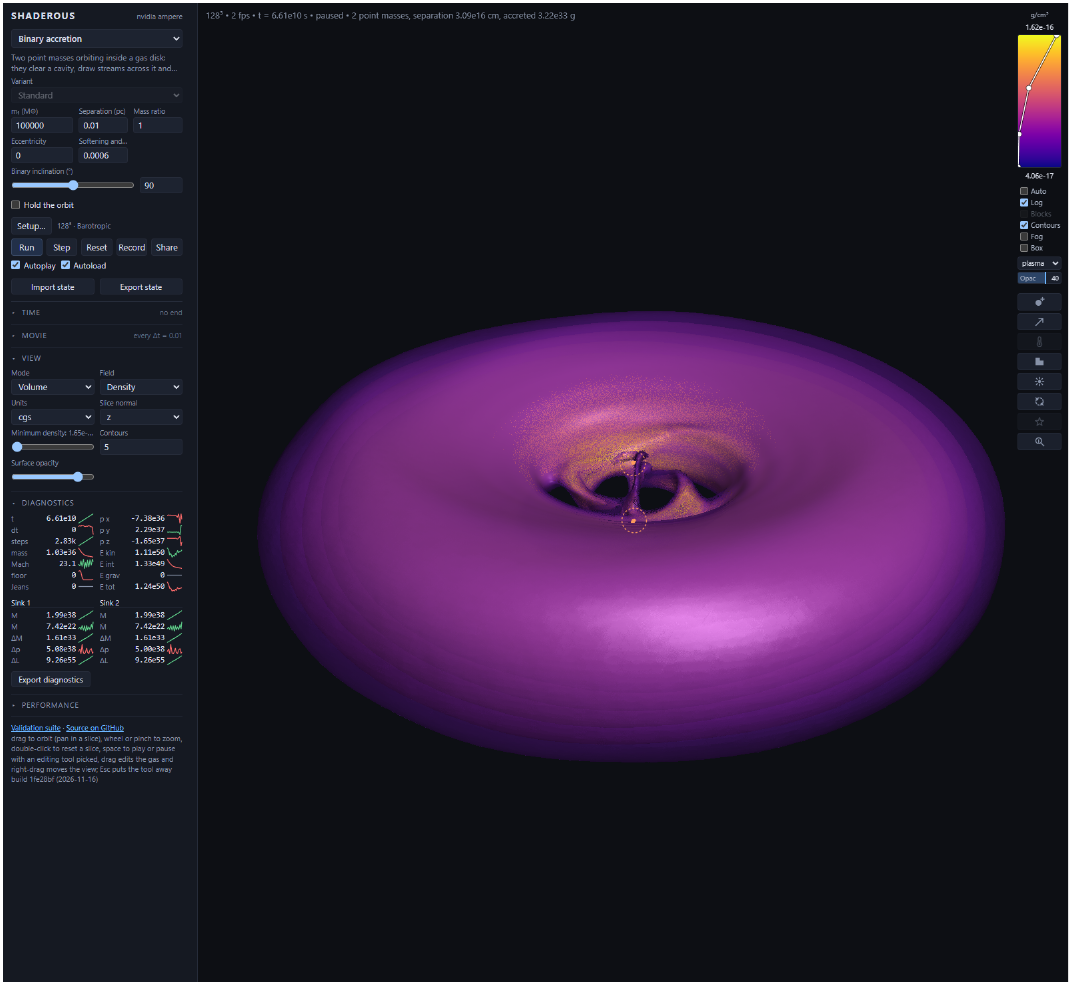}
\caption{The whole interface of Shaderous in the browser during a three-dimensional run of the binary problem, with the binary's orbit tilted $90^\circ$ to the disk. The volume view shows isosurfaces of the density. Beside it are the menu of problems with the main parameters of the binary, the controls of the run and of the view, and the diagnostics.}
\label{fig:interface}
\end{figure*}

Shaderous runs at a web address. A user picks a problem, changes its grid and its physics in a dialog that holds every setting (Section~\ref{sec:setups}), and watches the gas through a ray-marched volume rendering or an axis-aligned slice, both drawn on the same device from the simulation textures (Figure~\ref{fig:interface}). The view can show the density, temperature, pressure, a passive scalar, the speed and each velocity component, the Mach number, the entropy function $p/\rho^\gamma$, the net heating rate, the vorticity, the divergence of the velocity, the normalized density gradient $|\nabla\rho|/\rho$, the gravitational potential, the kinetic energy density, the Jeans length in cells and, on the adaptive mesh, the refinement level. The color range follows the field as it evolves, and for the derived fields it ignores the most extreme 0.5~percent of cells. Brushes add mass, velocity or heat, and in a magnetized run they strengthen, weaken or drag the field. Obstacles and sources can be drawn, and the sink tool places point masses. A query tool shows the state of the cell under the pointer and, on the adaptive mesh, the block that holds it, its level and its refinement indicator. The refinement thresholds can be changed during a run. The address bar records the problem, the grid and every setting that differs from the problem's defaults, so a link reproduces a setup.

With a magnetic field the view can also show the field strength, the plasma beta $\beta_{\rm M}$ and the divergence of the field, and it draws field lines. In two dimensions the lines are contours of the flux function $A_z$, integrated from the face field itself. Because the face field is free of divergence to round-off, the integral to a corner does not depend on its path, and the same magnetic flux passes between any two neighboring lines. In three dimensions, lines are traced both ways from seeds that move with the gas, so that the picture does not flicker. A seed that leaves the box or the field is placed again at random, with a probability proportional to the magnetic energy density. The lines are drawn as lit tubes inside the volume rendering, colored by the magnetic pressure, as \citet{GuillochonMcCourt2017} colored theirs, here on a logarithmic scale. A run can also be recorded as an MP4 movie, encoded frame by frame in the browser with a color bar, axes and the simulated time drawn on each frame. A frame is taken at the end of the step nearest to its time, and no step is shortened, so a recorded run is bit for bit the same as one that is not recorded.

The state can be exported as an HDF5 file in the layout of Athena++ \citep{Stone2020}, which yt reads directly \citep{Turk2011}, in two and three dimensions and on uniform and adaptive grids. Files written by Shaderous also carry the run's settings, time and point masses, so importing one continues the run on the same mesh. Each species is written as a variable named after it. Checkpoints write the same file each time the simulated time passes a multiple of a set interval.

The same page can also run without a person at it. A programming interface in the page loads a setup described by the keys of a link or by a setup document (Section~\ref{sec:setups}), advances the run by a number of steps or to a given time, and returns diagnostics, samples of the gas, images of the view and exported files as plain data. When the run advances to a given time, its last step is shortened so that it ends at that time exactly. With one parameter in the address, the page neither draws nor plays by itself, so the GPU takes only the steps that a script requests. Two tools in the repository build on this interface, and both open the page in a browser without a window. A command-line program runs a setup, or a sweep over the values of its parameters. It writes the diagnostics, images, slices, movies and HDF5 files at fixed intervals of simulated time, together with the setup, the version of the code and the GPU that produced them. It also runs on rented Linux servers with data-center GPUs, where the browser reaches the GPU through Vulkan, with a memory budget set on the command line. A run resumes from its checkpoint, which the runner writes and reads itself at any size, so a long run can proceed as a chain of jobs on the batch queue of a cluster. A server for the Model Context Protocol\footnote{\url{https://modelcontextprotocol.io}} offers the same operations to an AI agent as tools. Within one conversation the agent can load a problem, advance it, read the numbers back and inspect images of the flow.

Both tools are distributed as one package on npm, the public registry of JavaScript packages.\footnote{\url{https://www.npmjs.com/package/shaderous}} The package holds the built page together with the two tools, so the command \texttt{npx shaderous} runs either of them without a copy of the repository. It needs only Node.js and Google Chrome on a machine with a GPU. The server is listed in the official registry of the Model Context Protocol, from which clients that support the protocol can install it. The repository is also a plugin catalog for Claude Code. Its plugin installs the server together with written instructions for the agent: how to check that a run is healthy, and how to turn a description in words into a setup document. For agents that work through a browser instead, the site serves a short text file (\texttt{llms.txt}) that describes the programming interface and points to the documentation. A new version is released by pushing a tag to the repository. The tag starts a workflow that runs the tests, publishes the package with a signed record of the commit that built it, and updates the listing in the registry.

\begin{figure}[t]
\centering
\includegraphics[width=\columnwidth]{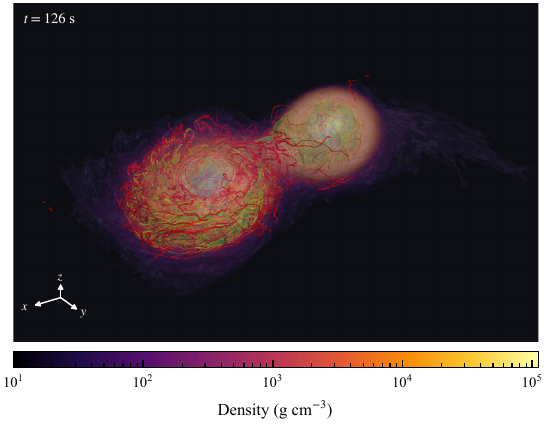}
\caption{Two magnetized white dwarfs during mass transfer, from a checkpoint opened for viewing only. A carbon-oxygen donor of $0.6\,\Msun$ that fills 97~percent of its Roche lobe orbits an accretor of $0.9\,\Msun$ in the frame that turns with the orbit. Each star is threaded by its own tangled field, with $\beta_{\rm M} = 10^5$ at its center, and the gas follows the equation of state for white dwarf matter (Section~\ref{sec:eos}). The run had $1024^3$ cells at the finest level of a box 80{,}000~km across, with five levels of refinement, and ran on an NVIDIA H100. Its checkpoint at $t = 126$~s, after 24{,}183 steps, holds 19{,}419 blocks. When we opened the checkpoint on a desktop GPU, with the face field kept for the field lines, the code merged the mesh to $512^3$ cells at its finest level. The density is drawn as fog and eight isosurfaces, and the 300 field lines are colored by the magnetic pressure on a logarithmic scale. The larger star is the donor. The arrows give the orientation of the axes.}
\label{fig:merger}
\end{figure}

A checkpoint too large to continue on the GPU at hand can still be opened for viewing. The page reads the file in parts and never holds it whole, so files larger than the 4~GB that WebAssembly can address open as well. It merges the finest levels of the mesh into coarser ones until the mesh fits the memory budget of the GPU. The merging conserves mass and momentum and keeps the magnetic field free of divergence. Such a view does not step, and it holds only what the display reads. If the view can keep more levels by holding only the field that is shown, it holds only that field and reads it again from the file when the shown field changes. Figure~\ref{fig:merger} shows a checkpoint of 3.5~GB that was written on a data-center GPU and opened on a desktop GPU.

\subsection{Problems}
\label{sec:problems}

The problems that ship with the code are the shock tubes of \citet{Sod1978} along each axis and the diagonal and a strong shock tube from \citet{Toro2009}, the blast of \citet{Taylor1950} and \citet{Sedov1959}, Kelvin-Helmholtz and Rayleigh-Taylor instabilities, the Mach 3 wind tunnel with a step of \citet{Emery1968} and \citet{Woodward1984}, smooth advection, the Jeans instability \citep{Jeans1902}, a polytrope of index 1 in equilibrium, a cold white dwarf in equilibrium, two white dwarfs that collide on a nearly parabolic orbit \citep{Rosswog2009, Raskin2009}, the collapse of \citet{Evrard1988}, a uniform sphere, an isothermal turbulent collapse, the thermal instability \citep{Field1965} with the cooling function of \citet{Koyama2002}, a tidal disruption of a polytropic star by a point mass, and a binary accreting from a viscous circumbinary disk \citep{Artymowicz1994, MacFadyen2008}.

With the ANEOS materials, a planet with a core of the iron alloy under a mantle of rock sits at rest, and Theia strikes the proto-Earth in the moon-forming impact \citep{Canup2004}. A shock in air hits a bubble of helium \citep{Haas1987, QuirkKarni1996}. Detonations of a single-step Arrhenius reaction start from the exact steady solution. Where the steady solution is unstable the front pulses \citep{Fickett1966}, and a pocket of unburned gas behind it breaks the front into cells. The network of Section~\ref{sec:burning} burns a detonation of carbon and oxygen at $10^7$~g~cm$^{-3}$, lit at the closed end of a tube 1000~km long, and the collision of two white dwarfs. It also burns the double detonation of a white dwarf of $1\,\Msun$ under a thin shell of helium \citep{Livne1990, Fink2010}. A hot spot at the base of the shell lights the helium. The helium detonation wraps around the star, and carbon ignites where its shock converges at the far pole. On cells of 98~km, the core burned within 2.2~s, and $0.51\,\Msun$ of $^{56}$Ni remained in the box. The detonation of a bare carbon-oxygen dwarf of the same mass makes $0.58\,\Msun$ \citep{Shen2018}.

With magnetic fields there are the shock tube of \citet{Brio1988}, the vortex of \citet{Orszag1979}, the rotor of \citet{Balsara1999}, the field loop of \citet{GardinerStone2005}, a circularly polarized Alfv\'en wave \citep{Toth2000}, a blast in a magnetized medium \citep{Stone2008}, polytropes threaded by a dipole or a random field, magnetized tidal disruptions, and a magnetized accretion disk. Following the WebGPU solver of \citet{Mallet2026}, further problems include isothermal turbulence driven in a box with and without a guide field (Section~\ref{sec:driving}), a Kelvin-Helmholtz layer threaded by a field, and Alfv\'en waves that travel in opposite directions along a guide field and collide, as packets \citep{Verniero2018} and also as periodic waves \citep{HowesNielson2013}. With resistivity, a Harris current sheet tears into islands \citep{Furth1963}, two islands coalesce \citep{FinnKaw1977}, and a stressed X-point collapses between line-tied walls \citep{Thurgood2018}.

\begin{figure*}[t]
\centering
\includegraphics[width=\textwidth]{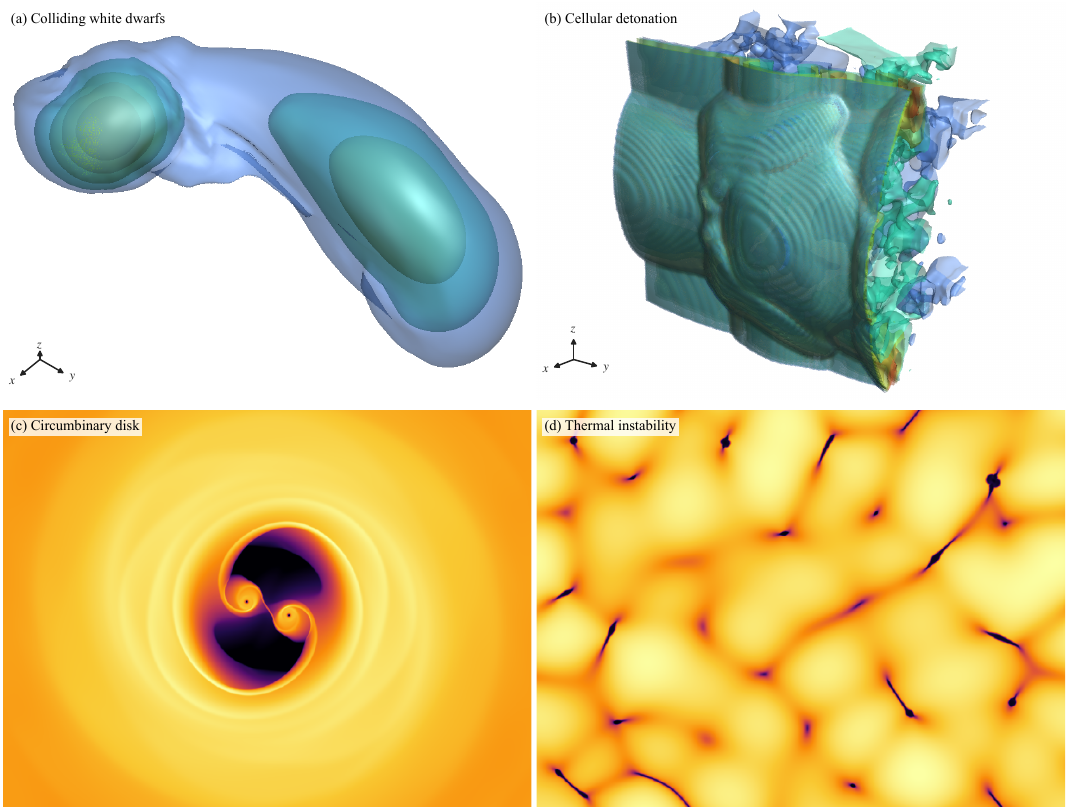}
\caption{Problems that ship with Shaderous, drawn by the code's own renderer. (a)~Isosurfaces of the density of two white dwarfs of 1.0 and $0.8\,\Msun$ after an off-center collision (pericenter 0.6 times the sum of their radii): the lighter one is torn into a tail of debris. (b)~Isosurfaces of the density of a detonation of the single-step Arrhenius reaction in three dimensions ($128^3$ cells, $t = 30$). A pocket of unburned gas behind the front breaks it into cells, and the burned gas behind it is turbulent. (c)~Density of gas around an accreting binary of equal masses after 20 orbits, with the cavity and the streams that cross it. (d)~Temperature in the thermal instability at $t = 0.6$, with cold clouds and filaments (dark) in a warm medium. The arrows in (a) and (b) give the orientation of the axes. The script that made the figure records each setup as the link that reproduces it.}
\label{fig:gallery}
\end{figure*}

The cavity of \citet{Ghia1982} tests the viscous fluxes against a standard benchmark. Nearly incompressible liquids flow past a cylinder, in the same cavity, in the vortices of Gresho and of Taylor and Green, and in a double shear layer (Section~\ref{sec:implicit}). A shearing box grows the magnetorotational instability and swings a shearing wave, a Zel'dovich pancake collapses in comoving coordinates, and blasts run in $(R, z)$ and in $(r, \theta)$ (Section~\ref{sec:coordinates}). In Roche lobe overflow, a Sun-like star that fills its lobe pours gas through the inner Lagrange point onto a black hole of $3\,\Msun$, in a stream that winds into a disk. A variant puts a resolved star of $0.6\,\Msun$ in place of the black hole. A polytrope spins rigidly in a rotating frame. A wind tunnel holds a bluff body shaped like a car, and a three-bladed wind turbine turns in its own frame, and the code reports its power coefficient (Section~\ref{sec:bodies}). Figure~\ref{fig:gallery} shows four of these problems.

Four problems use the equation of state for hydrogen and helium, without its radiation term. In a common envelope \citep{Ivanova2013}, a star of $0.5\,\Msun$, as a point mass, starts on a circular orbit just outside a red giant of $1\,\Msun$ and $50\,\Rsun$, and it spirals into the envelope of the giant. The giant is a polytrope of index 3/2 around a core of $0.4\,\Msun$ that is a point mass (Section~\ref{sec:settle}). Its envelope lies between $10^4$ and a few $10^5$~K, where hydrogen and helium are partly ionized, and the energy that their recombination returns as the envelope expands is a large part of what binds it. A variant runs the same problem with an ideal gas. The partial tidal disruption runs with this gas as well, at an impact parameter $\beta \equiv r_{\rm t}/r_{\rm p} = 0.6$, the ratio of the tidal radius to the pericenter distance, and the debris that the star loses recombines as it expands and cools. The thermal instability runs in physical units, in a box of 31~pc. The gas starts atomic near 1500~K, the cold clouds become molecular below 1000~K, and the warm medium ionizes above 4000~K, so the two phases differ fourfold in the number of particles per gram. The heat that dissociation and ionization take up makes the phases separate about three times later than in an ideal gas. The chemistry is that of equilibrium at the least density of the table, $10^{-20}$~g~cm$^{-3}$, and real interstellar gas is not in equilibrium.

The last problem is the second collapse of a protostellar first core \citep{Larson1969, Masunaga2000}, in $(r, \theta)$ on $128 \times 128$ cells spaced evenly in $\ln r$ from 0.008 to 6~au. The first core is molecular gas on one adiabat, that of gas that became opaque at $10^{-13}$~g~cm$^{-3}$ and 10~K, in hydrostatic balance computed with the equation of state of the run. Cores on this adiabat reach at most $0.040\,\Msun$, at a central density of $3\times10^{-9}$~g~cm$^{-3}$ and 1800~K. A denser center is hotter than 2000~K, where H$_2$ dissociates and $\Gamma_1$ falls below 4/3, and such a core is unstable. The core of the problem holds $0.036\,\Msun$ and is 2.6~au across. Its center starts at 2100~K and $1.8\times10^{-8}$~g~cm$^{-3}$, and the core starts with 5~percent less pressure than balance requires. Its center falls in along the adiabat until the hydrogen is atomic, and a second core, the protostar, forms there behind a shock. The validation test of Table~\ref{tab:validation-curvilinear} compares the temperature at the center with the adiabat while the density rises $10^4$-fold, and it checks the density and the temperature of the second core.

Two of these problems use the diffusive fluxes of Section~\ref{sec:diffusion} by default. The disk of the binary has $\nu = 10^{-3}$ in units of the square of the binary's separation times its orbital frequency. At the binary's separation this equals an $\alpha$ viscosity with $\alpha = 0.1$, the value that \citet{Farris2014} chose. The thermal instability has $\kappa = 3 \times 10^{-4}$ in units of the box size and the problem's unit of time, so that conduction rather than the cell size sets the size of the clouds. \citet{Koyama2004} showed that simulations of this instability converge with resolution only if they include conduction and resolve the Field length with at least three cells. With conduction the scale of the structure at $t = 0.6$ changes by 7.5~percent between $512^2$ and $1024^2$ cells. Without it the scale falls by 20~percent.

The disk of the binary also has a damping zone at its outer edge, of the kind that \citet{deValBorro2006} used to compare codes for disk-planet interaction. Beyond 4.5 binary separations from the center, the density, velocity and specific internal energy of the gas relax toward the initial disk. The rate of relaxation rises from zero there to three times per local orbital period at the edge of the box, 6 separations from the center, and both values can be changed during a run. Without the zone, gas flowed in through the outflow boundaries at the corners of the box, and the disk grew until its uneven pull pushed the binary out of the box. A zone that is fixed to the box brings a second problem. The outer disk then takes up none of the binary's recoil, and a fixed disk pulls an off-center binary further toward its near side. Making the zone follow the center of mass of the two masses (Section~\ref{sec:sinks}) removes this instability.

The magnetized accretion disk is a globally isothermal disk around one softened point mass with a sink, in exact vertical and radial balance with the softened potential. Four poloidal loops of field with no net flux thread it, whereas \citet{Hawley2000} laid theirs along the surfaces of constant density. Ours are laid along the radius, with $\beta_{\rm M} = 50$ in the midplane where two loops meet. Its damping zone acts on the gas alone and leaves the field to the induction equation, and a ceiling of 3 on the Alfv\'en speed keeps the nearly empty corona from setting the timestep. The shear winds the loops, and the magnetorotational instability \citep{Balbus1991} grows. The code reports the Maxwell and Reynolds stresses over the body of the disk as $\alpha$ parameters, and the quality factors $Q_z$ and $Q_\phi$ of \citet{Hawley2011}, the number of cells across the fastest-growing wavelength of the instability. On the default mesh ($256^3$ at the finest level, two levels of refinement), after three orbits at unit radius, $\alpha_{\rm M} = 0.026$, $\alpha_{\rm R} = 0.003$, $Q_z = 4.2$ and $Q_\phi = 15$, below the $Q_z \gtrsim 10$ with $Q_\phi \gtrsim 20$ that \citet{Hawley2011} found necessary for the turbulence to saturate as it does at higher resolution. The largest run of this disk on a desktop GPU is at $512^3$ with three levels on an NVIDIA RTX~4090. With room for 5832 blocks, about 24 million cells, it used 16.8~GB of GPU memory, and one orbit at unit radius (6359 steps) took 14~minutes. Over that orbit the magnetic energy grew 5.4 times, and the Maxwell stress reached $\alpha = 0.009$.

\begin{figure}[t]
\centering
\includegraphics[width=\columnwidth,height=\colfigheight,keepaspectratio]{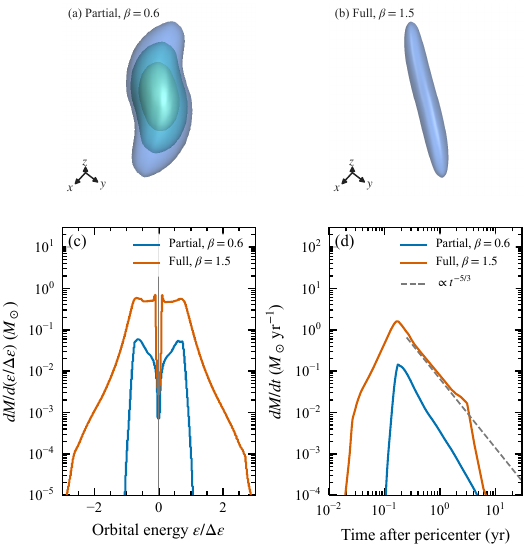}
\caption{Tidal disruptions of a polytrope of index 3/2 with the mass and radius of the Sun by a black hole of $10^6\,\Msun$, run with the command-line runner on the adaptive mesh in a box $16\,\Rsun$ across ($256^3$ at the finest level, three levels of refinement). (a, b)~Isosurfaces of the density drawn by the code's renderer at $t = 6$, about two dynamical times after pericenter, five levels evenly spaced in $\log\rho$ from 0.01 to 10~$\Msun\,\Rsun^{-3}$, for impact parameters $\beta = 0.6$ (the star keeps most of its mass) and $\beta = 1.5$ (it is torn apart). (c)~Mass per unit specific orbital energy about the black hole, in units of the spread $\Delta\epsilon = G M_\mathrm{bh} R_\star / r_\mathrm{t}^2$, where $M_\mathrm{bh}$ is the mass of the black hole, $R_\star$ the radius of the star and $r_\mathrm{t}$ the tidal radius, 40 dynamical times into the run: the gas that has left the box and the gas in it, without the surviving core. (d)~The rate at which the bound debris returns to pericenter, against time since pericenter. The dashed line falls as $t^{-5/3}$. The partial disruption returns its debris in a narrow peak and then falls more steeply.}
\label{fig:tde}
\end{figure}

For the tidal disruption the code also sorts the debris by its specific orbital energy about the black hole, from which follows the rate at which bound debris returns to pericenter. Gas that has left the box moves almost ballistically in the potential of the black hole, so its energy no longer changes. The update therefore adds the mass that crosses each open face of the box to a histogram of energy as it crosses. A census adds the gas that is still in the box. It sets apart a surviving core, which we find by iteration: gas belongs to the core if it is bound to the core's mean velocity by the gas's own potential, and the core's mean velocity is then recomputed from that gas. The histograms are sums of 64-bit integers, so they are exact and do not depend on the order in which the GPU adds them. In the validation runs, the mass in the box, the mass that has left and the mass that the sink has taken add up to the starting mass to within the rounding of the total. Kepler's third law turns the histogram of the bound debris into a rate of return (\mbox{Figure~\ref{fig:tde}}). When a star of index 3/2 passes a black hole of $10^6\,\Msun$ at $\beta = 1.5$, the rate peaks at $1.6\,\Msun\,\mathrm{yr}^{-1}$ 61~d after pericenter and then falls close to $t^{-5/3}$. At this resolution a clump stays bound, and it still sheds mass 40 dynamical times into the run, when it holds $0.12\,\Msun$. The late part of the rate fills in only as that gas leaves the clump. At $\beta = 0.6$ the star keeps $0.95\,\Msun$, and the rate falls faster than $t^{-5/3}$, as \citet{Guillochon2013} found for encounters that leave a core.

\begin{figure*}[t]
\centering
\includegraphics[width=\textwidth]{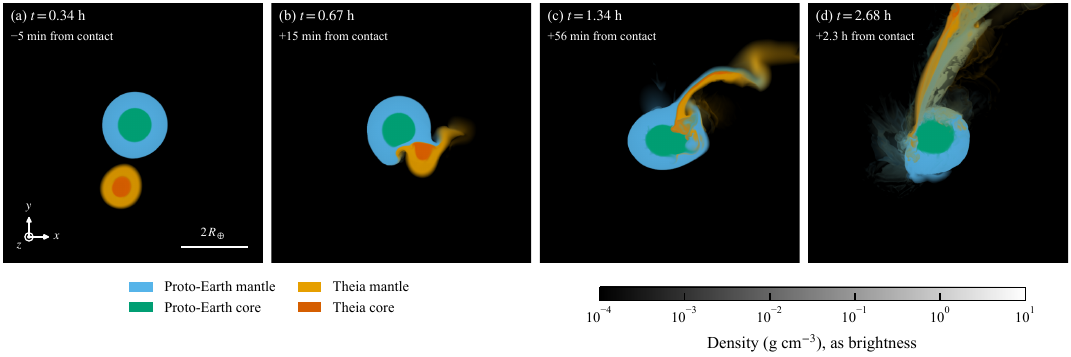}
\caption{The moon-forming impact, in slices through the orbital plane, the middle $8\,\Rearth$ of a box $10\,\Rearth$ across, at 102 cells to an Earth radius. Each panel gives the time since the run began and since contact. The color shows what the gas is made of: the mantle of pyrolite and the core of Fe85Si15 of the proto-Earth and of Theia, mixed in proportion where the gas is mixed, from the mass fraction of iron and from a passive scalar that marks Theia's gas. The brightness is the logarithm of the density. (a)~Theia approaches. (b)~Its mantle shears across the proto-Earth's. (c)~Its core is drawn out along the inside of an arm of its mantle and sinks toward the proto-Earth's core. (d)~The arm winds around the planet, which is wrapped in Theia's mantle mixed with its own. The gas that goes into orbit is mostly Theia's mantle, while most of Theia's iron joins the planet.}
\label{fig:impact}
\end{figure*}

A polytrope of index 3 is evolved with gas of $\Gamma = 5/3$, as in \citet{Guillochon2013}. In our runs it is destroyed for $\beta$ between 1.8, where its peak density collapses again, and 2, where the peak density keeps falling, against the 1.85 that they found. The magnetized disruptions use the field geometries of \citet{GuillochonMcCourt2017}, a dipole or a random field with a Kolmogorov spectrum. They set its strength by $\beta_{\rm M}$ at the center of the star. Ours follows the pressure of the star, so that $\beta_{\rm M}$ is about the same throughout it.

The moon-forming impact follows \citet{Canup2004}. Theia, of $0.12\,\Mearth$, falls from rest at infinity onto the proto-Earth, of $0.9\,\Mearth$, and the two touch at $45^\circ$. Both bodies have 30~percent of their mass in a core of Fe85Si15 under a mantle of pyrolite. As \citet{Canup2004} did, the code sorts the gas after the impact by the Kepler orbit that each cell's position and velocity give it about the planet. Gas on an unbound orbit escapes, gas whose pericenter lies inside the planet's radius belongs to the planet, and the rest makes up the disk. The planet's mass and radius are recomputed from the gas that falls to it until they stop changing. The disk's mass and angular momentum then give the mass of the moon that it would gather into \citep{Ida1997, Kokubo2000}.

We ran the impact on one NVIDIA A100 in a box $10\,\Rearth$ across with four levels of refinement, 102 cells to an Earth radius at the finest level (Figure~\ref{fig:impact}). The bodies touch 25~min into the run. Theia's mantle shears across the proto-Earth and is drawn out into an arm. Theia's core is stretched along the inside of the arm, and most of it sinks into the proto-Earth's core. From 1.8 to 2.3~h after contact, the disk held 0.031 to $0.035\,\Mearth$, of which 65~percent came from Theia, and it would gather into 1.1 to 1.3 lunar masses. For impactors of 0.11 to 0.13 of the total mass that strike at the mutual escape speed and at angles near $45^\circ$, \citet{Canup2004} found disks of 1.0 to 1.6 lunar masses (0.013 to 0.020$\,\Mearth$), of which 75 to 89~percent came from the impactor, and predicted moons of 1.0 to 1.6 lunar masses. Our moon lies in that range, Theia's share of our disk is below it, and our disk is heavier.

A tenth of the disk's mass is iron, at the top of the 2 to 10~percent of the disks of \citet{Canup2004}. In a box $24\,\Rearth$ across at 10.7 cells to an Earth radius, where Theia's core is six cells across, a quarter of the disk is iron, because the iron that the mesh mixes into Theia's mantle goes where the mantle goes. To 2.3~h after contact, total energy changed by at most $1\times10^{-3}$, and 0.15~percent of the mass left through the faces of the box. Soon after, the clump that remains of Theia, which is bound and would fall back, crosses a face of the box and is lost, so we stopped the run there. At 8 cells to an Earth radius, the resolution of the validation test, the disk would gather into 1.5 lunar masses.

\begin{figure*}[t]
\centering
\includegraphics[width=\textwidth]{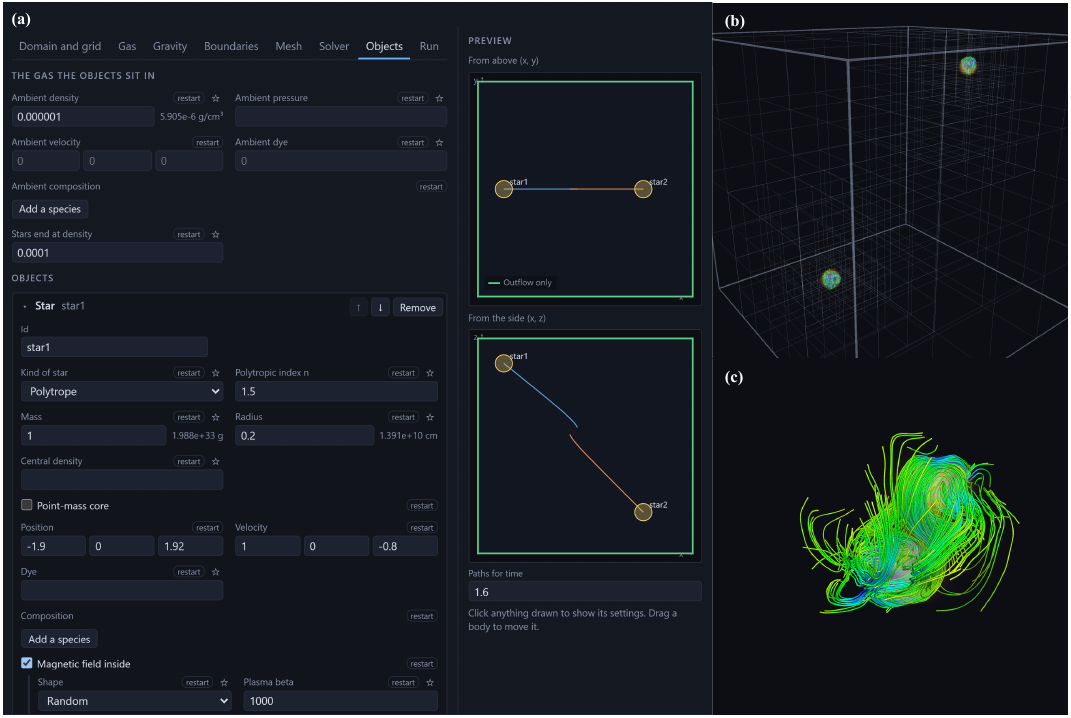}
\caption{A setup made in the page, and the run that it yields. (a)~The setup dialog on its page of objects: the ambient gas, the first of two polytropes with its mass, radius, position, velocity and magnetic field, and the preview, which draws both stars to scale in two projections with the paths that they would follow as point masses. (b)~The volume view after both stars have settled, with the leaf blocks of the adaptive mesh. The mesh is refined around each star. (c)~The magnetic field lines of the two stars as they merge, 46~minutes later, colored by the magnetic pressure. The script that made the figure keeps the setup as a file.}
\label{fig:setup}
\end{figure*}

\subsection{Setups}
\label{sec:setups}

Each of the 84 problems is a file, and a user can write another. A setup is one JSON document that holds every setting of a run: the grid, the physics, the boundaries and a list of parts. The parts are point masses, stars, disks, regions of gas with a state of their own, solid obstacles, inflows and uniform magnetic fields, and an orbit part places two bodies on a Kepler orbit. A named initial condition with parameters covers what shapes cannot express, such as a shear layer with its seeded modes or the profile of a steady detonation. The code compiles a document into the same internal description that the solvers have always used, so a setup that a user assembles runs through the same shaders as the problems that ship. We checked the conversion against the hand-written definitions that the files replaced. Every file reproduces the configuration of its problem and, at sampled points in two and three dimensions, an initial state, point masses and magnetic potential that are identical to the last bit.

A dialog in the page edits a copy of the document (Figure~\ref{fig:setup}a). Its forms are generated from one table that lists every option once, with its range, its default and the options that it requires or excludes. A star, for example, requires three dimensions or an axisymmetric $(R, z)$ run, and isolated self-gravity, and it excludes periodic boundaries. An option that cannot be combined with the rest of the setup is grayed out, and its tooltip gives the reason. A preview drawn from the document alone shows the box, the bodies to scale and the paths that they would follow as point masses, and bodies can be dragged to new positions. The table also records what a change costs a running simulation: a write to the uniform block, a rebuild of the shaders, or a fresh start. When the user applies an edit, the page does the least costly of these that the edit requires, so the viscosity or the Riemann solver can change in the middle of a run.

A setup file holds numbers and names and no code, so a file from a stranger is safe to open. The one exception to plain numbers is the density of a disk, which can be given as a formula of the radius and of the height above the midplane. The code parses a formula into arithmetic over a fixed list of functions and never executes it, and it generates the shader text for the disk's damping zone from the parsed formula. A setup can be saved as a file or kept in the browser. A link holds a setup that is not one of the shipped problems as a compressed copy of its document, and an exported state file contains the document as well, so the file restarts as the same setup after the shipped problems change. The programming interface and the command-line runner accept a document in place of the name of a problem.

Figure~\ref{fig:setup} shows a setup that no shipped problem contains. Two polytropes of index 3/2, each with a mass of $1\,\Msun$ and a radius of $0.2\,\Rsun$, approach each other in a box $5\,\Rsun$ across, and each is threaded by a tangled field with $\beta_{\rm M} = 1000$ at its center. The mesh has $256^3$ cells at its finest level and three levels of refinement. Both stars are settled into the balance of the mesh (Section~\ref{sec:settle}) before the first step. They collide 46~minutes of simulated time later, after 1304 steps, and the field lines of the two stars wind around each other as the stars merge.

\subsection{Setups from a description}
\label{sec:agents}

Because a setup is a document of numbers and names, an AI agent can write one. A user states what they want in words, for example a main-sequence star of $0.5\,\Msun$ colliding with a white dwarf of $1\,\Msun$. An agent connected to the server of Section~\ref{sec:interface} turns the request into a setup document, runs it and reports what it finds. The instructions that the plugin installs describe the steps. The agent starts from the shipped problem nearest to the request, whose floors, boundaries and refinement are already tuned, and it takes the meaning of each key from a reference and a JSON Schema that the site publishes.

Before the run, the server checks the document against the same table of options that drives the dialog. It returns the errors and warnings that the dialog would list: a broken rule with its reason, a softening length under two cells, bodies that overlap or leave the box, or a grid that exceeds the memory of the GPU. The agent then loads the setup on a coarse grid, checks that the mass equals the total of the bodies and that the stars hold still before they meet, and runs it at full resolution. Last, it tells the user each modeling choice that the request left open, such as the polytropic index, the radius of a star, or which bodies became point masses. Worked examples are in the repository. In the one for the collision above, the white dwarf is far smaller than a cell, so it becomes a softened point mass that passes through a polytrope of index 3. The result is an ordinary setup, which the user can open as a link, change in the dialog and run again. Because of the check, a request unlike any shipped problem fails with a reason rather than running a combination of options that the code does not support.

\section{Educational use}
\label{sec:education}

Several properties make the code practical in a classroom. It needs no installation, no account and no allocation of computing time. It runs on the laptops that students already own, and the two-dimensional problems run on phones. A link records a whole setup, so an instructor can hand a class a prepared run by posting its address. The brushes, the query tool and the choice of displayed field let students change a flow while it runs and read the state of any cell. In the Setup dialog they can build a run of their own from stars, orbits, obstacles and regions of gas, and share it as a link. The command-line runner writes slices that numpy reads, and an exported file opens in yt, so students who know Python can analyze runs with their own scripts. A recorded movie can be shown in a lecture. A guide in the repository gives each activity below with links that open its setup.\footnote{\url{https://github.com/guillochon/shaderous/blob/main/docs/teaching.md}} The code is too new to have been used in a course. The activities are proposals, and we invite instructors who try them to tell us what worked.

For primary and secondary school, the guide gives seven demonstrations that a teacher shows on a projector, each a few minutes long, in a routine of predicting, watching and explaining. Children see that air and water flow and swirl, that hot gas rises and spreads, and that a star is a ball of hot gas held together by gravity. Each demonstration is paired with something that students can see or do themselves, such as food coloring that curls in a glass of water. Older pupils work in pairs, read numbers with the query tool, and discuss how far a simulation can be trusted.

For undergraduates who are not majoring in physics or astronomy, the code works as a set of demonstrations that need no equations. A Kelvin-Helmholtz instability rolls up into the billows seen in clouds and in the bands of Jupiter. A Rayleigh-Taylor instability grows fingers like the filaments of the Crab Nebula. A Sedov-Taylor blast shows a thin, dense shell around a hot, thin interior, as in a supernova remnant, and opens a discussion of how \citet{Taylor1950b} estimated the energy of an explosion from photographs of its fireball. A polytrope at rest shows the balance between pressure and gravity in a star. A tidal disruption, in which students change the depth of the orbit, shows a star torn apart by a black hole. A wind tunnel shows the wake behind a body shaped like a car, and the code reports its drag, and students can drop a 3D model of their own into it. In each, students predict what will happen before they run it.

For physics and astronomy majors the same problems become quantitative. In a shock tube, students identify the rarefaction, the contact discontinuity and the shock, check the jump conditions with the query tool, and compare limiters by editing the link. From dimensional analysis they derive the growth of the Sedov-Taylor radius as $(E t^2/\rho)^{1/5}$ and fit it to a run. They measure the growth rates of the Rayleigh-Taylor and Jeans instabilities and compare them with linear theory. They compare the density profile of the polytrope with the Lane-Emden solution, and turn off the well-balanced reconstruction to see what it does. The thermal instability shows how the cooling curve sets the phases of the interstellar medium, and how conduction sets the size of the smallest clouds. Students also measure the Strouhal number of the vortex street behind a cylinder, follow radiation from diffusion through opaque gas to free streaming, and estimate the mass of the Moon that a giant impact would form. In a longer laboratory exercise, students run a problem with a known answer at four resolutions in one sweep of the command-line runner, compute the error against the exact solution from the slices that it writes, measure the order of convergence, and explain why shocks converge at first order. The validation page measures the order of convergence of smooth advection in the same way, so students can check their method against it.

For graduate students the code is a readable example of a modern Godunov method. It has about 23,000 lines of WGSL and 61,000 of TypeScript, of which 11,000 are the validation suite, and each physical module is a small file. Exercises compare the Riemann solvers on a contact discontinuity and a Gresho vortex, find the carbuncle by turning off the HLL blend at a grid-aligned shock, compare the flux and kick forms of the gravitational energy source in the Evrard collapse, measure when local timesteps make a run faster, and compare the edge electric fields of Section~\ref{sec:ct} on the field loop. Term projects include adding a Riemann solver or a test problem with its own validation test, an implicit solver for diffusion, a study of single against double precision using the CPU reference of Section~\ref{sec:validation}, and a performance study from the per-pipeline timing. We pair these with the introduction of \citet{Trac2003} and the texts of \citet{LeVeque2002} and \citet{Toro2009}.

\section{Validation}
\label{sec:validation}

Validation runs in the browser on the reader's own GPU, from a page next to the simulator, and from the command line through a browser driven by Playwright. The references are a CPU implementation of the same scheme in double precision and an exact Riemann solver \citep{Toro2009}. Appendix~\ref{sec:validationtests} lists nearly all of the 291 tests of the suite, with their pass criteria and the values that we measured. A float64 copy of the MHD scheme serves as the reference for the magnetic tests. Figure~\ref{fig:hydro} compares the hydrodynamics with exact solutions and shows how the error falls with resolution. Figure~\ref{fig:gravitytests} shows the energy budget of a gravitational collapse and the accuracy of the potential on the adaptive mesh.

Each test checks one thing that we expected to break. To find failures that we did not expect, we also ran the code on setups chosen at random: problems of Section~\ref{sec:problems} with a random grid, refinement and solver, setups assembled from random parts, and whole setup documents drawn from the table of options of Section~\ref{sec:setups}. A script flags a run whose values become non-finite, whose page reports an error, or whose mass changes in a closed box. In about 4,650 such setups, on an Apple M4 Pro and an NVIDIA RTX~3080, we found 20 failures that the validation suite had missed, most of them in combinations of features or regimes that no test had put together. Three examples show the kind of failure. The tabulated equation of state for white dwarf matter returned NaN in hot gas, but only on Apple GPUs, because the Metal compiler reordered a product of four temperatures until it overflowed single precision. Cold gas under a uniform gravitational field fell through the walls of a closed box, because that field alone did not limit the timestep. With local timesteps, the step of gas released from rest fell 30-fold within two steps, faster than a cycle could follow, so a run now starts with 16 single steps. The script is in the repository, so the same search can be repeated after any change.

\begin{figure*}[t]
\centering
\includegraphics[width=\textwidth]{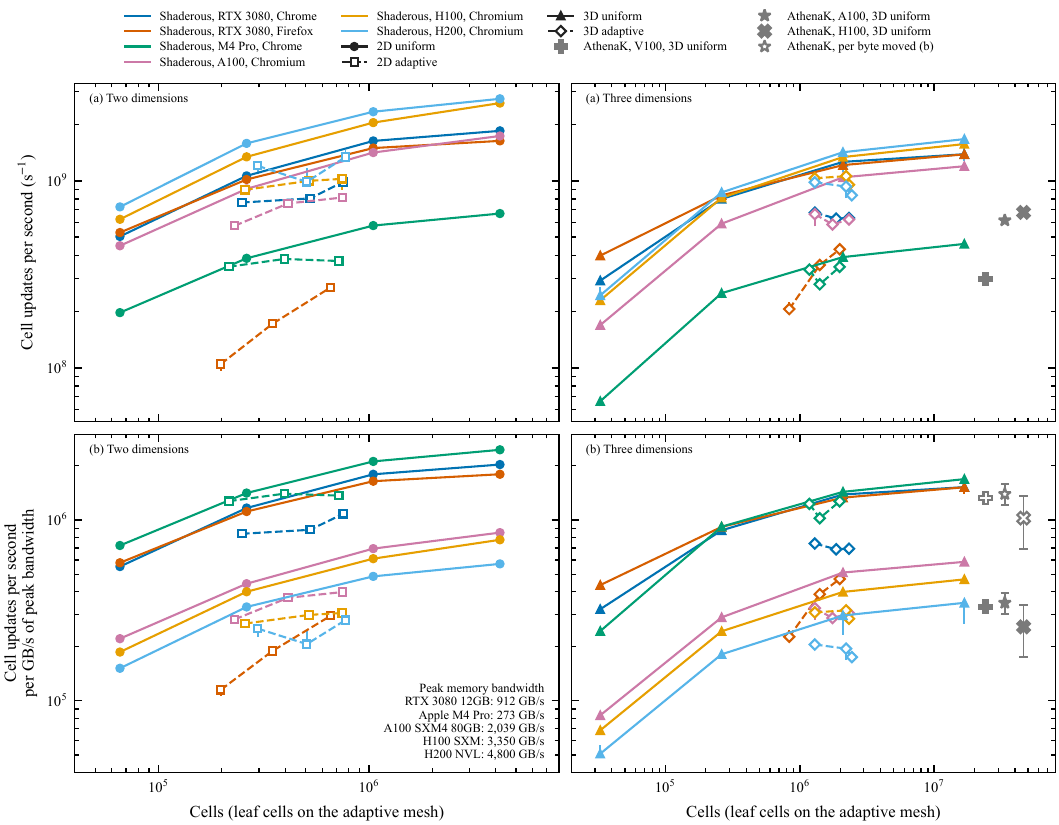}
\caption{Throughput on a discrete GPU (NVIDIA RTX~3080, 12~GB) in Chrome and in Firefox, on an integrated one (Apple M4 Pro) in Chrome, and on three data-center GPUs (NVIDIA A100 with 80~GB, H100 and H200) in the headless Chromium of Playwright. (a)~Cell updates per second against the number of cells, for a two-dimensional Kelvin-Helmholtz instability (left) and a three-dimensional Sedov blast (right), each on uniform grids and on adaptive meshes (two to four levels, local timesteps off). On the adaptive mesh every leaf cell counts as one update per step. Each point is the median of three runs, and the bars span all three. The gray points are the hydrodynamic throughput of AthenaK on one V100, one A100 and one H100 \citep{Stone2026}, on a uniform grid of $3.4 \times 10^7$ cells (drawn apart for legibility). (b)~The same throughputs divided by each device's peak memory bandwidth, from its datasheet. AthenaK's bars span the versions of the A100 (40 and 80~GB) and of the H100 (PCIe to NVL), since the device used is not stated. Shaderous runs in single precision with one stage per step. AthenaK's precision and integrator for this benchmark are not stated. The hollow gray markers are its values times four, which would be its throughput per byte moved if it ran in double precision with two stages per step.}
\label{fig:performance}
\end{figure*}

\section{Performance}
\label{sec:performance}

By default the uniform grid stores no primitive variables: the code computes them from the conserved variables as it loads each cell, which costs less than a pass that writes them out when the conversion is cheap. Where the equation of state solves for its temperature or reads a table, the pass that finds the timestep converts each cell once and keeps the result for the predictor and the update. On an A100 this makes a step of a white dwarf with the Helmholtz equation of state 3.0 to 3.1 times faster, and one of a planet with ANEOS 2.7 to 3.9 times faster, and it changes the result only at round-off.

In two dimensions a run uses $16 \times 16 \times 1$ workgroups, which more than halves the cost of the flux update against $8 \times 8 \times 4$ workgroups with three idle layers. The number of steps per frame adapts to the GPU. Results agree across browsers to round-off. On the adaptive mesh the passes that cover a block and its ghost layers, 18 or 20 cells across, run as rows of 64 threads per block, because $8 \times 8 \times 4$ tiles left about half of their threads idle there. In two dimensions the four layers of a workgroup take four blocks. These two changes made a step of a two-dimensional Kelvin-Helmholtz problem 2.8 times faster, and a step of a three-dimensional Sedov blast 19~percent faster. Taking the blocks of each pass in Morton order instead of slot order changed its time by less than 3~percent on meshes of up to 6063 blocks, because the slot order already keeps blocks that are close in space close together. The interface can also report the GPU time that each pipeline takes, from timestamp queries.

Figure~\ref{fig:performance} shows the throughput on a discrete GPU (an NVIDIA RTX~3080 with 12~GB) in Chrome 156 and Firefox 160, in Chrome on an integrated one (that of an Apple M4 Pro, which shares 273~GB\,s$^{-1}$ of memory bandwidth with the CPU), and on three data-center GPUs. We measured it with the scripting interface at normal priority, with the GPU otherwise idle, and each point is the median of three runs. In Chrome on the RTX~3080, a uniform grid of $2048^2$ cells takes $1.9 \times 10^9$ cell updates per second, and one of $256^3$ cells $1.4 \times 10^9$. On the adaptive mesh a leaf cell costs more than a cell of a uniform grid, because its ghost cells are computed as they are loaded and the refinement boundaries recompute fine fluxes. In two dimensions the mesh still takes $7.7$ to $9.9 \times 10^8$ updates per second, and in three, $6.3$ to $6.8 \times 10^8$ with two to four levels. The M4 Pro takes $6.7 \times 10^8$ updates per second on $2048^2$ cells and $4.6 \times 10^8$ on $256^3$, and 2.8 to $3.8 \times 10^8$ on the adaptive meshes. Firefox on the RTX~3080 is within 12~percent of Chrome on the largest uniform grids, with $1.6 \times 10^9$ updates per second on $2048^2$ cells and $1.4 \times 10^9$ on $256^3$. On the adaptive mesh it takes 1.5 to 7 times fewer: $1.1$ to $2.7 \times 10^8$ in two dimensions and $2.1$ to $4.3 \times 10^8$ in three. A step there takes about 2~ms in two dimensions and 4~ms in three, almost whatever the size of the mesh, against 0.3 to 0.8~ms and 1.9 to 3.7~ms in Chrome. That points to a fixed cost per step in Firefox, which we have not traced.

The data-center GPUs are an NVIDIA A100 with 80~GB, an H100 and an H200 of a computing cluster, with 2039, 3350 and 4800~GB\,s$^{-1}$ of memory bandwidth: 2.2, 3.7 and 5.3 times that of the RTX~3080. There the page ran in the headless Chromium 158 that Playwright supplies, on Linux with Vulkan, with one GPU to each job. On a uniform grid of $2048^2$ cells the three take 1.7, 2.6 and $2.7 \times 10^9$ updates per second, and on $256^3$ cells 1.2, 1.6 and $1.7 \times 10^9$. That is 0.86 to 1.5 times the throughput of the RTX~3080, and the A100 takes fewer updates than the RTX~3080 on both grids. On the adaptive meshes the A100 takes $5.7$ to $8.1 \times 10^8$ updates per second, and the H100 and H200 take $8.4 \times 10^8$ to $1.3 \times 10^9$. These GPUs gain so little because the update is not limited by memory traffic there. On the A100 a step moves about an eighth of the peak bandwidth, by our estimate of the bytes that it reads and writes. Doubling the texture loads of the flux update makes a step 4~percent slower, while the limiter of \citet{vanLeer1974} in place of the default reconstruction makes it 17~percent faster. The update is therefore limited by arithmetic, and the reconstruction is the largest part of it. Its cost comes mostly from its design, which limits each face of a cell on its own where a slope limiter takes one slope for both. After these measurements, we made the reconstruction skip the limiter where every quantity is smooth. This makes a step 7 to 9~percent faster on the A100 and changes the result only at round-off.

For comparison, AthenaK is the GPU version of Athena++, written with Kokkos. \citet{Stone2026} report that it takes $3.0 \times 10^8$ hydrodynamic cell updates per second on one NVIDIA V100, $6.1 \times 10^8$ on one A100 and $6.8 \times 10^8$ on one H100. All three were measured on a uniform $512 \times 256 \times 256$ grid. On the same kinds of GPU and on $256^3$ cells, Shaderous takes 2.0 times as many updates as AthenaK on the A100 and 2.3 times as many on the H100. Two differences account for much of this. Shaderous computes in single precision, which halves the bytes per cell, and it takes one MUSCL-Hancock step per update where a two-stage Runge-Kutta integrator does twice the work. \citet{Stone2026} do not state the precision or the integrator of their benchmark. AthenaK builds in double precision by default, and the linear-wave input in its repository takes two stages per step.\footnote{\url{https://github.com/IAS-Astrophysics/athenak}}

If the benchmark ran that way, AthenaK moved four times the bytes per update, and its throughput per byte moved is shown by the hollow markers of Figure~\ref{fig:performance}(b), where each throughput is divided by the device's peak memory bandwidth from its datasheet: 900~GB\,s$^{-1}$ for the V100, 1555 or 2039~GB\,s$^{-1}$ for the 40 and 80~GB A100, and 2000 to 3900~GB\,s$^{-1}$ for the H100. On the RTX~3080, whose bandwidth is close to the V100's, Shaderous on a $256^3$ grid then takes 1.0 to 1.3 times as many updates per byte as AthenaK. On the A100 it takes 0.4 to 0.5 times as many, because it does not use that GPU's bandwidth (above). The comparison therefore shows that a browser costs Shaderous little or no throughput against a native GPU code. It does not show that its scheme is faster, and a native code in single precision with one stage per step would likely match or exceed it.

The memory of the GPU, rather than the code, limits the size of a run. On a server, where the memory budget is set on the command line, the three-dimensional adaptive mesh holds up to $35^3$ blocks of $16^3$ cells with a magnetic field and $41^3$ with species. The field's limit comes from the largest storage buffer that one binding may hold, 2~GiB, which bounds its flux register. The mass fluxes of the species across faces fill one buffer of up to 4~GiB, bound as two windows of 2~GiB. Without a field or species, only the memory of the card bounds the mesh, because every pass over the blocks is dispatched in slabs that keep within the number of workgroups that one dispatch may launch. The magnetized disk at $1024^3$ with four levels ran on an NVIDIA H200 at $35^3$ blocks, about 175 million cells, using 117~GB, and took 10 steps in 10~s.

\section{Discussion}
\label{sec:discussion}

Shaderous is not a complete replacement for production codes. It runs on one GPU, within the memory that the browser will allocate, so one device rather than a cluster sets the size of a run. It works in single precision. That is adequate for the problems that ship with the code, but round-off is biased when a uniform background sits exactly on a power of two, and long integrations accumulate errors that double precision would not. Its resistivity, viscosity and conduction are explicit, isotropic and have constant coefficients. Its implicit pressure step is meant for slow flow. It costs more than the explicit scheme on small grids and above a Mach number of about 0.1 in two dimensions.

The limits in this section are the known shortcomings of the current code, and the list is not complete. They do not prevent accurate and useful results. The code passes the validation suite of Section~\ref{sec:validation}, and its tidal disruptions (Section~\ref{sec:problems}) show what earlier simulations found: the rate at which debris returns falls close to $t^{-5/3}$ after a full disruption, and faster when a core survives, as in \citet{Guillochon2013}.

The radiation transport (Section~\ref{sec:radiation}) has limits of its own. It is gray and takes one timestep for all blocks of the adaptive mesh. The radiation transport runs only in Cartesian coordinates, without magnetic fields, comoving coordinates or a shearing box, and it has no sources of radiation other than the gas. The sweeps are first order in time, as is their splitting from the hydrodynamic step, but neither limits our tests. A supercritical shock at a fifth of the speed of light is the same, to 0.05~percent, with the hydrodynamic step halved or doubled and with the sweeps' step halved. We therefore did not adopt the two-stage implicit-explicit scheme of \citet{He2024}, which would double the cost of the sweeps. Where the speed of light is reduced, a radiative structure that moves across the grid, such as the precursor of a supercritical shock, is misplaced (Section~\ref{sec:rsla}). Radiation pressure is a source term that the well-balanced reconstruction (Section~\ref{sec:wellbalanced}) does not account for, so a star held up mostly by radiation is not yet in discrete balance.

The design choices of Sections~\ref{sec:conservation} to~\ref{sec:amr} came from constraints of WebGPU, but they do not depend on it. Gathering fluxes and recomputing them across the boundaries of refinement levels and of time bins trades extra arithmetic for less storage and less memory traffic, which suits any processor whose throughput is limited by memory rather than by arithmetic. Computing ghost cells on load removes a synchronization step from every AMR cycle. The MHD solver does not gather: it stores its face fluxes and edge electric fields, and it keeps a flux register at refinement boundaries.

The burning network is a fit to a reference network. It covers the 13 isotopes of the $\alpha$-chain and the densities and temperatures of its training set, and on held-out burns it matches its float64 copy to 2~percent per step. Its accuracy against the reference network over a whole detonation remains to be measured.

The bodies made of ANEOS materials are fluids without strength, which is a fair approximation only above a few hundred kilometers. The water model of 2024 has one solid, with a density of 1.25~g~cm$^{-3}$ rather than that of ice~Ih, so a cold icy body is about a third too dense near its surface. At a body's surface, the pressure in the table rises from that of the vapor to gigapascals across the one cell of the table that holds the edge of the condensed phase, a cell 6~percent wide in density, so the outermost cells of a body are softer than ANEOS makes them.

\subsection{Future additions}
\label{sec:future}

We would like to add several capabilities. One is starting stars from the profiles of the stellar evolution code MESA \citep{Paxton2011}, with their density, temperature and composition mapped onto the grid and settled into the code's own balance. We would also like to set up a coronal mass ejection erupting from a magnetic flux rope in the corona of a star, and to compute light curves of shock breakout and of tidal disruptions with the radiation transport. We would like to test linear waves in comoving coordinates against the exact solutions of \citet{Berlok2022}, with pressure and self-gravity together. We aim to extend cut cells to moving bodies, to the adaptive mesh and to gases. The relativistic gas of Section~\ref{sec:relativity} has no magnetic field and no gravity, and we would like to add both.

\section{Conclusions}
\label{sec:conclusions}

Shaderous is a code for compressible hydrodynamics and MHD with self-gravity, adaptive mesh refinement, local timesteps, point masses, sinks, gray radiation transport, implicit pressure for nearly incompressible flow, tabulated equations of state from degenerate electrons to rock and iron, cooling, viscosity, conduction, resistivity, turbulence driving, mixtures of gases, nuclear burning, and solid bodies, fixed or moving, with the forces on them. It is written entirely in WebGPU compute shaders and runs from a web page. Its hydrodynamic solver conserves mass, momentum and energy to round-off on the adaptive mesh without storing any face flux, by having every cell recompute the fluxes that it needs, from the same states with the same arithmetic, across the boundaries of refinement levels and of time bins. It keeps every step on the GPU and computes ghost cells as it loads them. It works in single precision and holds hydrostatic stars and atmospheres in balance through a well-balanced reconstruction and a settled initial state, and it keeps magnetic fields free of divergence to round-off through compensated sums. The code and its validation suite run on the reader's own GPU, and a link reproduces any setup. Scripts and AI agents can drive the same page without a person at it, on a desktop or on a GPU server, and an agent can turn a simulation described in words into a checked setup and run it. The same properties make it a tool for teaching, from demonstrations for school pupils to term projects for graduate students.

We invite educators to try Shaderous in their classrooms. A class needs only a browser and a link, the activities of Section~\ref{sec:education} are ready to use, and a lesson can be shared as an address. We would like to hear what works, what confuses students and what is missing, through the issue tracker of the repository or by email, so that the next version of the code serves teaching better.

\begin{acknowledgments}
We thank Dan D'Orazio, Robert Fisher, Doron Kushnir and Anna Rosen for helpful discussions and feedback, and J. Xavier Prochaska for providing access to the Claude team through which all of the Claude models named below were used. The high-performance computing resources for this work, the data-center GPUs on which the largest simulations and the benchmarks of data-center hardware ran, were provided by the University of Massachusetts Dartmouth's partnership with the Unity Research Computing Platform, a multi-institutional cluster led by the University of Massachusetts and the University of Rhode Island.

The code, its validation suite, the manuscript source, the bibliography and the notes behind every number in the text are available online.\footnote{\url{https://github.com/guillochon/shaderous}} The author takes full responsibility for the contents of this paper and of the code. Large language models (Claude Opus 5.5 and Claude Fable 5.2) were used, under author supervision, to write and test the code, to search the literature, to draft and edit the text and the online documentation, and to measure the numbers that the text quotes. Every number in the text is traced to the code, to a measurement recorded in the repository, or to a cited paper. Every bibliography entry was checked against the NASA Astrophysics Data System, Crossref or arXiv, and all references and AI edits were hand-checked by the author. We aimed to be compliant with the AAS guidelines on generative AI in manuscript preparation.\footnote{\url{https://journals.aas.org/author-llm-guidelines}} Those guidelines do not treat such tools as authors. Responsibility for accuracy remains with the listed author.
\end{acknowledgments}

\software{Shaderous (\url{https://github.com/guillochon/shaderous}), TypeScript, Vite, Vitest, Playwright, h5wasm, Mediabunny, yt \citep{Turk2011}, pynucastro \citep{SmithClark2023}, PyTorch}

\appendix

\section{Validation tests}
\label{sec:validationtests}

Tables~\ref{tab:validation}, \ref{tab:validation-amr}, \ref{tab:validation-gravity}, \ref{tab:validation-mhd} and~\ref{tab:validation-mhd-amr} list the tests of the validation suite (Section~\ref{sec:validation}), each with its pass criterion and the value measured in one run of the suite in Chrome 156 on an NVIDIA RTX~3080. The report of that run is in the repository. The rows of tests added after it (driving, the many point masses, the orbit preview, moving walls, the replaced run, the near-vacuum corrections, field brushes, resistivity, line-tied walls, the equations of state and the ANEOS tests) were measured on an Apple M4 Pro, in the run whose report is also there. So were the rows that later changes to the code moved: the self-gravity of Section~\ref{sec:gravity}, the Zel'dovich pancake, and the implicit pressure. Tables~\ref{tab:validation-frames}, \ref{tab:validation-cosmo} and~\ref{tab:validation-curvilinear} list the tests of the frames and coordinate systems of Section~\ref{sec:coordinates}, with the values measured in one run of the suite in Chrome on an Apple M4 Pro. Tables~\ref{tab:validation-implicit}, \ref{tab:validation-implicit-gas} and~\ref{tab:validation-radiation} list the tests of the weakly compressible liquid and of implicit pressure (Section~\ref{sec:implicit}) and of radiation transport (Section~\ref{sec:radiation}), measured in the same way. Tables~\ref{tab:validation-relativity} and~\ref{tab:validation-relativity-2} list the tests of special relativity (Section~\ref{sec:relativity}), measured in one run on the RTX~3080 whose values are in the repository. Table~\ref{tab:validation-bodies} lists the tests of bodies in a stream (Section~\ref{sec:bodies}), and the rows marked $\dagger$ in the other tables are tests added since. All of these were measured in one later run on the RTX~3080, from the same build. The three tests of the rotor of a fast wind turbine with cut cells are left out: they measure a rotor that does not yet drive itself on these grids, and each takes up to several hours. Five more tests of cut cells have no rows yet. Tables~\ref{tab:validation-added}, \ref{tab:validation-spherical-3d}, \ref{tab:validation-exact}, \ref{tab:validation-particles} and~\ref{tab:validation-galaxies} list the tests added since the others: the adaptive mesh to seven levels, the hybrid burning step, spherical runs in three dimensions, the exact solutions of ExactPack, particles, sinks and galaxies. They were measured in one run on the RTX~3080 on 2 December 2026. The jet breakout with cooling, from the same run, is the last row of Table~\ref{tab:validation-relativity-2}.

\begin{table*}
\centering
\caption{The validation suite: hydrodynamics, species and burning on the uniform grid. Criteria are those coded in the suite. L1 errors are means over cells, relative to the mean density. The measured values are in the order of the criteria.}
\label{tab:validation}
\footnotesize
\begin{tabular}{lll}
\toprule
Test & Pass criterion & Measured \\
\midrule
Sod tube along $x$, $y$, $z$ & \tabpar{7.0cm}{L1 density error $< 0.01$; runs agree under axis permutation to $10^{-6}$} & \tabpar{4.4cm}{0.0079; 0} \\
Sod tube along the diagonal & \tabpar{7.0cm}{L1 density error $< 0.015$} & \tabpar{4.4cm}{0.011} \\
Strong shock tube (pressure ratio $10^5$) & \tabpar{7.0cm}{No floored cells; relative L1 density error $< 0.1$ at 128 cells} & \tabpar{4.4cm}{None; 0.075} \\
Sedov-Taylor blast & \tabpar{7.0cm}{Mean shock radius within 2\% of the similarity solution} & \tabpar{4.4cm}{1.6\%} \\
Kelvin-Helmholtz & \tabpar{7.0cm}{Transverse kinetic energy grows more than fivefold by $t = 1$} & \tabpar{4.4cm}{11-fold} \\
Low-Mach HLLC & \tabpar{7.0cm}{Gresho vortex loses less than half the energy lost with plain HLLC, and gains none} & \tabpar{4.4cm}{Loses 6\% against 15\%} \\
Well-balanced column & \tabpar{7.0cm}{Largest Mach number $< 10^{-3}$ after a sound crossing, uniform and block grids; at least 100 times below the plain scheme} & \tabpar{4.4cm}{$3.7\times10^{-5}$ and $8.6\times10^{-5}$; plain scheme 0.35} \\
Uniform advection, 10 crossings & \tabpar{7.0cm}{Mass and momentum drift $< 2 \times 10^{-6}$} & \tabpar{4.4cm}{Mass $2.4\times10^{-7}$, momentum $3.2\times10^{-7}$} \\
Species & \tabpar{7.0cm}{Fractions in $[0, 1]$ summing to 1 within $10^{-5}$; each species' mass to $10^{-5}$; a smooth species as accurate as the passive scalar} & \tabpar{4.4cm}{Sum $2\times10^{-7}$; mass $5\times10^{-7}$; error $2.5\times10^{-4}$ against $3.5\times10^{-4}$} \\
Species in the equation of state & \tabpar{7.0cm}{Helium in hydrogen at rest stays at rest to $10^{-5}$ under four equations of state; carried across, pressure balanced to $10^{-5}$ (ideal gas) or 0.05; helium in air, $\gamma$ of 5/3 and 1.4, to 0.01 (0.05 with the helium twice as hot)} & \tabpar{4.4cm}{At rest $6\times10^{-7}$; carried $1.4\times10^{-6}$ or 0.013; helium in air 0.002} \\
Shock and helium bubble & \tabpar{7.0cm}{Helium mass to $10^{-5}$; bit-identical faces; on the adaptive mesh, over 95\% of the helium at the finest level} & \tabpar{4.4cm}{$1.1\times10^{-6}$; no face differs; 99.3\%} \\
Arrhenius detonation & \tabpar{7.0cm}{Front within 1\% of the Chapman-Jouguet speed; fuel left behind $< 10^{-3}$} & \tabpar{4.4cm}{0.12\% slow; $6\times10^{-6}$} \\
Smooth advection & \tabpar{7.0cm}{Mean error $< 2 \times 10^{-4}$ at $256^2$; second order or better} & \tabpar{4.4cm}{$1.5\times10^{-4}$; order 2.0} \\
Flux determinism & \tabpar{7.0cm}{Bit-identical mass fluxes on every face, three flows} & \tabpar{4.4cm}{0 of 291,840 faces differ} \\
Blast in a reflecting box & \tabpar{7.0cm}{Mass to $10^{-5}$, total energy to $10^{-3}$} & \tabpar{4.4cm}{Mass $2.4\times10^{-7}$, energy $1.4\times10^{-6}$} \\
Two-dimensional grids & \tabpar{7.0cm}{Sod L1 error $< 0.003$ at $256^2$; bit-identical fluxes} & \tabpar{4.4cm}{0.0021; no face differs} \\
Rayleigh-Taylor & \tabpar{7.0cm}{At rest, Mach $< 10^{-3}$ at $t = 2$; perturbed, energy to $10^{-3}$} & \tabpar{4.4cm}{Mach $1.2\times10^{-6}$; energy $1.4\times10^{-7}$} \\
\bottomrule
\end{tabular}
\end{table*}

\begin{table*}
\centering
\caption{The validation suite, continued: the adaptive mesh, cooling, diffusion, driving and walls.}
\label{tab:validation-amr}
\footnotesize
\begin{tabular}{lll}
\toprule
Test & Pass criterion & Measured \\
\midrule
Block grid without refinement & \tabpar{7.0cm}{Density within $10^{-5}$ of the uniform grid in every cell; same number of steps} & \tabpar{4.4cm}{0; same} \\
Adaptive Sod tube & \tabpar{7.0cm}{L1 error $< 0.003$ on under 60\% of the cells; mass and energy to $10^{-6}$} & \tabpar{4.4cm}{0.0022 on 34\%; mass $1.1\times10^{-7}$, energy $3\times10^{-8}$} \\
Adaptive species & \tabpar{7.0cm}{As for species, with the mesh rebuilt; a smooth species within $1.3 \times 10^{-3}$ of the passive scalar} & \tabpar{4.4cm}{Sum $4\times10^{-8}$; mass $1.3\times10^{-6}$; $1.0\times10^{-3}$} \\
Adaptive species in the equation of state & \tabpar{7.0cm}{At rest to $10^{-5}$; carried across, pressure within 0.05; faces match across levels} & \tabpar{4.4cm}{0; 0.007; faces to $6\times10^{-8}$} \\
Adaptive detonation & \tabpar{7.0cm}{As the detonation, with the whole reaction zone at the finest level} & \tabpar{4.4cm}{0.12\% slow; no coarse cell in the zone} \\
Adaptive Sedov blast & \tabpar{7.0cm}{Radius within 2\%; mass and energy to $10^{-6}$} & \tabpar{4.4cm}{1.6\%; mass 0, energy $9\times10^{-8}$} \\
Adaptive instabilities & \tabpar{7.0cm}{Mass to $10^{-6}$, energy to $10^{-5}$ while the mesh follows the flow} & \tabpar{4.4cm}{Mass $1.6\times10^{-7}$, energy $2.4\times10^{-7}$} \\
Adaptive Poisson solve & \tabpar{7.0cm}{Sphere potential within 0.3\% without refinement and 0.1\% with 2 and 3 levels; Jeans rate within 5\%} & \tabpar{4.4cm}{0.15, 0.027 and 0.053\%; 0.02\%} \\
Adaptive Evrard and polytrope & \tabpar{7.0cm}{Energy to 2\% through bounce; central density within 10\%} & \tabpar{4.4cm}{0.19\%; 0.8\%} \\
Adaptive tidal disruption & \tabpar{7.0cm}{Pericenter between 2\% inside and 3\% outside; frame centered to 0.005 stellar radii} & \tabpar{4.4cm}{0.4\%; 0.0007} \\
Local timesteps & \tabpar{7.0cm}{Sod and Sedov as with one timestep; mass and energy to $10^{-6}$} & \tabpar{4.4cm}{Sod 0.0023, Sedov 1.6\%; mass $1.1\times10^{-7}$, energy $7\times10^{-8}$} \\
HDF5 export and import & \tabpar{7.0cm}{Restarted state equal to $10^{-5}$; continued runs agree to $10^{-4}$} & \tabpar{4.4cm}{0; $6\times10^{-7}$} \\
Cooling & \tabpar{7.0cm}{Power-law decay to 0.1\%; thermal balance to 1\% without overshoot} & \tabpar{4.4cm}{0.001\%; balance without overshoot} \\
Viscosity & \tabpar{7.0cm}{Energy of a shear wave decays as $\exp(-2\nu k^2 t)$ to 1\%, and to 3\% between no-slip walls; energy to $10^{-5}$ on the adaptive mesh} & \tabpar{4.4cm}{0.3\%; 0.05\%; $3\times10^{-8}$} \\
Conduction & \tabpar{7.0cm}{Temperature wave decays at the linear rate to 1\%; cloud scale of the thermal instability changes by $< 12\%$ when the resolution doubles} & \tabpar{4.4cm}{0.27\%; 7.5\%} \\
Turbulence driving & \tabpar{7.0cm}{Isothermal box: rms speed between 0.7 and 1.6 at $t = 3$, within 30\% of it at $t = 4$; the same seed gives the same run to $10^{-5}$ and another seed one more than 1\% apart; with the force frozen, unrefined blocks follow the uniform grid to $10^{-4}$; with a field in 2D, the field grows} & \tabpar{4.4cm}{1.04 and 1.00; 0 and 20\%; $1.0\times10^{-7}$; 12\%} \\
Moving walls & \tabpar{7.0cm}{Lid-driven cavity at Reynolds number 100: centerline velocity within 0.015 of \citet{Ghia1982}; a field through a moving wall carried along it to 1\% of its amplitude, the wall at rest opposite bit-identical, the flux through the moving one to $10^{-6}$; unrefined blocks follow the uniform grid to $10^{-4}$} & \tabpar{4.4cm}{0.0065; 0.25\%, no face differs, $5\times10^{-9}$; $1.0\times10^{-5}$} \\
\bottomrule
\end{tabular}
\end{table*}

\begin{table*}
\centering
\caption{The validation suite, continued: point masses, the reference checks, equations of state and gravity. The test marked $\dagger$ was measured in a later run on the same GPU.}
\label{tab:validation-gravity}
\footnotesize
\begin{tabular}{lll}
\toprule
Test & Pass criterion & Measured \\
\midrule
Point mass and sink & \tabpar{7.0cm}{Gas plus accreted mass constant to $10^{-5}$} & \tabpar{4.4cm}{$1.6\times10^{-7}$} \\
Many point masses & \tabpar{7.0cm}{Six masses with overlapping sinks on a ring: gas plus accreted mass to $10^{-5}$, every mass accretes, opposite masses take in the same to 1\% and stay opposite to $10^{-3}$ of the box, in 2D and 3D, uniform, adaptive and with per-block timesteps} & \tabpar{4.4cm}{$1.6\times10^{-7}$; all accrete; 0.04\%; $2.5\times10^{-6}$} \\
Orbit preview & \tabpar{7.0cm}{Gas read back as lumps holds the mass to $10^{-4}$ and the center of mass to $10^{-3}$ of the box; a light mass a quarter of an orbit on within 0.5\% of the box of its preview} & \tabpar{4.4cm}{$9\times10^{-6}$ and $4\times10^{-7}$; 0.23\%} \\
Binary orbit & \tabpar{7.0cm}{Separation within 0.5\% over one period; both masses accrete} & \tabpar{4.4cm}{0.016\%; both accrete} \\
Close encounter of two masses & \tabpar{7.0cm}{Orbit of eccentricity 0.9 over two periods: mean step within 10\% of a sixteenth of the Courant number times the lesser of $r/v$ and $\sqrt{r^3/GM}$; energy to 1\%} & \tabpar{4.4cm}{0.79 of the limit; 0.67\%} \\
GPU against CPU reference & \tabpar{7.0cm}{Density and energy differ by $< 10^{-4}$ of the mean after 20 steps} & \tabpar{4.4cm}{Density $8\times10^{-8}$, energy $2\times10^{-7}$} \\
Burning network & \tabpar{7.0cm}{One step on 48 held-out states within 2\% (L1) of the float64 network} & \tabpar{4.4cm}{0.10\%} \\
Equations of state$^\dagger$ & \tabpar{7.0cm}{Round trip to $10^{-5}$; WGSL matches float64 mirrors to $10^{-4}$, degenerate electrons and hydrogen ionization included; tabulated electrons and tabulated hydrogen and helium: round trip to $5\times10^{-5}$, energy and sound speed to $5\times10^{-4}$. Tabulated electrons also in hot thin gas, $2\times10^4$ to $10^{10}$~K and $6\times10^{-6}$ to 2~g~cm$^{-3}$. Hydrogen and helium from $10^{-18}$ to $10^{-2}$~g~cm$^{-3}$ and 20 to $10^7$~K, solar with and without radiation and pure hydrogen} & \tabpar{4.4cm}{Round trip $5.4\times10^{-6}$; mirrors $1.1\times10^{-6}$; tabulated electrons $1.3\times10^{-5}$ and $4.6\times10^{-4}$; hydrogen and helium $5.7\times10^{-6}$ and $3.3\times10^{-5}$} \\
ANEOS materials & \tabpar{7.0cm}{Four materials, and rock with iron at mass fractions 0.1 to 0.9, from $10^{-4}$ to 30~g~cm$^{-3}$ and 200 to $10^5$~K: WGSL matches the float64 mirror, the round trip from pressure to energy and back to $10^{-3}$, energy and sound speed to $10^{-2}$ (99th percentile)} & \tabpar{4.4cm}{Round trip $1.1\times10^{-4}$, energy $8.7\times10^{-5}$, sound speed $6\times10^{-5}$ (worst material or mixture)} \\
Uniform sphere potential & \tabpar{7.0cm}{Largest error $< 1\%$ (isolated boundaries)} & \tabpar{4.4cm}{0.15\%} \\
Jeans instability & \tabpar{7.0cm}{Growth rate within 5\% of linear theory} & \tabpar{4.4cm}{0.016\%} \\
Polytrope & \tabpar{7.0cm}{Central density within 10\% over 10 dynamical times} & \tabpar{4.4cm}{2.5\%} \\
Cold white dwarf & \tabpar{7.0cm}{Settled $0.6\,M_\odot$ dwarf: central density within 5\% over three dynamical times; energy to $10^{-3}$} & \tabpar{4.4cm}{3.3\%; $1.2\times10^{-5}$} \\
Evrard collapse & \tabpar{7.0cm}{Total energy to 0.6\%; thermal energy peaks at $0.9 < t < 1.3$} & \tabpar{4.4cm}{0.23\%; peak at $t = 1.12$} \\
Tidal disruption & \tabpar{7.0cm}{Pericenter within 2\%; star disrupted} & \tabpar{4.4cm}{0.14\%; peak density falls to 4\%} \\
Planet at rest & \tabpar{7.0cm}{Planet of $1\,\Mearth$ with a 30\% Fe85Si15 core under pyrolite, not settled, 16 cells to its radius: central density within 8\% over four dynamical times; mass within 1\% of $\Mearth$ and conserved to $10^{-4}$; energy to $10^{-3}$} & \tabpar{4.4cm}{Central density 3.4\%; mass $0.996\,\Mearth$, kept exactly; energy $2.2\times10^{-6}$} \\
Moon-forming impact & \tabpar{7.0cm}{Two levels, 8 finest cells to $\Rearth$, two hours after contact: mass to $10^{-3}$, energy to 1\%; planet of 0.93 to $1.01\,\Mearth$ with 28 to 32\% iron; disk of 0.005 to $0.08\,\Mearth$, over half from Theia; under $0.02\,\Mearth$ escaping} & \tabpar{4.4cm}{Mass exact, energy $8.7\times10^{-5}$; planet $0.98\,\Mearth$, 30\% iron; disk $0.035\,\Mearth$ (1.5 lunar masses), 80\% from Theia; $1.2\times10^{-4}\,\Mearth$ escaping} \\
\bottomrule
\end{tabular}
\end{table*}

\begin{table*}
\centering
\caption{The validation tests of MHD. Except where the field is the subject of the test or the comparison is bit for bit, each also requires $\max|\nabla\cdot\mathbf{B}|\,\Delta x/\max|B| < 10^{-6}$ and that faces shared by two blocks agree; the largest value measured was $2.8\times10^{-7}$. The test marked $\dagger$ was measured on an NVIDIA RTX~3080.}
\label{tab:validation-mhd}
\footnotesize
\begin{tabular}{lll}
\toprule
Test & Pass criterion & Measured \\
\midrule
Brio-Wu shock tube & \tabpar{7.0cm}{Compound wave, slow shock and least $v_y$ in ranges set around the solution of \citet{Brio1988}; mass and energy to $10^{-6}$; HLLD and HLLE agree to 0.01} & \tabpar{4.4cm}{Compound wave at $x = -0.029$, slow shock at 0.146, $v_y = -1.6$; energy $1.8\times10^{-7}$; 0.0017} \\
Orszag-Tang vortex & \tabpar{7.0cm}{Mass and energy to $10^{-6}$; magnetic energy and peak pressure within regression ranges} & \tabpar{4.4cm}{Mass $2\times10^{-7}$, energy $9\times10^{-8}$; 0.0617 and 0.511} \\
Field loop & \tabpar{7.0cm}{After one crossing, at least 90\% of the magnetic energy kept and the peak within 10\%} & \tabpar{4.4cm}{93.3\%; peak 8\% high} \\
Alfv\'en wave & \tabpar{7.0cm}{Mean error of $v_z$ $< 3 \times 10^{-5}$ at $256^2$; second order or better} & \tabpar{4.4cm}{$2.0\times10^{-5}$; order 2.0} \\
Blast in a field, 3D & \tabpar{7.0cm}{Mass to $10^{-6}$, energy to $3 \times 10^{-6}$; elongated along the field by at least 1.3} & \tabpar{4.4cm}{Mass $2.4\times10^{-7}$, energy $9\times10^{-7}$; 2.0} \\
Stepping in slabs & \tabpar{7.0cm}{State and faces bit-identical to one slab} & \tabpar{4.4cm}{No value differs} \\
GPU against CPU reference & \tabpar{7.0cm}{Density, energy and face field within $10^{-5}$ of the mean after 20 steps (Orszag-Tang) and 6 (blast)} & \tabpar{4.4cm}{$1.0\times10^{-7}$} \\
Field brushes & \tabpar{7.0cm}{Divergence below $10^{-6}$ after each brush. Orszag-Tang: the core of a stronger brush exactly 1.5 times stronger to $10^{-4}$; a pull raises $\max|B|$ by under 30\%, with the magnetic energy within 5\%. Pulls across refinement boundaries: under 50\%. A long sweep: at most 3 times the field it swept up. Total energy changed by the change in magnetic energy to 1\%} & \tabpar{4.4cm}{Below $2.7\times10^{-7}$. 1.5 to $4\times10^{-6}$; 10\%, 0.7\%. None. 2.6 times. $6\times10^{-5}$} \\
HDF5 restart & \tabpar{7.0cm}{Face field restored bit for bit; continued runs agree to $10^{-5}$} & \tabpar{4.4cm}{No face differs; $3.6\times10^{-6}$} \\
Run replaced while read & \tabpar{7.0cm}{No WebGPU error when a run is freed during a read of its diagnostics} & \tabpar{4.4cm}{No error, on either grid} \\
Near-vacuum correction$^\dagger$ & \tabpar{7.0cm}{Thin magnetized disk in $10^{-5}$ gas, $128^3$, 100 steps, uniform and blocks: finite, slower than 100, divergence below $10^{-6}$; slabs and chunks change no bit} & \tabpar{4.4cm}{Fastest 5.4 (blocks 5.9, local timesteps 8.1); $1.1\times10^{-7}$; none} \\
\bottomrule
\end{tabular}
\end{table*}

\begin{table*}
\centering
\caption{The validation tests of MHD, continued: the adaptive mesh and the other physics in a field. The requirements on the divergence and on shared faces are those of Table~\ref{tab:validation-mhd}. The test marked $\dagger$ was measured in a later run on the same GPU.}
\label{tab:validation-mhd-amr}
\footnotesize
\begin{tabular}{lll}
\toprule
Test & Pass criterion & Measured \\
\midrule
Beside slower blocks & \tabpar{7.0cm}{Blob through $10^{-6}$ gas, blocks with local timesteps: corrected while they save updates; mass to $10^{-6}$; chunks change no bit} & \tabpar{4.4cm}{1887 blocks, 30\% saved; mass exact; none} \\
Blocks without refinement & \tabpar{7.0cm}{Energies equal to the uniform grid's to $10^{-6}$ after 40 steps} & \tabpar{4.4cm}{0} \\
Refined mesh & \tabpar{7.0cm}{Mass to $3 \times 10^{-6}$, energy to $10^{-4}$; coarse face the mean of the fine to $10^{-6}$; magnetic energy within 1\% (rotor) and 0.5\% (blast) of the uniform grid} & \tabpar{4.4cm}{Mass $9\times10^{-8}$, energy $2.5\times10^{-5}$; $1.0\times10^{-7}$; 0.02\%} \\
Leaves in chunks & \tabpar{7.0cm}{Bit-identical to one chunk} & \tabpar{4.4cm}{No value differs} \\
Moving mesh & \tabpar{7.0cm}{Over at least 20 rebuilds, mass to $10^{-6}$; field loop keeps 94\% of its energy} & \tabpar{4.4cm}{144 rebuilds, mass $4\times10^{-7}$; 96.2\%} \\
Per-block timesteps & \tabpar{7.0cm}{Magnetic energy within 0.5\% of one timestep; never more block updates} & \tabpar{4.4cm}{0.013\%; 85 to 100\% of the updates} \\
Gravity, point masses and cooling & \tabpar{7.0cm}{Magnetized polytrope and column as still as without a field; cooling decay to $10^{-3}$; sink budget to $10^{-5}$} & \tabpar{4.4cm}{Central density of the star 0.1\%; cooling $9\times10^{-6}$; sink $7\times10^{-8}$} \\
The same on blocks & \tabpar{7.0cm}{As above, also with per-block timesteps} & \tabpar{4.4cm}{0.11\%; $3\times10^{-5}$; $3\times10^{-8}$} \\
Fields inside stars & \tabpar{7.0cm}{Dipole star: central density and magnetic energy within 5\% over two dynamical times} & \tabpar{4.4cm}{0.5\% and 2.3\%} \\
Species in a field & \tabpar{7.0cm}{As for species, in a field at $\beta_{\rm M} = 4$} & \tabpar{4.4cm}{Sum $2\times10^{-7}$; mass $6\times10^{-7}$} \\
Species that set the gas, on blocks$^\dagger$ & \tabpar{7.0cm}{Magnetized blast beside a carbon blob in helium: energies within $10^{-3}$ and largest Mach number within 2\% of the uniform grid; Mach number more than 10\% from the same run with tracer species, so each cell's composition is read} & \tabpar{4.4cm}{$\le 1.0\times10^{-4}$; 0.088\%; 40\%} \\
Viscosity and conduction in a field & \tabpar{7.0cm}{Shear waves to 1\%; viscously damped Alfv\'en wave to 2\%; temperature wave across the field to 1\%} & \tabpar{4.4cm}{0.31\%; 0.03\%; 0.10\%} \\
Resistivity & \tabpar{7.0cm}{A field wave at rest decays as $\exp(-2\eta k^2 t)$ to 1\%, energy to $10^{-6}$, the internal energy takes the loss to 2\%, in 2D and 3D, uniform and on blocks; refined mesh with per-block timesteps conserves mass to $10^{-6}$ and energy to $10^{-5}$} & \tabpar{4.4cm}{0.08 to 0.61\%; $4\times10^{-7}$; 0.26\%; mass $8\times10^{-8}$, energy $4\times10^{-7}$} \\
Line-tied walls & \tabpar{7.0cm}{Stressed X-point: every wall face bit-identical, mass and energy to $10^{-6}$, collapse; current-free null at rest to $10^{-4}$ of its magnetic energy; shear flow between no-slip walls decays at the viscous rate to 2\%} & \tabpar{4.4cm}{No face differs, energy $7\times10^{-8}$, peak density 3.5; $8\times10^{-6}$; 0.002\%} \\
Magnetized tidal disruption & \tabpar{7.0cm}{Field energy within 3\% of its integral; disruption follows the hydrodynamic run to 5\%} & \tabpar{4.4cm}{0.26\%; 0.56\%} \\
Magnetized disk & \tabpar{7.0cm}{Mass within 3\%, kinetic energy within 5\%; Alfv\'en ceiling held; Maxwell stress grows from zero} & \tabpar{4.4cm}{Mass 0.6\%, kinetic energy 0.6\%; $\alpha_{\rm M}$ from $3\times10^{-13}$ to $1.6\times10^{-3}$} \\
\bottomrule
\end{tabular}
\end{table*}

\begin{table*}
\centering
\caption{The validation tests of rotating frames and mirror symmetry (Section~\ref{sec:coordinates}), measured on an Apple M4 Pro, and those marked $\dagger$ on an NVIDIA RTX~3080.}
\label{tab:validation-frames}
\footnotesize
\begin{tabular}{lll}
\toprule
Test & Pass criterion & Measured \\
\midrule
Rotation of zero & \tabpar{7.0cm}{A frame that does not turn ends a Kelvin-Helmholtz run in the same bits as no frame} & \tabpar{4.4cm}{No value differs} \\
Coriolis force & \tabpar{7.0cm}{Uniform flow turns by $-2\Omega t$ to $10^{-4}$~rad; kinetic energy to $10^{-5}$; uniform and blocks} & \tabpar{4.4cm}{$2.7\times10^{-7}$~rad; $5.6\times10^{-7}$} \\
Centrifugal balance & \tabpar{7.0cm}{Isothermal gas held by the centrifugal force: Mach $< 10^{-4}$, and 100 times below the plain scheme} & \tabpar{4.4cm}{$8.3\times10^{-6}$, against 0.023} \\
Blob in the inertial frame & \tabpar{7.0cm}{After a quarter turn, centered on its inertial path to half a cell} & \tabpar{4.4cm}{0.085 of a cell} \\
Corotating binary & \tabpar{7.0cm}{Over two orbits, separation to $10^{-3}$ and each mass within 0.01 of its start} & \tabpar{4.4cm}{$6.9\times10^{-11}$; $5.9\times10^{-6}$} \\
Rotating frame with a field & \tabpar{7.0cm}{As the Coriolis test, with a field along $\boldsymbol\Omega$; magnetic energy to $10^{-6}$} & \tabpar{4.4cm}{$4.1\times10^{-7}$~rad; $1.9\times10^{-6}$; 0} \\
Frame that follows a binary$^\dagger$ & \tabpar{7.0cm}{Frame started 10\% slower than the orbit: line of centers within $2^\circ$ over the second orbit and $0.3^\circ$ at the end; frame's rate to 0.5\%; separation to $2\times10^{-3}$; uniform and blocks} & \tabpar{4.4cm}{$1.6^\circ$, $0.20^\circ$; 0.065\%; $9.0\times10^{-4}$; blocks the same} \\
Rotating polytrope$^\dagger$ & \tabpar{7.0cm}{$n = 1$ at $\Omega = 0.4$, four dynamical times: central density within 2\%, axis ratio within 0.005, dense gas slower than 0.01; settled without the spin, ten times the change; uniform and blocks} & \tabpar{4.4cm}{0.18\%, $9.5\times10^{-4}$, 0.0035; without the spin 0.15; blocks 0.20\%, $1.1\times10^{-3}$} \\
Flux through a mirror & \tabpar{7.0cm}{No mass crosses a plane of symmetry; both cells of every face agree} & \tabpar{4.4cm}{0; 0 of 95,200 faces} \\
Sedov blast on an octant & \tabpar{7.0cm}{Octant and 2D quadrant match the whole box in density to $10^{-5}$} & \tabpar{4.4cm}{$2.6\times10^{-6}$; $1.7\times10^{-6}$} \\
Half and quadrant grids & \tabpar{7.0cm}{Rectangular grids match the whole box: density to $10^{-5}$, potential to $10^{-4}$, blocks' mass and energy} & \tabpar{4.4cm}{$2.5\times10^{-6}$; $1.8\times10^{-6}$; 0 and $6\times10^{-8}$} \\
Sphere potential on an octant & \tabpar{7.0cm}{Within 1\% of the analytic potential and $10^{-4}$ of the whole box} & \tabpar{4.4cm}{0.15\%; $1.7\times10^{-6}$} \\
Ellipsoid on an octant & \tabpar{7.0cm}{Triaxial ellipsoid's potential within $10^{-4}$ of the whole box} & \tabpar{4.4cm}{$2.0\times10^{-6}$} \\
Polytrope on an octant & \tabpar{7.0cm}{Central density within 5\% over 3 dynamical times, as the whole box to 1\%} & \tabpar{4.4cm}{2.47\%, against 2.46\%} \\
Evrard collapse on an octant & \tabpar{7.0cm}{Energy to 0.6\%, thermal peak at $0.9 < t < 1.3$; also on blocks with local timesteps} & \tabpar{4.4cm}{0.13\%; blocks 0.08\%} \\
Field on an octant & \tabpar{7.0cm}{Density to $10^{-4}$ and magnetic energy to $10^{-5}$ of the whole box; div B at round-off} & \tabpar{4.4cm}{$3.6\times10^{-7}$; $1.1\times10^{-7}$; $1.3\times10^{-7}$} \\
Sink on an octant & \tabpar{7.0cm}{Mass accreted equals the whole box's to $10^{-5}$} & \tabpar{4.4cm}{$5.5\times10^{-8}$} \\
Two stars on a mirror$^\dagger$ & \tabpar{7.0cm}{Two settled polytropes above $z = 0$ against the whole box: mass to $10^{-5}$ and peak density to $10^{-4}$ when settled; at $t = 0.05$ density to $2\times10^{-3}$ and energy to $10^{-4}$; on blocks to $10^{-4}$ and $10^{-3}$} & \tabpar{4.4cm}{$1.9\times10^{-7}$, $8.8\times10^{-8}$; $7.4\times10^{-6}$, $3.6\times10^{-7}$; blocks $\le 9.0\times10^{-6}$} \\
\bottomrule
\end{tabular}
\end{table*}

\begin{table*}
\centering
\caption{The validation tests of the frames and coordinates, continued: comoving coordinates and the shearing box. Measured as in Table~\ref{tab:validation-frames}. The steady shear flow fails its mass criterion on the M4 Pro and passes elsewhere.}
\label{tab:validation-cosmo}
\footnotesize
\begin{tabular}{lll}
\toprule
Test & Pass criterion & Measured \\
\midrule
Scale factor & \tabpar{7.0cm}{$a$ from 0.01 to 1 in float32 within $10^{-5}$ of float64, matter only and with $\Lambda$} & \tabpar{4.4cm}{$3.4\times10^{-7}$; $2.9\times10^{-7}$} \\
Expansion drag & \tabpar{7.0cm}{Velocity as $1/a$ and pressure as $a^{3-3\gamma}$ to $10^{-5}$; uniform and blocks} & \tabpar{4.4cm}{$3.0\times10^{-7}$; $1.1\times10^{-8}$} \\
Cooling while expanding & \tabpar{7.0cm}{Internal energy within $10^{-3}$ of a float64 integration} & \tabpar{4.4cm}{$8.5\times10^{-6}$} \\
Gas and radiation expanding & \tabpar{7.0cm}{Temperature within $10^{-4}$ of a float64 integration} & \tabpar{4.4cm}{$2.2\times10^{-7}$} \\
Sound wave while expanding & \tabpar{7.0cm}{Phase within 1\% of $\int k c_s/a\,dt$ over 25~rad; one mode} & \tabpar{4.4cm}{0.07\%; 99.97\%} \\
Linear growth & \tabpar{7.0cm}{Amplitude grows as $a$ within 1\% from $a = 0.01$ to 0.1} & \tabpar{4.4cm}{0.8\%} \\
Zel'dovich pancake & \tabpar{7.0cm}{At $a = 0.05$ density and velocity within 2\%, on the uniform grid, on a mesh that refines as it runs and on a fixed mesh of two levels} & \tabpar{4.4cm}{0.59\%, 0.25\%; 0.66\%, 0.18\%; 0.67\%, 0.29\%} \\
Jeans length while expanding$^\dagger$ & \tabpar{7.0cm}{The Jeans floor, refinement and diagnostic in proper lengths: a wave of the box's length grows alike from $a = 0.1$ and 0.5, to $10^{-3}$, and not under the floor; a box refines where 16 cells per Jeans length are asked for and not where 4 are} & \tabpar{4.4cm}{Growth 2.435 from both, none under the floor; refined as asked; every unresolved cell counted} \\
Steady shear flow & \tabpar{7.0cm}{Over 10 orbits, velocity within $2\times10^{-3}$, density $10^{-3}$, mass $10^{-6}$} & \tabpar{4.4cm}{$8.8\times10^{-4}$; $2.4\times10^{-4}$; $8.5\times10^{-8}$} \\
Blob across the shearing boundary & \tabpar{7.0cm}{Mass and $x$ momentum to $10^{-6}$ over two orbits; density in $[0.9, 2.1]$} & \tabpar{4.4cm}{$2.2\times10^{-7}$; $3.1\times10^{-7}$; 0.97 to 2.07} \\
Shearing wave & \tabpar{7.0cm}{Amplitude within 3\% of the linear solution at eight times to $t = 4$; phase to 0.05~rad} & \tabpar{4.4cm}{0.33\%; $2\times10^{-5}$~rad} \\
Magnetized shearing box & \tabpar{7.0cm}{Net $b_z$ and $b_x$ to $10^{-6}$ of the field over two orbits; div B $< 10^{-5}$} & \tabpar{4.4cm}{$1.6\times10^{-9}$, $8\times10^{-10}$; $8.8\times10^{-8}$} \\
Magnetorotational instability & \tabpar{7.0cm}{Channel mode grows at $0.75\Omega$ within 5\%; $Q_z$ within 2\%; Maxwell stress positive} & \tabpar{4.4cm}{1.6\%; 0\%; $3.8\times10^{-6}$} \\
Epicycles & \tabpar{7.0cm}{After one epicycle the kick returns to $10^{-3}$; epicyclic energy to $10^{-4}$} & \tabpar{4.4cm}{$5.4\times10^{-5}$; $5.2\times10^{-5}$} \\
\bottomrule
\end{tabular}
\end{table*}

\begin{table*}
\centering
\caption{The validation tests of axisymmetric $(R, z)$ and spherical $(r, \theta)$ runs (Section~\ref{sec:curvilinear}), measured on an Apple M4 Pro, and those marked $\dagger$ on an NVIDIA RTX~3080.}
\label{tab:validation-curvilinear}
\footnotesize
\begin{tabular}{lll}
\toprule
Test & Pass criterion & Measured \\
\midrule
Gas at rest in $(R, z)$ & \tabpar{7.0cm}{Mach $< 10^{-5}$ after a sound crossing} & \tabpar{4.4cm}{$5.9\times10^{-7}$} \\
Swirling ring & \tabpar{7.0cm}{Mass, energy, $z$ momentum and angular momentum to $10^{-5}$ over a crossing} & \tabpar{4.4cm}{$1.0\times10^{-7}$ to $1.1\times10^{-7}$} \\
Sedov blast on the axis & \tabpar{7.0cm}{Radius along $R$ and $z$ within 2\% of the similarity solution} & \tabpar{4.4cm}{1.5\% along both} \\
Sink on the axis & \tabpar{7.0cm}{Accretion within 5\% of the same sink in 3D} & \tabpar{4.4cm}{0.31\%} \\
Ring and blast on blocks & \tabpar{7.0cm}{As the ring and the blast, through re-meshing} & \tabpar{4.4cm}{$\le 8.7\times10^{-7}$; 1.5\%} \\
Sphere potential in $(R, z)$ & \tabpar{7.0cm}{Within 1\% of the analytic potential} & \tabpar{4.4cm}{0.073\%} \\
Evrard collapse in $(R, z)$ & \tabpar{7.0cm}{Energy to 0.1\%, thermal peak at $0.9 < t < 1.3$} & \tabpar{4.4cm}{0.037\%; $t = 1.125$} \\
Sphere potential on blocks & \tabpar{7.0cm}{Within 1\% in every leaf, three levels} & \tabpar{4.4cm}{0.026\%} \\
Evrard collapse on blocks & \tabpar{7.0cm}{As in $(R, z)$, with self-gravity on blocks} & \tabpar{4.4cm}{0.12\%; $t = 1.125$} \\
Mirrored across $z = 0$$^\dagger$ & \tabpar{7.0cm}{Upper half against the whole run: blast density to $10^{-5}$; Evrard potential to $10^{-3}$, density $2\times10^{-3}$ and energy $10^{-4}$ at $t = 0.75$; sink to $10^{-4}$; blast on blocks to $10^{-4}$} & \tabpar{4.4cm}{$3.8\times10^{-6}$; $7.0\times10^{-6}$, $8.7\times10^{-4}$, $4.6\times10^{-5}$; $3.1\times10^{-10}$; $6.0\times10^{-8}$} \\
Species in $(R, z)$$^\dagger$ & \tabpar{7.0cm}{Point blast in three nuclides that set the gas: each species' mass to $10^{-5}$, uniform carbon to $10^{-4}$, fractions summing to 1 within $10^{-5}$; uniform and blocks} & \tabpar{4.4cm}{$6.3\times10^{-8}$; $4.6\times10^{-7}$; $2.0\times10^{-7}$; blocks $\le 2.1\times10^{-7}$} \\
Noh implosion & \tabpar{7.0cm}{Post-shock density within 5\% of 64; pre-shock within 5\% of 9} & \tabpar{4.4cm}{62.2; 9.01} \\
Bondi accretion on the axis & \tabpar{7.0cm}{Rate within 5\% of Bondi's} & \tabpar{4.4cm}{3.5\%} \\
Torus around a point mass$^\dagger$ & \tabpar{7.0cm}{Over five orbits, peak density, mass and center within 1\%} & \tabpar{4.4cm}{$2.9\times10^{-3}$; $1.5\times10^{-4}$; $1.9\times10^{-5}$} \\
Field in $(R, z)$ & \tabpar{7.0cm}{Field loop: div B $< 10^{-5}$, mass and energy to $10^{-5}$, energy kept within 1\% of Cartesian; field at rest, Mach $< 10^{-5}$} & \tabpar{4.4cm}{$2.9\times10^{-7}$; $\le 2.3\times10^{-7}$; 0.25\%; $2.9\times10^{-7}$} \\
Column under gravity & \tabpar{7.0cm}{Mach $< 10^{-4}$ after a sound crossing} & \tabpar{4.4cm}{$8.6\times10^{-5}$} \\
Star on the axis & \tabpar{7.0cm}{Settled polytrope: central density within 3\%, mass $10^{-4}$, to $t = 2$; uniform and blocks} & \tabpar{4.4cm}{1.3\%; $1.7\times10^{-5}$; blocks the same} \\
Swirl reaching the axis & \tabpar{7.0cm}{300 steps finite, none floored, mass to $10^{-5}$; uniform and blocks} & \tabpar{4.4cm}{Finite; 0; $7.5\times10^{-8}$} \\
Gas at rest in $(r, \theta)$ & \tabpar{7.0cm}{Mach $< 10^{-5}$, uniform and logarithmic in $r$} & \tabpar{4.4cm}{$3.6\times10^{-6}$; $5.7\times10^{-6}$} \\
Blob across the poles & \tabpar{7.0cm}{Mass, energy and angular momentum to $10^{-5}$ over a crossing} & \tabpar{4.4cm}{$\le 4.7\times10^{-7}$} \\
Sedov blast at the origin & \tabpar{7.0cm}{Radius within 1\% on every ray; spread over $\theta$ below 0.5\%} & \tabpar{4.4cm}{0.74\%; $1.3\times10^{-7}$} \\
Polytrope in its monopole & \tabpar{7.0cm}{Innermost density within 1\% over 10 dynamical times} & \tabpar{4.4cm}{0.020\%} \\
Bondi accretion through the inner sphere & \tabpar{7.0cm}{Rate through every shell within 2\% of Bondi's} & \tabpar{4.4cm}{1.9\%} \\
Blast and blob on blocks & \tabpar{7.0cm}{Radius within 2\% on 64 rays; blob to $10^{-5}$; one and two time bins} & \tabpar{4.4cm}{0.74\%; $\le 1.2\times10^{-6}$} \\
Polytrope on blocks & \tabpar{7.0cm}{Innermost density within 1\%; one and two time bins} & \tabpar{4.4cm}{0.053\%; 0.056\%} \\
Species in $(r, \theta)$$^\dagger$ & \tabpar{7.0cm}{As the species in $(R, z)$} & \tabpar{4.4cm}{$4.6\times10^{-6}$; $6.5\times10^{-6}$; $1.9\times10^{-7}$; blocks $\le 1.9\times10^{-6}$} \\
Swirl around the origin & \tabpar{7.0cm}{As the swirl in $(R, z)$, reaching both poles} & \tabpar{4.4cm}{Finite; 0; 0} \\
Second collapse of a first core & \tabpar{7.0cm}{Hydrogen and helium, $128 \times 128$, center starting at 2000 to 2200~K: central temperature within 1\% of its adiabat from 2 to $10^4$ times the starting density; second core denser than $5\times10^{-4}$~g~cm$^{-3}$ and hotter than $1.5\times10^4$~K, behind a shock inside 0.3~au; mass to 0.5\%} & \tabpar{4.4cm}{Adiabat $5.9\times10^{-4}$; core $7.5\times10^{-3}$~g~cm$^{-3}$, hottest $2.9\times10^4$~K, shock at 0.089~au; mass $6.9\times10^{-4}$} \\
\bottomrule
\end{tabular}
\end{table*}

\begin{table*}
\centering
\caption{The validation tests of the weakly compressible liquid and of implicit pressure (Section~\ref{sec:implicit}), measured on an Apple M4 Pro. $\mathcal{M}$ is the Mach number. Tests marked implicit use the implicit pressure step. The others use explicit pressure. The test marked $\dagger$ was measured on an NVIDIA RTX~3080.}
\label{tab:validation-implicit}
\footnotesize
\begin{tabular}{lll}
\toprule
Test & Pass criterion & Measured \\
\midrule
Liquid equation of state & \tabpar{7.0cm}{Against float64, for departures from $\rho_0$ of $10^{-6}$ to 0.1: pressure to $10^{-6}$ of $\rho_0 c_0^2$ times the departure, sound speed exactly $c_0$, elastic energy to $10^{-5}$} & \tabpar{4.4cm}{Pressure $4.8\times10^{-8}$; sound speed exact; energy $2.6\times10^{-7}$} \\
Liquid Gresho vortex & \tabpar{7.0cm}{Mach 0.1, 0.03 and 0.01 to $t = 2$: kinetic energy kept as by the ideal gas to 0.002, none gained; density within $2\mathcal{M}^2$; blocks as the uniform grid to $10^{-4}$} & \tabpar{4.4cm}{Kept 0.990, 0.975 and 0.940, as the ideal gas; density $0.83\mathcal{M}^2$; blocks $2.7\times10^{-7}$} \\
Liquid Taylor-Green vortex & \tabpar{7.0cm}{Decay rate within 1\% of $4\nu k^2$, uniform and blocks; density within $2\mathcal{M}^2$} & \tabpar{4.4cm}{Within 0.05\%, uniform and blocks; density 0.0023} \\
Liquid lid-driven cavity & \tabpar{7.0cm}{Reynolds number 100, $128^2$: centerline velocity within 0.015 of \citet{Ghia1982}; density within 5\%} & \tabpar{4.4cm}{0.0045; density 3.6\%} \\
Liquid vortex on blocks & \tabpar{7.0cm}{Three levels through a Gresho vortex, to $t = 0.5$: mass to $2\times10^{-6}$, momentum to $10^{-6}$, kinetic energy kept above 99\%, none gained} & \tabpar{4.4cm}{Mass $1.2\times10^{-7}$, momentum $5.9\times10^{-8}$; 99.9\% kept} \\
Pressure multigrid & \tabpar{7.0cm}{Manufactured solution with an obstacle, coefficients varying by 3 and each kind of box face: within $10^{-4}$ and residual below $10^{-4}$ in eight cycles; 2D and 3D} & \tabpar{4.4cm}{Error $1.9\times10^{-5}$, residual $3.1\times10^{-5}$} \\
Gresho vortex, implicit & \tabpar{7.0cm}{Mach 0.1, 0.01 and 0.001 to $t = 2$: steps alike to 2\%, under a tenth of the explicit scheme's; kinetic energy at least the explicit scheme's, alike to 0.002; density within $2\mathcal{M}^2$; mass to $2\times10^{-6}$, momentum to $10^{-6}$; under 2\% of solves above tolerance} & \tabpar{4.4cm}{212 steps at each Mach number against 3410; kept 0.993 against 0.990; density $0.83\mathcal{M}^2$; mass $1.2\times10^{-7}$, momentum $4.7\times10^{-8}$; none above tolerance} \\
Two and three dimensions & \tabpar{7.0cm}{A vortex in 2D and as a column in 3D: kinetic energy alike to $10^{-4}$ in the same steps; motion along $z$ below $10^{-6}$ of the flow} & \tabpar{4.4cm}{Alike, 0.986 in 26 steps; $2.8\times10^{-7}$} \\
Taylor-Green vortex, implicit & \tabpar{7.0cm}{Decay rate within 1\% of $4\nu k^2$, in at most a fifth of the explicit scheme's steps} & \tabpar{4.4cm}{Within 0.05\%, 237 steps against 1840} \\
Lid-driven cavity, implicit & \tabpar{7.0cm}{As the liquid cavity; mass to $2\times10^{-6}$} & \tabpar{4.4cm}{0.0047; mass $3.6\times10^{-7}$} \\
Cylinder wake & \tabpar{7.0cm}{Reynolds number 100, $256^2$, inlet and open faces, to $t = 20$: finite, density within 5\%, none floored; sheds with a period within 10\% of the explicit scheme's 1.43} & \tabpar{4.4cm}{Density 0.8\%, none floored; period 1.43} \\
Liquid column in a field & \tabpar{7.0cm}{In trapezoid balance: Mach $< 10^{-5}$ after 60 sound crossings; started uniform: Mach $< 10^{-3}$ by $t = 20$, contrast within 2\% of balance; mass to $2\times10^{-6}$} & \tabpar{4.4cm}{$1.0\times10^{-6}$; $6.5\times10^{-5}$, contrast as in balance; mass $6.2\times10^{-8}$} \\
Sound wave, implicit & \tabpar{7.0cm}{Acoustic Courant number 0.5: back after a period to 0.5\% of a wavelength, amplitude between 97\% and 100\%; Courant number 10: amplitude below 20\%, never above the start} & \tabpar{4.4cm}{Shift 0.18\%, amplitude 98.4\% (backward Euler 86\%); 12.5\%} \\
Blocks without refinement & \tabpar{7.0cm}{Kinetic energy as the uniform grid to $10^{-3}$, steps alike to one; mass to $2\times10^{-6}$} & \tabpar{4.4cm}{Alike to $10^{-3}$, 107 steps each; mass $1.2\times10^{-7}$} \\
Gresho vortex on blocks & \tabpar{7.0cm}{Three levels, Mach 0.1, 0.01 and 0.001 to $t = 1$: steps alike to 2\%, under a tenth of the explicit scheme's; kinetic energy above 99.8\%, none gained; density, mass and momentum as on the uniform grid; residual below $3\times10^{-3}$} & \tabpar{4.4cm}{273 or 274 steps; 99.9\% kept; density $0.83\mathcal{M}^2$, mass $2.4\times10^{-7}$, momentum $4.4\times10^{-8}$; residual $1.1\times10^{-3}$} \\
Regridding & \tabpar{7.0cm}{Fine blocks merged during the run: density within $2\mathcal{M}^2$; mass to $2\times10^{-6}$; kinetic energy above 99\%, none gained} & \tabpar{4.4cm}{Three levels to one; density $0.78\mathcal{M}^2$; mass $2.4\times10^{-7}$; 99.9\% kept} \\
Blocks in 3D and with viscosity & \tabpar{7.0cm}{Gresho column to $t = 0.25$: mass to $2\times10^{-6}$, motion along $z$ below $10^{-5}$, kinetic energy above 98\%; Taylor-Green decay rate on blocks within 1\%} & \tabpar{4.4cm}{Mass exact, $5.1\times10^{-7}$, 99.2\% kept; rate within 0.05\%} \\
Refinement on the velocity & \tabpar{7.0cm}{Double shear layer refined to three levels by the velocity alone, at the start and to $t = 0.3$; mass to $2\times10^{-6}$; density within 5\%} & \tabpar{4.4cm}{Three levels throughout; mass exact; density 0.5\%} \\
Species that set the gas, implicit$^\dagger$ & \tabpar{7.0cm}{Carbon blob in helium at Mach 0.05: pressure uniform to 0.03; fractions sum to 1 within $10^{-5}$; masses to $10^{-6}$; uniform and blocks} & \tabpar{4.4cm}{0.014 (explicit 0.017); $1.9\times10^{-7}$; $1.1\times10^{-8}$; blocks 0.0079} \\
\bottomrule
\end{tabular}
\end{table*}

\begin{table*}
\centering
\caption{The validation tests of implicit pressure, continued: the ideal gas, refinement boundaries along a field, species, stars and point masses. Measured as in Table~\ref{tab:validation-implicit}, and the test marked $\dagger$ on an NVIDIA RTX~3080.}
\label{tab:validation-implicit-gas}
\footnotesize
\begin{tabular}{lll}
\toprule
Test & Pass criterion & Measured \\
\midrule
Gas Gresho vortex, implicit & \tabpar{7.0cm}{Mach 0.1 and 0.01: steps alike to 2\%, under a tenth of the explicit scheme's; kinetic energy at least the explicit scheme's less 0.002; density within $2\times10^{-3}$; mass and energy to $2\times10^{-6}$. Mach 0.001: kinetic energy above 95\%, density within 5\%} & \tabpar{4.4cm}{219 steps against 3420; 0.992 against 0.990; density $3.2\times10^{-4}$; mass $6\times10^{-8}$, energy $8\times10^{-8}$. Mach 0.001: 98.5\%, density 1.4\%} \\
Gas Gresho vortex on blocks & \tabpar{7.0cm}{Three levels, Mach 0.1 and 0.01, to $t = 1$: kinetic energy above 99.5\%; density within $5\times10^{-3}$; mass and energy to $2\times10^{-6}$; residual below $10^{-4}$ early in the run} & \tabpar{4.4cm}{99.9\% kept; density $3.3\times10^{-3}$; mass $6\times10^{-8}$, energy $4.6\times10^{-8}$; residual $2.8\times10^{-5}$} \\
Sod tube, implicit & \tabpar{7.0cm}{Shock within one cell; mass and energy to $10^{-6}$; under 0.4 of the explicit scheme's steps; L1 density error below 0.012, and 0.009 at half the Courant number} & \tabpar{4.4cm}{0.59 cell; mass and energy below $10^{-7}$; 0.34 of the steps; 0.0097 and 0.0078 (explicit 0.004)} \\
Sod tube on blocks & \tabpar{7.0cm}{Two levels following the waves: mass and energy to $10^{-6}$; steps within 10\% of the uniform grid's; L1 density error below 0.008; every solve within tolerance} & \tabpar{4.4cm}{Mass exact, energy $7\times10^{-8}$; 73 steps against 75; 0.0061; all within tolerance} \\
Hot spot in moving gas & \tabpar{7.0cm}{Uniform and blocks to $t = 0.1$: none floored; density within a factor of 6 of 30; first step moves the gas less than 0.6 of a cell; mass to $10^{-6}$} & \tabpar{4.4cm}{None floored; 8.6 to 58; 0.50 of a cell; mass exact} \\
Liquid column across levels & \tabpar{7.0cm}{Refinement boundary across the field: Mach $< 10^{-3}$ at $t = 2$; mass to $2\times10^{-6}$} & \tabpar{4.4cm}{$2.5\times10^{-6}$; mass exact} \\
Boundaries along a field & \tabpar{7.0cm}{Mach below $10^{-4}$ for the liquid with a refined patch; for isothermal gas, below $10^{-4}$, $5\times10^{-4}$ and $2\times10^{-3}$ with a boundary across the field, a strip and a patch; energy to $10^{-6}$} & \tabpar{4.4cm}{$8.3\times10^{-6}$; $3.1\times10^{-6}$, $4.8\times10^{-5}$ and $7.5\times10^{-4}$; energy $1.0\times10^{-7}$} \\
Species, implicit & \tabpar{7.0cm}{Liquid vortex and gas bump, and the liquid on blocks: fractions sum to 1 within $10^{-5}$; uniform species to $10^{-4}$; each species' mass to $10^{-6}$} & \tabpar{4.4cm}{Sum $1.9\times10^{-7}$; uniform $5.8\times10^{-7}$; mass $9.6\times10^{-8}$, on blocks $3.3\times10^{-7}$} \\
Radiation with implicit pressure$^\dagger$ & \tabpar{7.0cm}{Equilibrium of Table~\ref{tab:validation-radiation}; trapped pulse at Mach 0.08 on its path to 0.005 of the box, width 7\%; blocks 0.004, 5\%} & \tabpar{4.4cm}{$1.9\times10^{-7}$; $7.4\times10^{-4}$, 5.5\%; blocks $5.5\times10^{-4}$, 1.2\%} \\
Gas columns in a field & \tabpar{7.0cm}{Density falling by $e$ every 16 and every 4 cells: Mach $< 10^{-3}$ after a sound crossing; energy to $10^{-5}$} & \tabpar{4.4cm}{$1.6\times10^{-6}$ and $3.0\times10^{-6}$; energy $9.4\times10^{-8}$} \\
Polytrope, implicit & \tabpar{7.0cm}{$64^3$, two dynamical times: central density within 2\%, mass 0.5\%, energy 2\%, in under a fifth of the explicit scheme's steps} & \tabpar{4.4cm}{Central density 0.3\% (explicit 2\%); mass 0.24\%, energy 0.80\%; 29 steps against 436} \\
Star in near-vacuum & \tabpar{7.0cm}{Surroundings at $10^{-7}$ of the central density, one dynamical time: no cell above 1.5 times the hottest at the start, none floored, central density within 5\%} & \tabpar{4.4cm}{Hottest unchanged; none floored; central density 0.5\%} \\
Moving star & \tabpar{7.0cm}{At 0.3 of its central sound speed: central density 2\%, momentum 1\%, energy 1\%; on two levels, central density 2\%, momentum 4\%} & \tabpar{4.4cm}{Central density 0.5\%, momentum 0.2\%, energy 0.34\%; on two levels 0.5\% and 1\%} \\
Polytrope on blocks & \tabpar{7.0cm}{Star one level finer than its surroundings: as the polytrope above, two levels throughout} & \tabpar{4.4cm}{Central density 0.5\%; mass 0.23\%, energy 0.56\%; 62 steps against 737} \\
Point masses, implicit & \tabpar{7.0cm}{Atmosphere around a mass: Mach $< 2\times10^{-3}$ and at most the explicit scheme's plus a tenth, in under a fifth of its steps, energy to $10^{-5}$; on blocks Mach $< 10^{-2}$. Two moving masses: momentum of gas and masses to $10^{-4}$ of the exchange; velocity within 4\% of the explicit scheme's at the default Courant number and 1\% at an eighth of it, on the uniform grid and on blocks} & \tabpar{4.4cm}{Mach $1.2\times10^{-3}$ against $1.3\times10^{-3}$, 40 steps against 380, energy $3.3\times10^{-6}$; blocks $4.7\times10^{-3}$. Moving: $3.6\times10^{-7}$; 2.8\% and 0.47\%, on blocks 0.37\% (backward Euler 7.9\%, 1.0\% and 1.3\%)} \\
Cylinder wake on blocks & \tabpar{7.0cm}{Three levels, 32 cells across the cylinder, to $t = 16$: finite, density within 5\%; sheds with a period within 10\% of 1.43} & \tabpar{4.4cm}{Density 0.8\%; period 1.40} \\
\bottomrule
\end{tabular}
\end{table*}

\begin{table*}
\centering
\caption{The validation tests of radiation transport (Section~\ref{sec:radiation}), measured on an Apple M4 Pro. The first 12 run on the uniform grid and the last five on the adaptive mesh. Times in the Marshak wave are in units of the absorption time $1/(c\kappa_{\rm P}\rho)$. The test marked $\dagger$ was measured on an NVIDIA RTX~3080.}
\label{tab:validation-radiation}
\footnotesize
\begin{tabular}{lll}
\toprule
Test & Pass criterion & Measured \\
\midrule
Radiation off & \tabpar{7.0cm}{Off, no radiation passes and the equation of state keeps its radiation; on, $E_{\rm r}$ carries it, split from the pressure to $10^{-5}$} & \tabpar{4.4cm}{No passes; split $2.8\times10^{-8}$} \\
Equilibrium in a closed box & \tabpar{7.0cm}{With $\xi = 1$, temperature to $10^{-4}$ of the root of $e(T) + a_{\rm r}T^4 = e_0 + E_{\rm r,0}$ and energy to $10^{-5}$; $\hat{c} = c/10$ everywhere misses by over 5\%; per-cell $\xi$ in opaque gas to $10^{-4}$} & \tabpar{4.4cm}{$1.9\times10^{-7}$, energy $5.2\times10^{-8}$; misses by 26\%; $5.0\times10^{-8}$} \\
Equilibrium in single precision & \tabpar{7.0cm}{At $\tau_{\rm c} = 10^{10}$ over $10^4$ sweeps, $E_{\rm r}/a_{\rm r}T^4$ within $10^{-5}$ of 1; energy to $10^{-6}$ at rest and $10^{-4}$ in moving gas} & \tabpar{4.4cm}{$7.7\times10^{-7}$; energy $3\times10^{-14}$ at rest, $4.0\times10^{-5}$ moving} \\
Streaming pulse & \tabpar{7.0cm}{A pulse with $|\mathbf{f}| = 1$ moves at $\hat{c}$, and at $c$ with $\hat{c} = c$, to 2\%; energy to $10^{-5}$; $|\mathbf{f}| > 0.95$ at its peak} & \tabpar{4.4cm}{Both to 0.5\%; energy $7.5\times10^{-8}$; $|\mathbf{f}| = 1.00$} \\
Diffusing pulse & \tabpar{7.0cm}{Variance grows as $2Dt$, $D = c/(3\kappa_{\rm R}\rho)$, to 10\% at $\tau_{\rm c} = 1$, 100 and $10^4$} & \tabpar{4.4cm}{1.01, 1.00 and 1.00 of $2Dt$} \\
Shadow of an opaque disk$^\dagger$ & \tabpar{7.0cm}{Behind a scattering disk of optical depth about 1000, $E_{\rm r}$ under 10\% of that beside it, where the beam keeps $E_{\rm r} > 0.85$ and $|\mathbf{f}| > 0.9$} & \tabpar{4.4cm}{3.4\%; beside it 0.92 and 0.93} \\
Trapped radiation advected & \tabpar{7.0cm}{Gas at $10^{-2}c$, box of optical depth $10^6$, $\hat{c} = c/100$: centroid within 1\% of the box of the gas's path, width to 10\%} & \tabpar{4.4cm}{Centroid on the path; width 3.4\%} \\
Marshak wave & \tabpar{7.0cm}{$E_{\rm r}$ and $a_{\rm r}T^4$ match the float64 reference to $2\times10^{-3}$ of the incident energy at times 1, 3.16 and 10, and the non-equilibrium diffusion solution to 0.03 at the last two} & \tabpar{4.4cm}{$1.6\times10^{-3}$; 0.0064 and 0.023} \\
Star held up partly by radiation & \tabpar{7.0cm}{$64^3$ octant. At rest for a dynamical time: central density to 10\%, rms and largest speeds under twice those of the star with radiation in its equation of state, energy to 3\%. At Mach 10: central density within 2\% of that star's, rms speed under 1.5 times, energy to 1\%} & \tabpar{4.4cm}{At rest: 8\%; 1.4 and 1.3 times; 1.4\%. At Mach 10: 1.1\%, 1.0 times, 0.1\%} \\
Radiative shocks & \tabpar{7.0cm}{Against the jump with radiation. Subcritical at $\hat{c} = c/5$: density, temperature and shock position to 1\%, a temperature spike, preheating under 5\%. Supercritical: at $c$ to 1\%, gas ahead heated over threefold; at $c/5$ and $c/2.5$ to 5\% and 2\%, the error falling as $\hat{c}$ grows} & \tabpar{4.4cm}{Subcritical 0.4\%; supercritical at $c$ 0.4\%, gas ahead 6.3 times hotter; 4.3\% and 1.4\%} \\
Hot core in an opaque shell & \tabpar{7.0cm}{Luminosity steady to 2\% over 30 steps and given to 0.1\% by the energy counted out; the same to 5\% at $\hat{c} = c/20$ and $c/10$} & \tabpar{4.4cm}{Steady to $3\times10^{-4}$; counted luminosity equal to 0.1\%; reduced speeds within 0.3\%} \\
Restart & \tabpar{7.0cm}{Radiation restored from a snapshot bit for bit; ten steps later $E_{\rm r}$ and the gas energy to $10^{-5}$} & \tabpar{4.4cm}{Bit for bit; $9.5\times10^{-7}$ and $3.7\times10^{-7}$} \\
Diffusion across levels & \tabpar{7.0cm}{Pulse on the face between coarse and fine blocks: energy to $10^{-4}$ and within $10^{-5}$ of the uniform grid's change; centroid to $2\times10^{-3}$ of the box and widths to 2\% of the fine uniform grid} & \tabpar{4.4cm}{$1.3\times10^{-5}$, within $7\times10^{-7}$ of the uniform grid's; centroid alike, widths 0.4\%} \\
Trapped radiation on blocks & \tabpar{7.0cm}{As advected above, through at least two re-meshes: centroid to $2\times10^{-3}$ of the box, width to 5\%, $E + E_{\rm r}/\xi$ to $10^{-5}$} & \tabpar{4.4cm}{Centroid on the path, width 2.1\%, $1.7\times10^{-6}$; four re-meshes} \\
Hot core on blocks & \tabpar{7.0cm}{Only the shell refined: luminosity within 1\% of the fine uniform grid's, steady to 1\% over 20 steps, given to 0.1\% by the energy counted out} & \tabpar{4.4cm}{Within 0.2\%; steady to $3\times10^{-5}$; equal to 0.1\%} \\
Star on blocks & \tabpar{7.0cm}{Three levels. At rest: central density to 5\%, rms speed under 3\% and largest under 10\% of the central sound speed, energy to 2\%. At Mach 10 as on the uniform grid} & \tabpar{4.4cm}{At rest: 2\%; 1.1\% and 2.9\%; 0.6\%. At Mach 10: 0.1\%, 0.94 times, 0.1\%} \\
Restart on blocks & \tabpar{7.0cm}{The same mesh and radiation bit for bit; ten steps later $E_{\rm r}$ to $10^{-5}$} & \tabpar{4.4cm}{Bit for bit; $2.4\times10^{-7}$} \\
\bottomrule
\end{tabular}
\end{table*}

\begin{table*}
\centering
\caption{The validation tests of bodies in a stream (Section~\ref{sec:bodies}), measured on an NVIDIA RTX~3080. $C_d$ and $C_l$ are the drag and lift coefficients and Re the Reynolds number. The liquid runs at Mach 0.1. The comparison of a moving body with one at rest fails, as the text describes. The test marked $\dagger$ was measured in a later run.}
\label{tab:validation-bodies}
\footnotesize
\begin{tabular}{lll}
\toprule
Test & Pass criterion & Measured \\
\midrule
Force balance & \tabpar{7.0cm}{Cylinder in a periodic box: gas momentum change is minus the force's impulse to $10^{-4}$ of it; liquid and gas, explicit and implicit, uniform and blocks} & \tabpar{4.4cm}{$2.4\times10^{-7}$ to $7.1\times10^{-6}$ in the eight runs} \\
Force on a moving body & \tabpar{7.0cm}{A cylinder sliding through liquid at rest: the same, with the momentum handed over where its cells change kind, to $10^{-3}$; explicit and implicit pressure} & \tabpar{4.4cm}{$1.5\times10^{-7}$; $1.9\times10^{-7}$} \\
Piston & \tabpar{7.0cm}{A slab sliding at Mach 0.085 through a closed box of gas: mass to $10^{-6}$; pressure ahead within 5\% of the adiabat; energy gained equal to the work done to 5\%} & \tabpar{4.4cm}{Exact; 1.443 against 1.456; 0.18745 against 0.18751} \\
Moving body against a stream$^\dagger$ & \tabpar{7.0cm}{A cylinder 16 cells across moving through still liquid, against liquid streaming past it at rest, Re 100: mean drag alike to 5\%} & \tabpar{4.4cm}{Fails: $C_d$ 1.32 at rest, 2.46 moving; implicit 1.32 and 2.12} \\
Sphere from a mesh & \tabpar{7.0cm}{An icosphere of 5120 triangles against the built-in sphere: solid cells and drag alike to 3\%, at 8 and 16 cells across} & \tabpar{4.4cm}{The same cells, and the same drag to the bit} \\
Paddle in its own frame & \tabpar{7.0cm}{Torque about the axis from explicit and implicit pressure alike to 5\%, and not zero} & \tabpar{4.4cm}{$-4.09$ and $-4.01$ (2.0\%)} \\
Thin plate on blocks & \tabpar{7.0cm}{A plate thinner than a coarsest cell: refined from its shape and whole, mass on each side to $10^{-6}$, before and after the mesh merges and refines again} & \tabpar{4.4cm}{Whole; $1.3\times10^{-8}$; whole again (unrefined, 3.4\% crosses)} \\
Turbine on blocks & \tabpar{7.0cm}{Every blade of the preset of the wind turbine at the finest level, cell for cell the shape's, at the start and after 100 steps} & \tabpar{4.4cm}{All three blades exact} \\
Cylinder in a tunnel & \tabpar{7.0cm}{Re 100, 3\% blockage: $C_d$ extrapolated from 16 and 32 cells across within 1.30 to 1.45; at 32 cells, Strouhal number within 0.160 to 0.178 and lift amplitude within 0.25 to 0.50} & \tabpar{4.4cm}{1.528 and 1.453, extrapolated 1.378; 0.170; 0.43} \\
Sphere in a tunnel & \tabpar{7.0cm}{Re 100, 16 cells across on two levels of blocks: $C_d$ within 0.95 to 1.35} & \tabpar{4.4cm}{1.253} \\
Cut cells at rest & \tabpar{7.0cm}{Liquid around a cut cylinder and sphere: after 50 steps speeds below $10^{-5}$, mass to $10^{-6}$, force below $10^{-6}$} & \tabpar{4.4cm}{$3.5\times10^{-8}$; exact; $3.2\times10^{-11}$} \\
Force balance with cut cells & \tabpar{7.0cm}{As the force balance, with the body cut into the cells, to $10^{-4}$; 2D and 3D} & \tabpar{4.4cm}{$2.8\times10^{-7}$; $4.0\times10^{-6}$} \\
Cut cylinder against float64 & \tabpar{7.0cm}{Re 100, 16 cells across: $C_d$ within 5\% of the float64 prototype's 1.220} & \tabpar{4.4cm}{1.204 (staircase 1.300)} \\
Airfoil & \tabpar{7.0cm}{NACA 0012 at $5^\circ$, Re 1000, 16, 32 and 64 cells on the chord: at 64, $C_l$ within 0.05 of 0.24 and $C_d$ 0.02 of 0.133 (float64 prototype)} & \tabpar{4.4cm}{$C_l$ 0.273, 0.319, 0.246 (staircase 0.164, 0.183, 0.234); $C_d$ 0.125, 0.118, 0.130} \\
Cylinder at Re 40 & \tabpar{7.0cm}{Steady wake, 8, 16 and 32 cells across: $C_d$ with cut cells at 16 within 3\% of that at 32} & \tabpar{4.4cm}{1.689, 1.768, 1.757 (staircase 1.835, 1.772, 1.753)} \\
Sphere at Re 100 & \tabpar{7.0cm}{Uniform grid, 8 and 16 cells across: $C_d$ with cut cells at 16 within 0.95 to 1.25} & \tabpar{4.4cm}{0.985, 1.073 (staircase 1.210, 1.139)} \\
\bottomrule
\end{tabular}
\end{table*}

\begin{table*}
\centering
\caption{The validation tests of special relativity (Section~\ref{sec:relativity}), measured on an NVIDIA RTX~3080. The blast waves are those of \citet{MartiMuller2003}. L1 errors are means over cells of the absolute difference in proper density from the exact solution.}
\label{tab:validation-relativity}
\footnotesize
\begin{tabular}{lll}
\toprule
Test & Pass criterion & Measured \\
\midrule
Recovery against float64 & \tabpar{7.0cm}{Above the float32 limit of the specific internal energy, $W$ and $\rho$ within $2\times10^{-3}$ and $p$ within 1\% of the float64 recovery; below it $W$ and $\rho$ within 1\%; closed form within $10^{-3}$} & \tabpar{4.4cm}{$1.2\times10^{-3}$; 0.3\%; 0.8\%; $2.8\times10^{-4}$} \\
Sod tube along $x$, $y$, $z$ & \tabpar{7.0cm}{L1 density error $< 0.015$ at $64^3$; runs equal under axis permutation to $10^{-6}$} & \tabpar{4.4cm}{0.0097; 0} \\
Blast wave 1 & \tabpar{7.0cm}{L1 density error $< 0.03$ at 512 cells; shock within a cell; no floored cells} & \tabpar{4.4cm}{0.022; 0.58 cells; none} \\
Blast wave 2 & \tabpar{7.0cm}{L1 density error $< 0.1$; shock within 2.5 cells; shell above 7 of its 10.42} & \tabpar{4.4cm}{0.076; 1.9 cells; 8.17} \\
Blast wave 1 along the diagonal & \tabpar{7.0cm}{L1 error along $(1, 1)$ within twice that along $x$, at $256^2$} & \tabpar{4.4cm}{0.64 times} \\
Tangential velocity 0.9 & \tabpar{7.0cm}{L1 density error $< 0.13$; shock speed within 1\%} & \tabpar{4.4cm}{0.098; 0.48\%} \\
Newtonian limit & \tabpar{7.0cm}{Sod tube at sound speeds near 0.01 against the Newtonian scheme: mean density difference $< 3\times10^{-4}$, largest $< 3\times10^{-3}$} & \tabpar{4.4cm}{$1.1\times10^{-4}$; $1.4\times10^{-3}$} \\
Advection at 0.9 and 0.99 & \tabpar{7.0cm}{Order $> 1.8$; L1 $< 3\times10^{-3}$ at 128 cells; pressure uniform to $2\times10^{-3}$} & \tabpar{4.4cm}{2.02, 2.09; $1.1\times10^{-3}$, $9.7\times10^{-4}$; $2.2\times10^{-5}$, $2.2\times10^{-4}$} \\
Conservation at $W = 5$ & \tabpar{7.0cm}{$D$, $\mathbf{S}$ and $\tau$ to $10^{-5}$ over 300 steps, periodic, $64^2$} & \tabpar{4.4cm}{$3.1\times10^{-9}$, $4.6\times10^{-10}$, $8.0\times10^{-9}$} \\
GPU against float64 & \tabpar{7.0cm}{Density, $\mathbf{u}$ and pressure within $10^{-5}$ of the float64 step after 20 steps of a 2D blast} & \tabpar{4.4cm}{$1.4\times10^{-6}$, $5.9\times10^{-6}$, $5.3\times10^{-7}$} \\
2D blast symmetry & \tabpar{7.0cm}{Mirror images in $x$ and $y$ to $10^{-5}$ of the peak density, the diagonal image to $10^{-2}$} & \tabpar{4.4cm}{0; $2.8\times10^{-5}$} \\
Cold beam & \tabpar{7.0cm}{$W = 10$, specific internal energy $10^{-6}$, once round a periodic box: pressure uniform to 1\%} & \tabpar{4.4cm}{$3.2\times10^{-3}$} \\
HLLC & \tabpar{7.0cm}{Contact at rest held to $10^{-5}$; sheared at $\pm0.5$ to $10^{-4}$ and at $\pm0.9$ to 0.1; blast wave 1 within 5\% of HLLE} & \tabpar{4.4cm}{$1.1\times10^{-6}$; $7.9\times10^{-6}$, 0.041; 0.99 times} \\
Blocks without refinement & \tabpar{7.0cm}{$D$ of every cell within $10^{-5}$ of the uniform grid, in as many steps} & \tabpar{4.4cm}{0; 199 steps each} \\
Blast wave 2 on three levels & \tabpar{7.0cm}{L1 $< 0.1$, shell above 7, shock within 2.5 finest cells, on under 90\% of the cells; $D$ and $\tau$ to $10^{-6}$} & \tabpar{4.4cm}{0.076, 8.17, 1.9 cells, 88\%; $6.0\times10^{-8}$, $8.1\times10^{-8}$} \\
Square across two levels & \tabpar{7.0cm}{Moving at 0.8: $D$, $\mathbf{S}$, $\tau$ to $10^{-6}$; faces equal within a level, to $10^{-6}$ across levels} & \tabpar{4.4cm}{$1.6\times10^{-7}$ to $2.0\times10^{-7}$; 0; $4.8\times10^{-7}$} \\
2D blast on three levels & \tabpar{7.0cm}{In the emptied center, largest pressure under 20 times the median; $D$ to $10^{-5}$} & \tabpar{4.4cm}{2.4; 0} \\
Rest in $(r, \theta)$ & \tabpar{7.0cm}{$|\mathbf{u}| < 10^{-6}$ after 200 steps, uniform and logarithmic in $r$} & \tabpar{4.4cm}{$2.6\times10^{-7}$, $3.7\times10^{-7}$} \\
Spherical blast & \tabpar{7.0cm}{Shell in 1D, in $(r, \theta)$ and in 3D Cartesian coordinates at one radius to 2 Cartesian cells; the same on every ray in $(r, \theta)$} & \tabpar{4.4cm}{0.280, 0.277, 0.270; 0} \\
Spherical blast on blocks & \tabpar{7.0cm}{Two levels: shell within a finest cell of the uniform grid's on 64 rays; $D$ to $10^{-6}$} & \tabpar{4.4cm}{0 cells; $1.4\times10^{-8}$} \\
Jet breakout & \tabpar{7.0cm}{Jet out of the star along both poles by $t = 2$, reaches within 10\%} & \tabpar{4.4cm}{1.29, 1.27} \\
Jet breakout on blocks & \tabpar{7.0cm}{The same on two levels, reaches within 25\%} & \tabpar{4.4cm}{1.29, 1.41} \\
Reflecting walls & \tabpar{7.0cm}{Blast wave 1 reflected off both walls: $D$ and $\tau$ to $10^{-5}$} & \tabpar{4.4cm}{$7.7\times10^{-8}$, $6.8\times10^{-8}$} \\
\bottomrule
\end{tabular}
\end{table*}

\begin{table*}
\centering
\caption{Table~\ref{tab:validation-relativity}, continued: special relativity with species, in $(R, z)$, with local timesteps and with cooling.}
\label{tab:validation-relativity-2}
\footnotesize
\begin{tabular}{lll}
\toprule
Test & Pass criterion & Measured \\
\midrule
Species & \tabpar{7.0cm}{A pulse carried at 0.9 in gas with $h \approx 3.5$: L1 $< 3\times10^{-3}$, within $10^{-3}$ of the passive scalar; species mass in the 2D blast to $10^{-5}$, on the grid and on blocks} & \tabpar{4.4cm}{$2.4\times10^{-4}$, $1.5\times10^{-6}$; $1.6\times10^{-7}$} \\
Rest in $(R, z)$ & \tabpar{7.0cm}{$|\mathbf{u}| < 10^{-6}$ after 200 steps} & \tabpar{4.4cm}{$8.3\times10^{-7}$} \\
Blast in $(R, z)$ & \tabpar{7.0cm}{Shell within 2 cells of the 1D spherical run's; on blocks within a cell of the grid's; $D$ to $10^{-5}$} & \tabpar{4.4cm}{0.2783, 0.2799; 0 cells; $3.6\times10^{-9}$} \\
Swirl in $(R, z)$ & \tabpar{7.0cm}{$v_\phi$ up to 0.9, 300 steps: $D$, $R S_\phi$ and $\tau$ to $10^{-5}$} & \tabpar{4.4cm}{$1.0\times10^{-8}$, $8.1\times10^{-8}$, $3.6\times10^{-8}$} \\
Rotating column & \tabpar{7.0cm}{In balance, $W$ to 1.2, $t = 1$: $|u_R|$ and the change of $u_\phi$ below $10^{-3}$} & \tabpar{4.4cm}{$1.2\times10^{-4}$, $5.2\times10^{-5}$} \\
Round jet & \tabpar{7.0cm}{$W = 7$ in $(R, z)$: head speed within 15\% of one-dimensional momentum balance} & \tabpar{4.4cm}{0.443, 0.424} \\
Local timesteps & \tabpar{7.0cm}{Bins up to 8 apart: blast wave 2 on three levels as with one step, $\tau$ to $10^{-6}$; 2D blast center under 20 times its median pressure; spherical shell within a cell of one step's} & \tabpar{4.4cm}{L1 0.079; $8.1\times10^{-8}$; 4.6; 0 cells} \\
Cooling & \tabpar{7.0cm}{$\Lambda = \rho^2 T^{1/2}$: at rest the Newtonian decay to $10^{-3}$; at 0.9, $e$ within 1\% and $u$ within $10^{-3}$ of a float64 integration} & \tabpar{4.4cm}{$5.8\times10^{-6}$; $7.0\times10^{-4}$, $6.1\times10^{-4}$} \\
Jet breakout with cooling & \tabpar{7.0cm}{The preset with species for the jet, the star and the medium, on $256^2$ cells in $(r, \theta)$ to $t = 3.5$: species sum to 1 within $10^{-5}$ and lie in $[-10^{-4}, 1 + 10^{-4}]$; the medium's mass to $10^{-4}$ and the star's to $10^{-3}$; the jet's gas out of the star within $10^\circ$ of both poles; the star holds under 0.8 of the heat that it has without cooling; no cell above the speed limit or at a floor} & \tabpar{4.4cm}{$2.3\times10^{-7}$; $3.1\times10^{-7}$, $4.0\times10^{-4}$; reach 2.5, 2.7; 0.65; $W \le 21.8$, none floored} \\
\bottomrule
\end{tabular}
\end{table*}

\begin{table*}
\centering
\caption{The validation suite, continued: the adaptive mesh in depth, the burning network and its hybrid step, the Jeans floor and radiation with a magnetic field. Measured on an NVIDIA RTX~3080.}
\label{tab:validation-added}
\footnotesize
\begin{tabular}{lll}
\toprule
Test & Pass criterion & Measured \\
\midrule
Seven levels of refinement & \tabpar{7.0cm}{A 2D Sedov blast on one root block of 16 cells and seven levels (2048 cells finest): shock radius within 2\% of the cylindrical similarity solution and within 0.5\% of the same blast on four levels; mass and energy to $10^{-6}$; fluxes consistent across levels. Again with per-block timesteps in cycles of 32: the same, with fewer block updates} & \tabpar{4.4cm}{0.25\%, and 0.25\% with timesteps; within $5\times10^{-6}$ of the four-level radius; $6.0\times10^{-8}$, $8.9\times10^{-8}$; 29\% of the block updates} \\
Depth of the multigrid & \tabpar{7.0cm}{An Evrard collapse to $t = 0.3$ on 2, 3 and 4 levels below roots of 64, 32 and 16 cells: last relative residual below $10^{-3}$ at every depth; no more than 2 more V-cycles a step at four levels than at two; energy lost within 1.5 times that at two levels} & \tabpar{4.4cm}{Residual $6.5\times10^{-5}$ to $6.9\times10^{-5}$; 2.03 cycles at every depth; $4.3\times10^{-5}$ lost at every depth} \\
Local timesteps up to 32 & \tabpar{7.0cm}{Blocks stepping up to 32 times less often (five bins) on meshes deep enough to reach them: L1 density error of the Sod tube $< 0.003$, shock radius of the Sedov blast within 2\%, mass and energy $< 10^{-6}$, Sedov under 85\% of the block updates; the Evrard collapse, given four bins (self-gravity stops at 16), loses no more than 1.5 times the energy of one timestep} & \tabpar{4.4cm}{0.0014; 1.6\%; 0, $2.0\times10^{-7}$ and 0, $6.0\times10^{-8}$; 74\%; $6.1\times10^{-4}$ against $6.6\times10^{-4}$} \\
Batched burning network & \tabpar{7.0cm}{48 held-out states, in f32 and with f16 weights: change in composition within 2\% (L1) of the float64 mirror. On 131,072 states, each batched kernel within 2\% of the kernel with one thread per cell and no slower, for the shipped network and a $128\times10$ residual one; f16 activations are timed, not judged} & \tabpar{4.4cm}{0.37\%, 0.37\%; shipped network 36.6~ms a step against 7.2~ms in f32 (5.1 times), 4.9 times with f16 weights; residual network 5.8 times in both} \\
Hybrid burning step & \tabpar{7.0cm}{Against the float64 prototype with the same network and the run's equation of state (Helmholtz with the ions' Coulomb terms): single steps of $10^{-7}$ to $10^{-2}$~s, L1 error $< 0.002$ at the 99th percentile; rollouts to 1~ms of C/O, Si/S ash, hot C/O and NSE with steps of 3~$\mu$s to 1~ms, L1 $< 0.01$ at the 90th percentile; the spread of $X(^{56}{\rm Ni})$ across steps within 0.01 of the prototype's} & \tabpar{4.4cm}{$6.4\times10^{-5}$ (99th), $1.6\times10^{-4}$ at most; $8.8\times10^{-5}$ at most; within $10^{-4}$} \\
Burning watch & \tabpar{7.0cm}{The energy rate that the watch reads, in single precision, within 1\% of the float64 mirror on 107 states of helium, carbon-oxygen and their mixtures, $10^4$ to $10^8$~g~cm$^{-3}$ and $10^8$ to $5\times10^9$~K, where the rate exceeds 1~erg~g$^{-1}$~s$^{-1}$} & \tabpar{4.4cm}{$3.3\times10^{-5}$ at most} \\
Jeans floor on a planet & \tabpar{7.0cm}{The planet of the planet-at-rest test with the Jeans floor on (8 cells to the Jeans length), to $t = 1$: every value finite, mass to $10^{-3}$, central density within 8\%} & \tabpar{4.4cm}{Finite; $6.0\times10^{-4}$; 4.6\%} \\
Radiation in a magnetic field & \tabpar{7.0cm}{Gas and radiation relax in a periodic box with a field whose energy is ten times the gas pressure: temperature at the root of $e(T) + aT^4 = e_0 + E_0$ to $10^{-4}$ on the uniform grid and $10^{-3}$ on blocks; also with gas moving along the field and across a field on blocks} & \tabpar{4.4cm}{$5.2\times10^{-7}$ uniform; $1.9\times10^{-5}$ blocks; 0 moving; 0 across; field energy changes $\le 1.6\times10^{-5}$} \\
\bottomrule
\end{tabular}
\end{table*}

\begin{table*}
\centering
\caption{The validation tests of spherical runs in three dimensions, $(r, \theta, \phi)$ (Section~\ref{sec:curvilinear}), measured on an NVIDIA RTX~3080. All use 64 cells along each axis unless stated.}
\label{tab:validation-spherical-3d}
\footnotesize
\begin{tabular}{lll}
\toprule
Test & Pass criterion & Measured \\
\midrule
Rest in $(r, \theta, \phi)$ & \tabpar{7.0cm}{$|\mathbf{u}| < 10^{-6}$ after 100 steps, uniform and logarithmic in $r$} & \tabpar{4.4cm}{$4.4\times10^{-7}$, $3.4\times10^{-7}$} \\
Spherical blast in 3D & \tabpar{7.0cm}{Shell at one radius on every $(\theta, \phi)$ ray to half a cell, and within half a cell of the $(r, \theta)$ run's} & \tabpar{4.4cm}{0.308 on every ray; $-1.5\times10^{-12}$ cells} \\
Off-center ball & \tabpar{7.0cm}{A ball of radius 0.1 at $r = 0.5$ on the equator, walls on every face: mass and energy to $10^{-6}$ at $\phi = 0$ and $\phi = 0.4$; mirror symmetry under $\phi \to -\phi$ within twice that under $y \to -y$ on a Cartesian grid} & \tabpar{4.4cm}{$3.3\times10^{-9}$, $1.3\times10^{-8}$; $3.1\times10^{-9}$, $1.8\times10^{-8}$; $2.7\times10^{-3}$ against $1.8\times10^{-3}$} \\
Blocks in 3D & \tabpar{7.0cm}{The blast on two levels, 64 cells finest: shell on every leaf ray within a finest cell of the uniform grid's; the off-center ball on two levels, with one timestep and with blocks up to four times apart: mass and energy to $10^{-6}$} & \tabpar{4.4cm}{0.308 on 4100 rays, spread 0; $2.1\times10^{-9}$, $3.0\times10^{-8}$; with bins $2.3\times10^{-9}$, $3.9\times10^{-8}$} \\
Relativistic blast in 3D & \tabpar{7.0cm}{$p = 100$ inside $r = 0.1$, $\gamma = 5/3$, $t = 0.2$: shell on every ray within half a cell and within half a cell of the 2D run's; no cell above the speed limit} & \tabpar{4.4cm}{0.259 on every ray; $-7\times10^{-14}$ cells; Lorentz factor at the limit of 50} \\
Jet breakout in 3D & \tabpar{7.0cm}{The two presets on 128 cells finest and two levels, $t = 2$: the jet's gas out of the star within $10^\circ$ of $+x$; no floored cell, none above the speed limit. With cooling, the species sum to 1 within $10^{-5}$ and lie in $[-10^{-4}, 1 + 10^{-4}]$, the medium's mass to $10^{-4}$, and the star holds less heat than without} & \tabpar{4.4cm}{Reach 1.39, 1.45 with cooling; none floored; $W \le 18.9$, 19.3; $4.5\times10^{-8}$; $2.9\times10^{-8}$; 0.63 of the heat} \\
\bottomrule
\end{tabular}
\end{table*}

\begin{table*}
\centering
\caption{The validation tests against the exact solutions of ExactPack, measured on an NVIDIA RTX~3080. Errors are relative L1 errors, $\sum|{\rm run} - {\rm exact}| / \sum|{\rm exact}|$.}
\label{tab:validation-exact}
\footnotesize
\begin{tabular}{lll}
\toprule
Test & Pass criterion & Measured \\
\midrule
Riemann problems & \tabpar{7.0cm}{Einfeldt 1-2-3, stationary contact, slow shock, shock-contact-shock and LeBlanc at $256^2$, along the middle row: density and pressure errors below 0.025, 0.05, 0.01, 0.01 and 0.02; no floored cell} & \tabpar{4.4cm}{0.0096, 0.015; 0.037, 0.0094; 0.0039, 0.0047; 0.0026, 0.0026; 0.0085, 0.0078; none} \\
Noh implosion & \tabpar{7.0cm}{Planar, cylindrical and spherical, $\gamma = 5/3$, unit inflow, $t = 0.6$, $256^2$: post-shock density and pressure within 1\%, 3\% and 6\% of the exact 4, 16 and 64 (and their pressures); shock within 1.5 cells of $r = 0.2$; pre-shock density within $10^{-4}$ in L1} & \tabpar{4.4cm}{Density 4.00, 15.7, 61.3; pressure 1.33, 5.26, 20.6; shock 0.11, 0.65, 0.84 cells; $3.0\times10^{-9}$, $8.7\times10^{-6}$, $8.7\times10^{-6}$} \\
Noh implosion on a Cartesian grid & \tabpar{7.0cm}{Gas falling in from every direction in the plane, $256^2$: post-shock density within 2\% of 16; shock along the axes and along the diagonal within a cell of 0.2; pre-shock density within $10^{-4}$ of $1 + t/r$ in L1} & \tabpar{4.4cm}{15.9; 0.33 and 0.07 cells; $7.7\times10^{-6}$} \\
Uniform collapse & \tabpar{7.0cm}{Planar, cylindrical and spherical at $256^2$, outer cells held at the exact solution, to $t = 0.6$ (density 2.5, 6.25 and 15.6 times the start): density, pressure and velocity within $10^{-3}$ of the exact solution in L1} & \tabpar{4.4cm}{At most $2.5\times10^{-6}$, $6.6\times10^{-6}$ and $5.5\times10^{-5}$} \\
Coggeshall similarity solutions & \tabpar{7.0cm}{Solutions 1, 2 and 6 (free expansion, expansion against pressure, Kidder compression) on shells from $r = 0.1$ to 1, both radial faces held at the exact solution: density, pressure and velocity within $10^{-3}$ in L1} & \tabpar{4.4cm}{At most $9.9\times10^{-5}$, $1.5\times10^{-4}$ and $4.9\times10^{-5}$} \\
Piston-driven detonations & \tabpar{7.0cm}{Mader and EHEP at $512^2$, $\gamma = 3$, started from the exact solution. Mader: speed within 1\% of CJ; density and pressure within 3\% and velocity within 1\% of the CJ speed. EHEP: density within 2.5\%, pressure within 5\%, velocity within 1\% of the detonation speed} & \tabpar{4.4cm}{Mader: $+0.66$\%; 1.8\%, 1.9\%; 0.42\%. EHEP: 1.3\%, 3.2\%, 0.17\%} \\
\bottomrule
\end{tabular}
\end{table*}

\begin{table*}
\centering
\caption{The validation tests of particles, measured on an NVIDIA RTX~3080. We use isolated self-gravity and near-empty gas where the gas is not the subject.}
\label{tab:validation-particles}
\footnotesize
\begin{tabular}{lll}
\toprule
Test & Pass criterion & Measured \\
\midrule
Tracers on an orbit & \tabpar{7.0cm}{A ring of 64 massless particles at $r = 1$ about a fixed softened mass, over one period: radius within 0.5\% and angle within 0.02~rad of the start, none lost; in 2D and 3D, on the uniform grid and on blocks} & \tabpar{4.4cm}{Radius $\le 8.7\times10^{-6}$; angle $\le 6.9\times10^{-5}$~rad; 64 live in every run} \\
Tracers at the faces & \tabpar{7.0cm}{Tracers crossing 1.25 boxes: in a periodic box all 16 where uniform motion wrapped into the box puts them, to $10^{-4}$ of the box; with open faces all 16 left and counted as lost} & \tabpar{4.4cm}{$1.5\times10^{-6}$; 16 lost} \\
Tracer in a star & \tabpar{7.0cm}{A tracer released at rest at half the radius of an $n = 1$ polytrope, over half an oscillation through its center: within 2\% of the starting distance of the exact orbit, on the uniform grid and on blocks} & \tabpar{4.4cm}{$3.1\times10^{-3}$; $5.2\times10^{-4}$} \\
Two particles & \tabpar{7.0cm}{Two particles of mass 0.5, 16 cells apart, on the circular orbit of their pull, over one period: separation within 2\%, energy within 1\%, momentum below $10^{-3}$ of either one's} & \tabpar{4.4cm}{0.58\%; 0.19\%; $2.2\times10^{-5}$} \\
A point mass and a particle & \tabpar{7.0cm}{The same with a point mass and a particle: separation within 2\%, momentum below $10^{-3}$, each back within 2\% of where it started} & \tabpar{4.4cm}{0.12\%; $6.2\times10^{-4}$; 0.41\%} \\
Plummer sphere & \tabpar{7.0cm}{20,000 particles, mass 1, scale radius 0.1 (6.4 cells), for ten crossing times at $128^3$: the radii holding 10\%, 50\% and 90\% of the mass change by less than 10\%, the energy by less than 2\%} & \tabpar{4.4cm}{2.1\%, 2.0\%, 0.45\%; 0.050\%} \\
Gas and particles falling together & \tabpar{7.0cm}{A cold ball of gas and a Plummer cluster of 5,000 particles of the same mass, at rest 0.6 apart, until they meet: the total momentum below 1\% of the momentum that either gains} & \tabpar{4.4cm}{0.88\%} \\
Across refinement levels & \tabpar{7.0cm}{The Plummer sphere on blocks (128 finest, one level), its inner part refined: radii within 10\%; and the particle pair on blocks, inside the refined level: separation within 2\% and momentum below $10^{-3}$ over a period} & \tabpar{4.4cm}{2.1\%, 2.1\%, 0.41\%; 1.5\%, $2.0\times10^{-6}$} \\
Self-force at a level change & \tabpar{7.0cm}{A particle at rest one coarse cell outside a refined region, for the time in which a pull of $Gm/(4\Delta x_{\rm c}^2)$ would carry it a coarse cell: it moves less than a tenth of a coarse cell} & \tabpar{4.4cm}{$2.4\times10^{-5}$ coarse cells} \\
Restart & \tabpar{7.0cm}{2,000 particles with mass, and two that have left, run to $t = 0.1$, saved to an \texttt{.athdf} file and started again: the live particles identical bit for bit and the dead ones not kept; at $t = 0.2$ the two runs within $10^{-4}$ of the cluster's radius; uniform grid and blocks} & \tabpar{4.4cm}{Identical; $1.2\times10^{-6}$, $3.0\times10^{-6}$} \\
\bottomrule
\end{tabular}
\end{table*}

\begin{table*}
\centering
\caption{The validation tests of sinks, star formation and galaxies (Section~\ref{sec:sinks}), measured on an NVIDIA RTX~3080. The galaxy tests run for one orbit at 8~kpc.}
\label{tab:validation-galaxies}
\footnotesize
\begin{tabular}{lll}
\toprule
Test & Pass criterion & Measured \\
\midrule
Collapse into a sink & \tabpar{7.0cm}{A singular isothermal sphere ($c_{\rm s} = 0.1$) with a central sink, sinks at the Jeans density of four finest cells, to $t = 3$: the central sink holds over 90\% of the sinks' mass within two finest cells of the center; accretion rate from $t = 1$ to 3 between 0.8 and 1.4 times Shu's $0.975\,c_{\rm s}^3/G$; mass to $10^{-4}$, momentum below $3\times10^{-2}$ of mass times $c_{\rm s}$; $64^3$ and blocks (128 finest)} & \tabpar{4.4cm}{One sink with all the mass, 0.04 and 0.6 cells; 1.33, 1.22; $5.3\times10^{-7}$, $2.1\times10^{-7}$; $1.6\times10^{-2}$, $1.5\times10^{-2}$} \\
Accretion luminosity & \tabpar{7.0cm}{One step with and without the luminosity: the gas gains $f\,G M_0\,{\rm d}M/R_*$ more, to 1\%, and the sink takes the same mass to $10^{-4}$} & \tabpar{4.4cm}{Ratio 1.00 ($3.07\times10^{-7}$); the same mass} \\
Molecular cloud & \tabpar{7.0cm}{The preset on 128 cells finest and one level: without turbulence the first sink within 0.7 to 1.4 free-fall times; with $\alpha = 1$, sinks within one free-fall time holding 10\% to 95\% of the mass by two, gas and sinks keeping their mass to 3\%; the same seed makes the same sinks to $10^{-5}$} & \tabpar{4.4cm}{1.0; 11 sinks at one free-fall time, 13 at two, holding 88\%; $-2.5$\%; the same 11 sinks} \\
Tracers in the galaxy & \tabpar{7.0cm}{Rings of 32 tracers at 2, 8 and 14~kpc on the circular orbits of the reference galaxy's potential: radius within 0.3\% and angle within 0.01~rad of uniform circular motion, in 2D on the uniform grid and in 3D on blocks} & \tabpar{4.4cm}{$4.9\times10^{-5}$, $1.3\times10^{-4}$~rad; $1.2\times10^{-3}$, $6.3\times10^{-4}$~rad} \\
Gas disk at rest in the potential & \tabpar{7.0cm}{Two orbits on 512 cells: angular momentum within 1\%, surface density at 4, 8 and 12~kpc within 5\%; the disk in 3D on blocks: angular momentum within 2\%, surface density within 10\%} & \tabpar{4.4cm}{$2.0\times10^{-4}$; $2\times10^{-5}$, $4\times10^{-5}$, 0.42\%; 3D 0.52\%, 5.5\%, 0.49\%, 0.19\%} \\
Spiral shocks & \tabpar{7.0cm}{Two arms of pitch $15^\circ$ and 5\% of the radial pull, two turns of the pattern at 512 cells: at 6~kpc the $m = 2$ amplitude over 0.1 and a thousand times $m = 1$ and 3; largest density over 1.5 times the least; axisymmetric preset varying by under 2\%; mass kept to 2\%} & \tabpar{4.4cm}{0.20, $6\times10^{-6}$; 2.1; 1\%; $-0.7$\%, $-1.4$\%} \\
Stable self-gravitating disk & \tabpar{7.0cm}{Toomre $Q = 3$, 156~pc, three levels: angular momentum within 12~kpc kept to 2\%, surface density at 5 to 11~kpc to 10\%, rotation speed to 3\%; $Q$ near 3 (2.5 to 3.5) at 8~kpc} & \tabpar{4.4cm}{$-0.44$\%; 8.9\%; 0.78\%; 3.14} \\
Growth of arms with $Q$ & \tabpar{7.0cm}{With 1\% seed perturbations: at $Q = 0.7$ the non-axisymmetric amplitude of the surface density ($m = 1$ to 6) grows at least 2.5-fold and at least 1.5 times as much as at $Q = 2$} & \tabpar{4.4cm}{3.6-fold; 1.8-fold at $Q = 2$ (1.95 times)} \\
Live stellar disk & \tabpar{7.0cm}{500,000 disk and 125,000 bulge particles in the stable gas disk. At $Q = 2$: angular momentum within 2\%; at 4, 8 and 12~kpc surface density within 10\%, mean height 25\%, radial dispersion 25\%; rotation at 8~kpc within 5\%. At $Q = 1.2$: modes $m = 2$ to 6 grow over 2-fold and over 1.5 times as much as at $Q = 2$} & \tabpar{4.4cm}{0.077\%; 3.1\%, 7.9\%, 4.2\% at most; $-1.0$\%; 5.7-fold, 2.8-fold at $Q = 2$ (2.0 times)} \\
Stars by the Schmidt law & \tabpar{7.0cm}{A uniform periodic box, efficiency 0.1 per free-fall time, for 0.01: the stars' mass within 5\% of $\epsilon M t/t_{\rm ff}$ (about 4000 stars); with supernovae of 10 per unit mass, the gas gains their energy less the share that the stars took, to 5\%} & \tabpar{4.4cm}{3870 stars, 0.973 of the mass; energy 1.01 of that expected} \\
Delayed supernovae & \tabpar{7.0cm}{Supernovae spread evenly from 0.002 to 0.006 after each star's birth: at $t = 0.0018$ none has released energy (under 1\% of the energy due by the end); at $t = 0.008$ the gas has gained the sum over the stars of the energy released so far, to 5\%} & \tabpar{4.4cm}{$8.7\times10^{-6}$ of the energy due; 0.989 of that due} \\
Supernova mass return & \tabpar{7.0cm}{Eight stars in near-empty gas return 10\% of their mass with their supernovae: the gas gains what the stars lose, to $10^{-3}$, and between $1 - e^{-0.1}$ and 0.1 of their mass; the hottest gas stays at or below the ejecta's own temperature} & \tabpar{4.4cm}{Balance exact; 0.095 of the mass; maximum 50 against 11,700 without the return} \\
\bottomrule
\end{tabular}
\end{table*}

\clearpage
\bibliographystyle{aasjournalv7}
\bibliography{references}

\end{document}